\pdfoutput=1
\documentclass[journal=iecred,manuscript=article]{achemso}
\usepackage[utf8]{inputenc}
\usepackage[T1]{fontenc}
\usepackage{lmodern}
\usepackage{amsmath,amssymb}
\usepackage{array}
\usepackage{booktabs}
\usepackage{threeparttable}
\usepackage{makecell}
\usepackage{multirow}
\usepackage{adjustbox}
\usepackage{graphicx}
\usepackage{xcolor}
\usepackage{xurl}
\usepackage{siunitx}
\usepackage{placeins}
\usepackage{algorithm}
\usepackage{algpseudocode}
\graphicspath{{figs/}}
\setkeys{acs}{maxauthors = 0, articletitle = true, chaptertitle = true}
\SectionNumbersOn
\AtBeginDocument{\singlespacing}

\author{Erbet Almeida Costa}
\email{erbetcosta@puffin-hub.com}
\affiliation{Puffin Hub AS, Sluppenvegen 6, Trondheim, Norway}
\author{Galdir D. Reges}
\affiliation{CTAI, Federal University of Bahia (UFBA), R. Aristides Novis 2, 40210-630 Salvador, Bahia, Brazil}
\author{Te\'ofilo Paiva Guimar\~aes Mendes}
\affiliation{Federal University of Rec\^oncavo da Bahia (UFRB), Cruz das Almas, Bahia, Brazil}
\author{Carine Menezes Rebello}
\affiliation{Puffin Hub AS, Sluppenvegen 6, Trondheim, Norway}
\author{Odilon Santana Luiz de Abreu}
\affiliation{CTAI, Federal University of Bahia (UFBA), R. Aristides Novis 2, 40210-630 Salvador, Bahia, Brazil}
\author{Vin\'icius Viena Santana}
\affiliation{Department of Chemical Engineering, Norwegian University of Science and Technology (NTNU), 7491 Trondheim, Norway}
\author{Francisco Raphael Ribeiro Bacellar}
\affiliation{Petrobras, Av. Hor\'acio Macedo 950, Cidade Universit\'aria, Rio de Janeiro, Brazil}
\author{Gabriel Shen}
\affiliation{Petrobras, Av. Hor\'acio Macedo 950, Cidade Universit\'aria, Rio de Janeiro, Brazil}
\author{M\'arcio Fontana}
\affiliation{CTAI, Federal University of Bahia (UFBA), R. Aristides Novis 2, 40210-630 Salvador, Bahia, Brazil}
\author{Marcos Pellegrini Ribeiro}
\affiliation{Petrobras, Av. Hor\'acio Macedo 950, Cidade Universit\'aria, Rio de Janeiro, Brazil}
\author{Leizer Schnitman}
\affiliation{CTAI, Federal University of Bahia (UFBA), R. Aristides Novis 2, 40210-630 Salvador, Bahia, Brazil}
\author{Idelfonso B. R. Nogueira}
\email{idelfonso.b.d.r.nogueira@ntnu.no}
\affiliation{Department of Chemical Engineering, Norwegian University of Science and Technology (NTNU), 7491 Trondheim, Norway}
\alsoaffiliation{Puffin Hub AS, Sluppenvegen 6, Trondheim, Norway}

\title{Data-Driven Multi-Rate Predictive Control with Echo State Networks for Safe Production Maximisation of a Subsea Electrical Submersible Pump: From Historical Data to Field Deployment}

\keywords{Data-driven model predictive control, echo state networks, electrical submersible pump, subsea production optimisation, multi-rate control, real-time optimisation}

\begin{document}

\begin{abstract}
Subsea electrical submersible pumps must be driven towards their productive limit while kept away from trip conditions, several of which move with the operating frequency. This work presents a hybrid data-driven model predictive controller in which an Echo State Network propagates ten measured variables and a phenomenological map estimates the unmeasured oil flow rate. The controller is multi-rate, with inference every \SI{10}{s} and optimisation every \SI{30}{s}. The frequency-dependent alarm limits enter as time-varying constraints, and a proximity penalty prices the approach to each limit so that the well can operate on an active constraint. The path from historical data to the field is documented: data engineering, model screening under free-run validation, operating maps, and tuning in a simulation environment. Six field campaigns were carried out over eighteen months. The final test delivered a same-day oil production gain of \SI{6.2}{\percent}, with every measured variable inside its limits.
\end{abstract}

\section{Introduction}
\label{sec:intro}

Oil production faces persistent challenges of well stabilisation, production optimisation and recovery-factor maximisation \cite{Foss2012,Foss2018}, and deep-water environments add energy management and asset integrity to them. Electrical submersible pumping (ESP) has consolidated itself as one of the most effective artificial-lift methods for deep-water fields, and the subsea variant places the pump on the seabed, where any intervention requires a vessel and any failure is expensive \cite{Takacs2018}. Operating such a well is therefore an exercise in restraint. The pump must be pushed towards its productive limit and, at the same time, kept away from a set of trip conditions that would shut the well in.

These limits are not few. Motor temperature, torque and total current, intake pressure (to avoid cavitation), vibration, and the admissible axial-thrust range each define a boundary, and several of them move with the operating frequency. The result is a dynamic safety window inside which the well must be operated. Conventional decentralised proportional--integral--derivative (PID) loops handle each variable in isolation and cannot anticipate a constraint violation in one channel caused by a move in another. Model predictive control (MPC) \cite{Qin2003} with a first-principles well model \cite{Pavlov2014} would address the coupling, but multiphase well models are computationally heavy for real-time use and their calibration inherits the geological uncertainty of the reservoir.

Data-driven MPC (DDMPC) offers a different route: identify the fast dynamics directly from operational data and keep the physics where it is reliable, in the static production and thrust relationships. The term is used here for an MPC whose prediction model is identified from data, and not in the behavioural sense of data-enabled predictive control \cite{Coulson2019,Berberich2021}. Echo State Networks (ESNs) \cite{Jaeger2004,Lukosevicius2009} are attractive for the dynamic part because their training reduces to linear regression and inference reduces to one matrix--vector product per step, properties that make them viable inside a real-time optimisation loop \cite{Jordanou2018,Jordanou2022}. A network identified for one-step prediction is, however, not necessarily fit for use along a prediction horizon \cite{Rebello2022guideline,Santana2025multistep}, and this distinction shapes the model selection reported here.

This article extends the conference paper by the same group \cite{Reges2026}, which reported the controller concept and the December 2025 field result in eight pages. What was compressed there is developed here: the data engineering that separated three years of exploratory historian records from the uniform-rate acquisition used for identification, the model benchmarking and selection protocol, the operating maps, the complete optimisation formulation, the simulation environment, the tuning study, and the full multi-campaign field validation, including seven weeks of sustained operation after the production-gain test.

\subsection{Related work}
\label{sec:intro:related}

The use of ESNs in the predictive control of oil wells was proposed by Jordanou et al. \cite{Jordanou2018}, who applied an ESN-based MPC to a simulated well. The same group extended it to a practical MPC framework for unknown dynamic systems \cite{Jordanou2022} and studied online-learning ESN control of an oil production platform \cite{Jordanou2019}. Physics-informed neural networks were later proposed by them as control models \cite{Antonelo2024}. Armenio et al. \cite{Armenio2019} designed an MPC for dynamical systems learned by ESNs, and the use of recurrent networks in learning-based control was surveyed by Bonassi et al. \cite{Bonassi2022}; learned models inside MPC have also been treated for process applications \cite{Ren2022,Wu2019} and for safety \cite{Hewing2020}. Data-driven models of ESPs have been built with Bayesian \cite{Costa2021Bayes} and deep neural network \cite{AlmeidaCosta2024} approaches, and soft sensors based on nonlinear state estimators were evaluated experimentally \cite{deAbreu2024}. On operational data of an offshore field, ESNs were compared with long short-term memory (LSTM) networks \cite{Rebello2025RNN}, and LSTM networks were trained for multi-step-ahead prediction \cite{Santana2025multistep}. What a controller needs from such a model is accuracy along the horizon, when the network is fed its own predictions, and this is the criterion adopted here for the selection of the network (Section~\ref{sec:evalproto}).

Predictive control of ESP-lifted wells has a literature of its own. Pavlov et al. \cite{Pavlov2014} proposed a dynamic model of the well together with an MPC tested on a large-scale test facility, and a robustness analysis was later presented for heavy oil wells \cite{Krishnamoorthy2016}. Binder et al. \cite{Binder2014} implemented a linearised ESP controller on a programmable logic controller (PLC), and Patel et al. \cite{Patel2019} reported advanced process control automating conventional ESP operations across an onshore field. An MPC with an adaptive strategy was applied by Delou et al. \cite{Delou2019} to a model of an ESP in a subsea environment. Closer to the constraint-handling question addressed here, Santana et al. applied an adaptive infinite-horizon MPC with zone control \cite{Santana2021} and a zone nonlinear MPC \cite{Santana2022} to ESP-lifted wells, the latter coupling a state estimator to cope with the absence of measurements. On the ESP rig of the Artificial Lift Laboratory at UFBA, a standard MPC and an input-target MPC ran in real time \cite{Costa2025IFACMPC} inside a legacy supervisory architecture \cite{Costa2026legacy}. In simulation, a nonlinear MPC was tested on a pilot-scale ESP \cite{Rebello2025NMPC}, formulations that guard against model uncertainty with sets of symbolic-regression models were tested on ESP systems \cite{MenezesRebello2026,Rebello2026minmax}, and a real-time optimisation layer on steady-state surrogate models was placed above an advanced regulatory control structure \cite{Costa2026RTO}. Apart from the onshore deployment of Patel et al., these controllers were evaluated in simulation or on test rigs.

The way a limit enters the optimisation problem is the second line of work this article draws on. Softening a constraint with a penalised slack variable is the usual remedy for infeasibility \cite{deOliveira1994,Scokaert1999}, and zone control penalises an output only once it leaves its admissible band \cite{Gonzalez2009,Ferramosca2010}, which is the tradition the ESP formulations above follow. In barrier-function MPC the cost itself grows as a constraint is approached \cite{Wills2004,Feller2017}. The proximity penalty of Section~\ref{sec:mpc:cost} is a bounded relative of this last idea, used alongside hard constraints: no decision variable is added, the cost is written on the percentage distance to a limit that can move with the frequency, and one function serves every variable and both alarm sides. A slack-softened constraint placed in front of a hard one would give a comparable two-tier structure; what is specific here is the shape of the cost inside the band and its single percentage-based form for eleven variables.

Krishnamoorthy et al. \cite{Krishnamoorthy2019} showed that optimal operation of oil and gas production can be achieved with simple feedback control structures, a line of work built on self-optimising control \cite{Skogestad2000} and on advanced control from simple elements \cite{Skogestad2023}; feedback control as a route to optimal operation is discussed by Engell \cite{Engell2007}. In the well studied here eleven variables carry limits, several of which move with the frequency, and the objective is not measured, and this is where an online constrained optimiser earns its cost. In economic MPC \cite{Ellis2014,Angeli2012} the economic objective enters the controller directly, which has been demonstrated with a learnt model in simulation on a pressure swing adsorption unit \cite{Martins2021}. The controller reported here is closer to the two-layer practice of real-time optimisation and MPC \cite{Darby2011}: it tracks a production target computed from a static map, and the active constraints decide where the well settles.

The oil flow rate of the well is not measured online. Estimating it from the measured variables is the virtual flow metering problem \cite{Bikmukhametov2020}, addressed with Bayesian neural networks \cite{Grimstad2021} and gray-box models \cite{Hotvedt2022} and, for ESP wells, by moving horizon estimation \cite{Binder2015}. A simpler route is taken here, a static map recalibrated against periodic production tests.

Any data-driven controller also depends on how its data were acquired. Thornhill et al. \cite{Thornhill2004} quantified the impact of historian compression on data-driven analyses, and Section~\ref{sec:data} shows what it does to dynamic identification in this well.

A different route is to learn the decision policy itself. Reinforcement learning has been reviewed for process control \cite{Badgwell2018,Nian2020} and demonstrated as a controller by Spielberg et al. \cite{Spielberg2019}; a control-informed structure was proposed by Bloor et al. \cite{Bloor2025}. A producing well does not allow training by exploration, and its limits move with the operating point. For this reason the learned component is confined here to the model, and the decision is left to a constrained optimiser.

The architecture adopted here also sits inside a longer tradition. Combining a
first-principles description of what is well understood with a data-driven description of
what is not goes back to the first hybrid neural network models \cite{Psichogios1992,Thompson1994} and has since been the subject of the hybrid semi-parametric modelling literature \cite{vonStosch2014}, and the ways of combining the two components, that is,
where the boundary between them is placed, were recently reviewed by Schweidtmann et al. \cite{Schweidtmann2024}.  What
distinguishes the present case is which side of the boundary the economics falls on. In this
application the objective is carried by the physics, as a static map recalibrated against
production tests, and the network is confined to the dynamics. Apart from the conference paper that precedes this one \cite{Reges2026}, the authors know of no MPC built on a learned dynamic model that has been run in closed loop on a producing subsea ESP well.

\subsection{Relation to previous work of the authors}
\label{sec:intro:group}

This article belongs to a line of research of the authors on learned models for the optimisation and control of process units, and four of its design choices come from that experience.

The network is selected in free run because earlier studies separated models meant for simulation from models meant for prediction \cite{Rebello2022guideline} and examined the gap between one-step training and multi-step use of a network \cite{Costa2024multistep,Santana2025multistep}. The uncertainty of such models was evaluated for the pump \cite{AlmeidaCosta2024} and within a general learning methodology \cite{Costa2022robust}; in the controller, the uncertainty of the flow estimate is what sizes the margin applied to that channel. The physics was placed outside the network. In earlier hybrid models the learned part sits inside the differential equations, as universal differential equations for multicomponent adsorption \cite{Nogueira2022UDE} and for batch crystallisation \cite{Lima2023UDE}, or corrects the phenomenological model of an ESP prototype \cite{Rebello2024interp}; here the network carries the dynamics and a static map carries the production estimate. The optimisation is kept online, in the supervisory layer, because the limits and the map change during operation. Explicit and approximate MPC \cite{Bemporad2002,Karg2020}, which some of the authors have embedded in a PLC before \cite{Mendes2022TSK}, would fix the control law offline, and a reinforcement learning policy of the kind used where candidate solutions can be evaluated at will, as in the design of flavour molecules \cite{Queiroz2023RL}, would have to be trained by exploration.

Finally, the operating map is read as a region and not as a point, as in the feasible operating regions drawn for a true moving bed \cite{Nogueira2019SPSO} and for the simultaneous screening of adsorbents and optimisation of the process \cite{Nogueira2022nested}: the engineer sees where the well may be placed and how far the current point is from each boundary before any optimiser is run (Section~\ref{sec:maps}). The dependence of data-driven operation on how a pumping system is instrumented, reviewed by one of the authors for water distribution networks \cite{Sanchez2020water}, is taken up for the well in Section~\ref{sec:data}.

\subsection{Contributions}
\label{sec:intro:contrib}

The main contributions of this work are:
\begin{itemize}
\item[C1] A data-engineering account that separates two sources usually conflated in the applied literature. Three years of plant-information (PI) historian records, 2.96 million samples per variable, are used for exploratory characterisation only, because the compression applied by the historian reorders events and can invert cause and effect between coupled variables, which disqualifies the record for dynamic identification. The identification data sets come instead from direct acquisition through the Open Platform Communications (OPC) interface at \SI{10}{s}, the fastest rate the supervisory platform sustains, and the article documents the removal of frozen readings from the historian record and the use of acceleration and deceleration ramps down to \SI{40}{Hz} as excitation in the OPC record (Section~\ref{sec:data}).
\item[C2] A multi-input multi-output nonlinear autoregressive model with exogenous inputs (NARX) in which two manipulated inputs drive ten fed-back process channels, so that every measured variable is simultaneously a predicted output and a regressor of all the others. The design rationale is that this redundancy, rather than regressor pruning, supports the free-run accuracy the controller depends on: an error in one channel is constrained by the remaining nine through the couplings the network has learnt (Section~\ref{sec:model}).
\item[C3] A large-scale ESN screening and selection protocol combining one-step-ahead and free-run (output-feedback) validation against field data, free-run error histograms, and a justified adoption of the mean absolute percentage error (MAPE) as the ranking metric across ten heterogeneous outputs; the operating-point histograms (Figure~\ref{fig:datahist}) show where the identification data are dense and where the well had little operating history, which includes the maximum-flow region the controller was required to reach; the field campaigns of Section~\ref{sec:field} operated in that region with every monitored variable inside its limits. Model ranking is reported under two aggregations, the mean across channels and the worst channel, and the selection follows the worst channel, with the vibration channel treated as described in Section~\ref{sec:eval:vibration} (Sections~\ref{sec:evalproto} and~\ref{sec:eval}).
\item[C4] The reconstruction and exploration of the operating maps. The maps themselves are Petrobras engineering artefacts and are not claimed here. What is contributed is their reconstruction from a sensitivity analysis of the multiphase well simulator, their interrogation through iso-flow curves, the flow surface and a steepest-ascent path over it, and their presentation in normalised form with percentage margin lines to the thrust boundaries, together with an explicit account of the flow-estimation uncertainty that motivates those margins (Section~\ref{sec:maps}).
\item[C5] The complete optimisation formulation as implemented in the field, including elements absent from the conference paper: prediction-error compensation of the set-point, move blocking and the pressure-target band (Section~\ref{sec:mpc}).
\item[C6] A constraint-handling scheme in which the process limits, derived from the platform cause-and-effect matrix that separates trip-inducing from alarm-only variables, enter the controller as time-varying, frequency-dependent constraints rather than fixed limits; a two-tier margin structure, configured per process variable rather than as a single global tolerance, has its inner tier enforced as hard constraints and its outer tier priced by a proximity penalty on the approach of the plant to its constraints. The penalty is a fourth-order polynomial fitted to an exponential of the percentage distance to the limit, which lets one single function serve all eleven variables and both the high and low alarms. Unlike slack-variable and zone formulations, which charge for a violation once the softened limit is crossed \cite{Gonzalez2009,Santana2021,Santana2022}, it charges for proximity before the limit is reached, while the limit itself stays hard (Sections~\ref{sec:system} and~\ref{sec:mpc}).
\item[C7] A reproducible description of the simulation environment (plant emulation by ESN, protection layer, noise and instrument resolution) that supported tuning and commissioning (Section~\ref{sec:hil}).
\item[C8] A systematic tuning study covering weights, soft margins, targets, prediction horizon and initialisation, whose central finding is that the active constraints and the target definition, not the weight tuning, determine the achievable production (Section~\ref{sec:tuning}).
\item[C9] The eighteen-month, six-campaign field validation history with its analyses of set-point variability and disturbance response, of which \cite{Reges2026} reported only the final production-gain test (+6.2\,\%) (Section~\ref{sec:field}).
\end{itemize}

The remainder of the article separates the methodology from the results. Section~\ref{sec:method} is methodological: it describes the asset and formalises the control problem (Section~\ref{sec:system}), the data engineering (Section~\ref{sec:data}), the dynamic modelling (Section~\ref{sec:model}), the model evaluation protocol (Section~\ref{sec:evalproto}), the production estimate and the operating maps (Section~\ref{sec:maps}), the controller formulation (Section~\ref{sec:mpc}), and the simulation environment together with the experimental and field protocols (Section~\ref{sec:hil}). Section~\ref{sec:results} reports and discusses the results: the benchmarking and selection of the dynamic model (Section~\ref{sec:eval}), the closed-loop tuning study (Section~\ref{sec:tuning}) and the field validation (Section~\ref{sec:field}). Section~\ref{sec:conc} concludes.

\section{Methodology}
\label{sec:method}

\subsection{System description and control problem}
\label{sec:system}

\subsubsection{The subsea ESP well (7-JUB-27, P-57)}
\label{sec:system:well}

\begin{figure}[htbp]
  \centering
  \includegraphics[width=\linewidth]{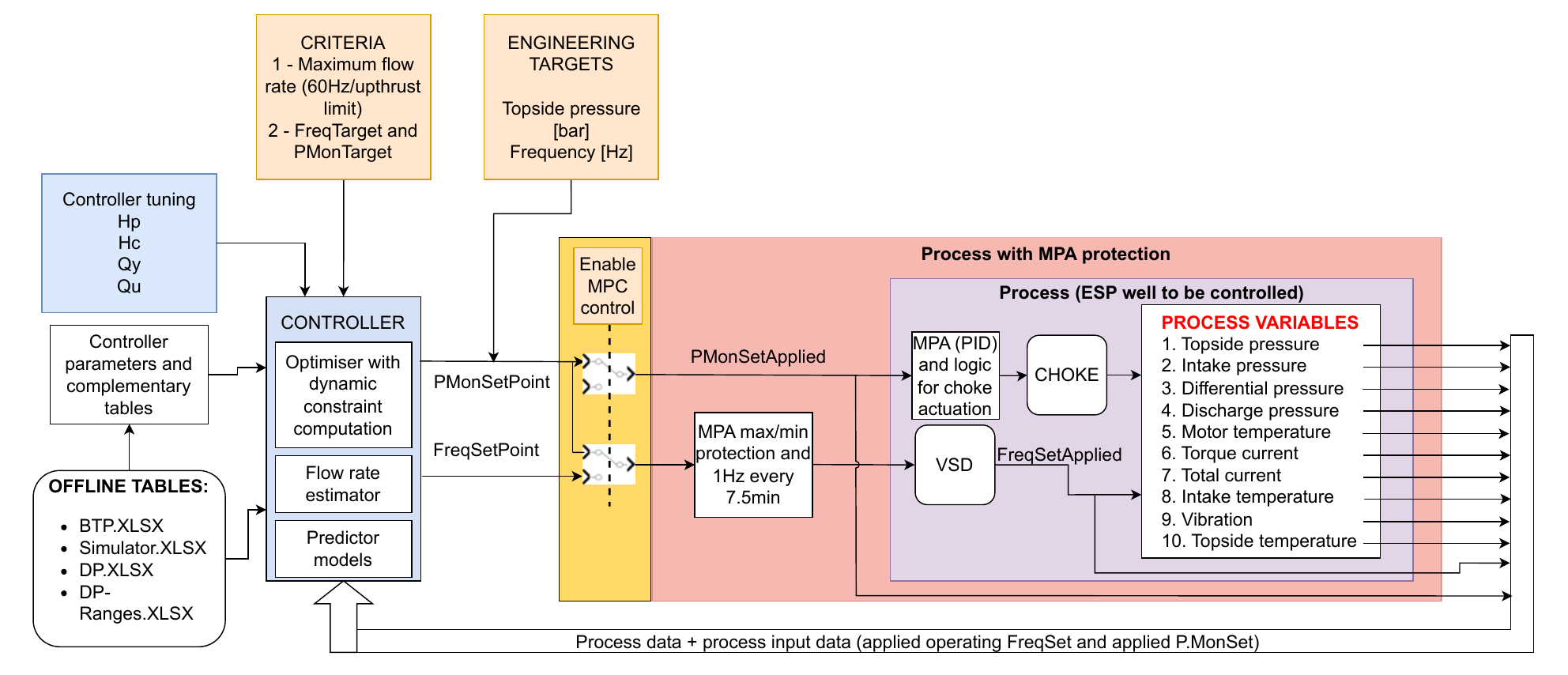}
  \caption{General architecture of the DDMPC system: the hybrid ESN-plus-phenomenological model and the multi-rate supervisory optimiser interacting with the platform regulatory layer.}
  \label{fig:arch}
\end{figure}

The well is equipped with a subsea ESP of the MOBO type, installed on the seabed and fed from the P-57 platform. Two degrees of freedom are available to the operation: the frequency set-point of the variable-speed drive (VSD) and the topside pressure target, the latter realised by the production choke. The well produces with a low free-gas fraction at the pump intake and has no record of gas locking over the period analysed, which is why the gas-handling limits that often dominate ESP operation are not among the binding constraints of Section~\ref{sec:system:vars}. This is a characteristic of this well. Figure~\ref{fig:arch} places the controller in this context: it acts as a supervisory layer that issues both set-points to the platform regulatory loops, and does not manipulate the drive or the choke directly.

\subsubsection{Monitored variables and the dynamic safety window}
\label{sec:system:vars}

Eleven quantities define the well's state for control purposes. Ten are measured:
the intake pressure, the topside pressure, the differential pressure, the discharge
pressure, the motor temperature, the torque and total motor currents, the intake
temperature, the pump vibration and the topside temperature. The eleventh, the oil flow
rate, is not measured online and is estimated from the maps of Section~\ref{sec:maps},
but it is measured during the monthly tests performed in the well. In this order, the
eleven quantities form the state vector $x \in \mathbb{R}^{11}$ used from
Section~\ref{sec:mpc} onward, so that the topside pressure is its second component and
the estimated flow rate its eleventh; Table~\ref{tab:margins} lists the variables in the
same order.

Each of these carries a protection band, and the bands do not all come from the same
place. Three variables have limits that do not move with the operating point, because
the equipment tolerance that sets them does not: the motor temperature, the intake
temperature and the vibration. Seven others are protected by a table that Petrobras
maintains as a reference value per frequency together with the percentage offsets that
raise the low-low (LL), low (L), high (H) and high-high (HH) alarms, so their limits are recomputed whenever the frequency
set-point changes. The torque current is the clearest case. Its reference grows
monotonically with frequency across the whole operating range, which means that a
current perfectly acceptable at \SI{57}{Hz} would raise an alarm at \SI{45}{Hz}.
Table~S1 reports the resulting bands, min--max normalised, at three
representative set-point frequencies.

Two variables have limits from a third source, and this is where the axial-thrust
requirement enters the controller. The admissible oil flow rate and the admissible
topside pressure are read off the operating maps of Section~\ref{sec:maps}, which encode
the upthrust and downthrust boundaries of the pump. For the topside pressure the two
sources are combined conservatively, the enforced band being the intersection of the
alarm band with the thrust band. The intake pressure is protected only by its alarm band
in the implementation reported here, even though the maps also return bounds for it. That
was a commissioning choice: the map-based intake bound was found to be the
more conservative of the two by a margin that cost production without a corresponding
integrity benefit, so it was left out of the enforced set and kept as a monitored
quantity. Two further channels of the Petrobras protection table, the christmas-tree
pressure measured by the TPT and the downhole pressure measured by the PDG, are not used
as constraints, because neither is available to the supervisory platform at the
sampling rate the controller needs.

The distinction that shapes the horizon of Section~\ref{sec:tuning:hp} is between the two
alarm tiers. The platform cause-and-effect matrix separates variables whose HH or LL
alarm commands a shutdown from variables that only annunciate. The controller is
constrained at the H and L levels, one tier below the trip, so the enforced limit always
leaves the shutdown logic untouched. What the controller buys with that arrangement is
time: a variable that crosses H is still recoverable, and the horizon is long enough for
the optimiser to see the crossing coming. Section~\ref{sec:mpc:cost} describes how the
margin is applied on top of the H and L values, and Table~\ref{tab:margins} gives the
margin adopted for each variable.

The union of these bands over the admissible frequency range is what this article calls
the dynamic safety window. It is not a box. Its faces move with the frequency, some of
them come from a static equipment tolerance and others from a thrust map that itself
depends on the water cut, and one of its faces, the flow rate, is bounded on a quantity
that is estimated rather than measured. Operating close to that face is exactly what the
economic objective demands, and it is why the margin structure of
Section~\ref{sec:mpc:cost} is configured per variable rather than as one global
tolerance.

\subsubsection{Control objective}
\label{sec:system:obj}

The objective is to maximise the oil flow rate of the well while keeping every one of the
eleven variables of Section~\ref{sec:system:vars} inside the dynamic safety window.

Stated in those terms the problem sounds like a standard constrained economic
optimisation, and in one sense it is. Two features of this particular well make the
formulation of Section~\ref{sec:mpc} less obvious than the statement suggests. The
quantity being maximised is not measured, so the objective is evaluated on an estimate
whose bias is discussed in Section~\ref{sec:maps} and whose uncertainty sizes the margin
applied to that channel. And the maximum does not lie in the interior of the admissible
region. Following the flow gradient from any admissible starting point leads to a
boundary, as {Section~\ref{sec:maps}} shows, so the economic optimum sits on
an active constraint by construction and the controller has to be able to operate there
without provoking a trip.

The two degrees of freedom available to reach it are the VSD frequency set-point and the
topside pressure target, both delivered to the platform regulatory layer rather than to
the actuators. The operation retains authority over the frequency it wants, and the
controller treats that value as an engineering preference instead of a set-point to be
tracked, for the reason Section~\ref{sec:maps} explains: the operating point the well
settles on follows from the active constraints and from the relative weight on the flow
channel. Where the two conflict, the constraints win.

\subsection{Historical data engineering and exploratory analysis}
\label{sec:data}

This section describes how the industrial record was converted into data sets suitable for identification, and why two different acquisition paths had to be kept apart.

The exploratory starting point was the PI historian of the well, covering 2021 to 2023 and totalling 2\,963\,529 samples per variable. The record is dominated by steady operation. The well spent most of the period at two frequency plateaus, around \SI{43}{Hz} and \SI{55}{Hz}, with the production choke close to \SI{30}{\percent} opening. The record also contains 18 shutdown and restart events. A stationarity screening classified 97.79\,\% of the record as steady, and 14\,\% of all samples, counted separately, were discarded as frozen readings from instrumentation or historian faults.

Although the PI record informed the exploratory analysis, it cannot be used to identify dynamics, and the reason deserves to be stated plainly. A PI historian stores samples under a compression policy that keeps a value only when it departs from the previous one by more than a configured deadband. The timestamps that survive are therefore not a uniform sampling of the process: two coupled variables are compressed independently, each according to its own noise and deadband, so the recorded order of two events can be the reverse of the physical one. In the records examined here the arrival pressure occasionally appears to move before the choke that caused it. A model fitted to such a series learns the wrong causal direction, and no amount of resampling repairs it, because the information was discarded at acquisition; compression is known to interfere with data-driven analyses \cite{Thornhill2004}.

The identification data therefore come from direct OPC acquisition at \SI{10}{s},
described in Section~\ref{sec:mpc:multirate}, collected during continuous operation conditions in
which frequency ramps, choke moves, shutdowns and restarts were commanded.
The acquisition ran from 19 January to 19 July 2024, 182 days, and retained
\num{1419416} samples of each of 31 tags. Two properties of this record matter more than
its size. It was acquired at a uniform period rather than under a deadband, so the
timestamps are a faithful ordering of the events. And it contains transients that were
commanded rather than waited for, including deceleration ramps down to \SI{40}{Hz},
which no amount of passive observation of a well held at two frequency plateaus would
have produced. Windows containing genuine transients were selected manually, and
training, validation and test sets were split at window level rather than sample level,
so that no window contributes to more than one set.

\begin{figure}[htbp]
  \centering
  \includegraphics[width=\linewidth]{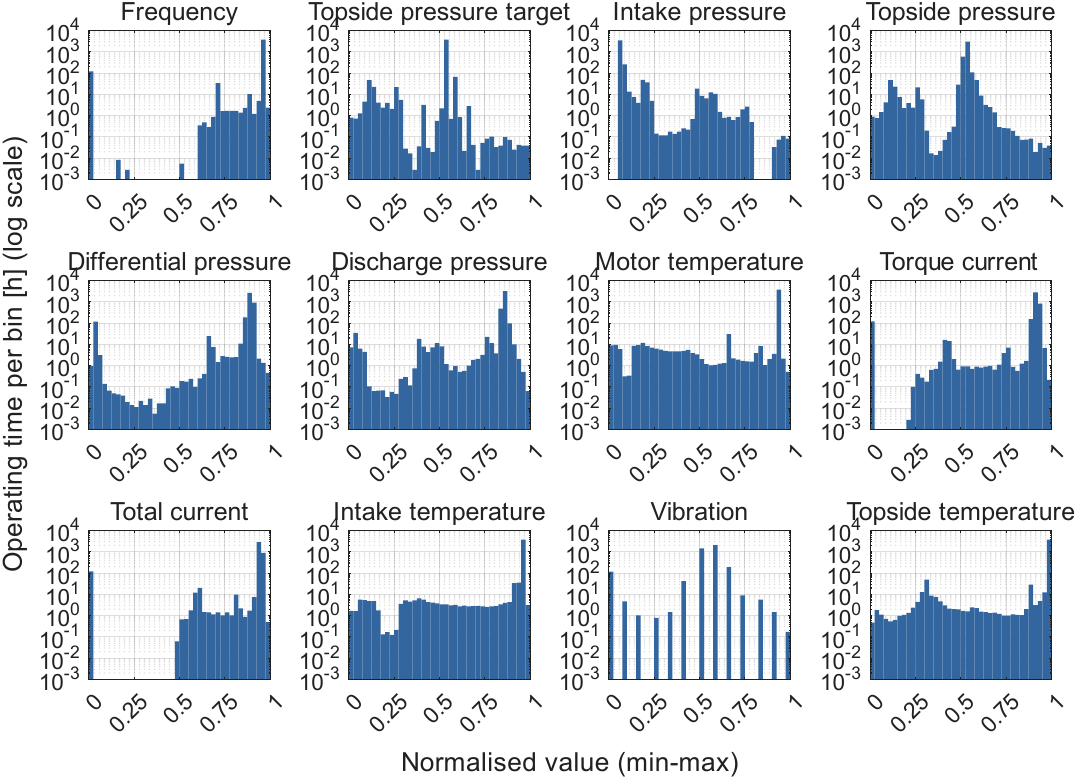}
  \caption{Operating-point histograms of the OPC record, used to identify the predominant steady-state operating points.}
  \label{fig:datahist}
\end{figure}

A stratified form of the data is presented in Figure~\ref{fig:datahist} to identify the predominant steady-state operating points. The figure shows the operating-point histograms on which the first OPC data analysis was based. Figure~\ref{fig:datahist} is particularly interesting because it shows the system operated in the same steady state for most of the time. Its ordinate is the operating time accumulated in each bin, on a logarithmic scale: the dominant plateaus reach the order of \SI{e3}{\hour} while the rarely visited regions amount to a few minutes each (\SI{e-3}{\hour}).

\begin{figure}[htbp]
  \centering
  \includegraphics[width=\linewidth]{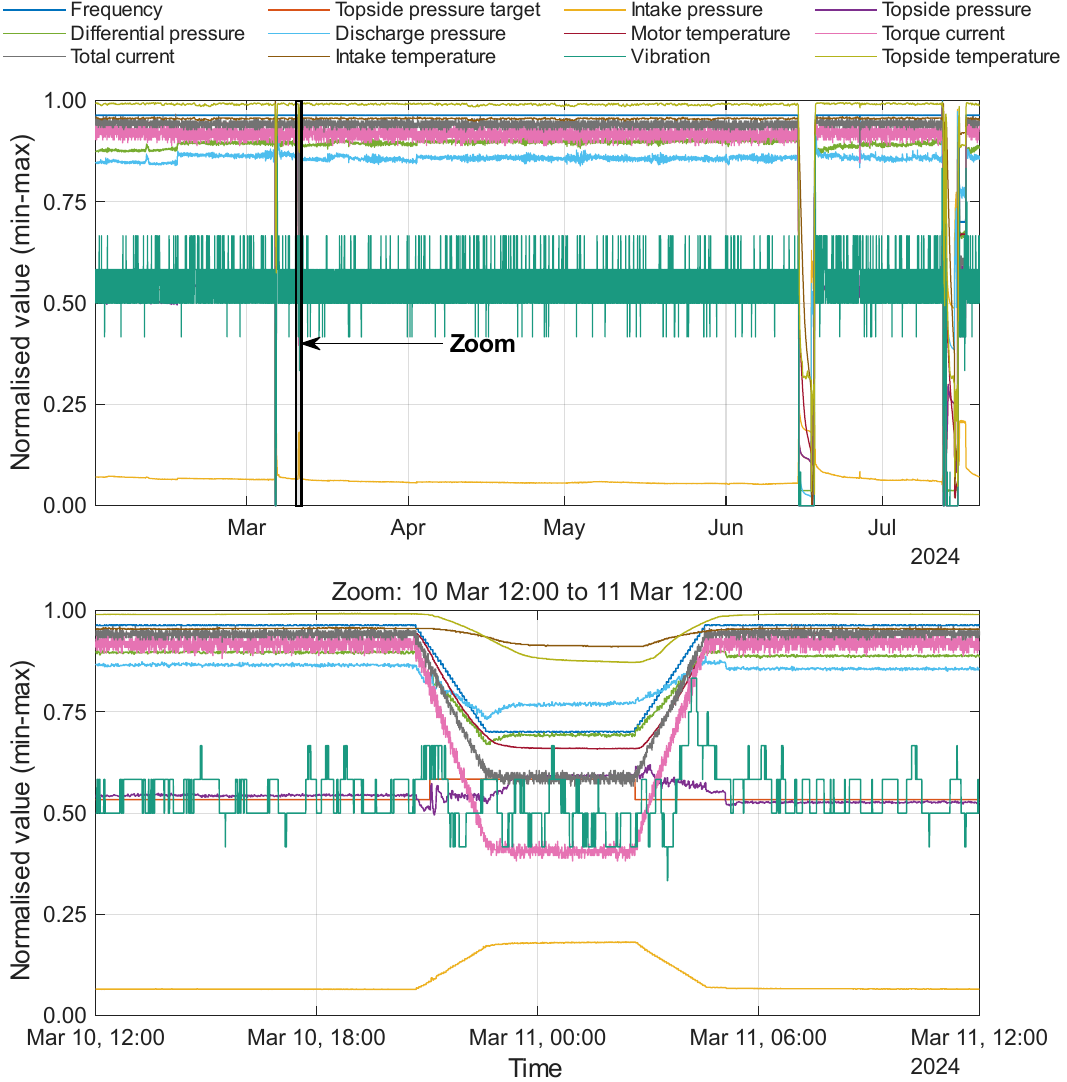}
  \caption{Normalised OPC record of the process variables. Top: the full series, with the
  dynamic stretch used for model validation shaded. Bottom: the same stretch expanded
  (10--11 March 2024), where the commanded ramp-down to shutdown and the subsequent
  restart are visible.}
  \label{fig:windows}
\end{figure}

\begin{figure}[htbp]
  \centering
    \includegraphics[width=\linewidth]{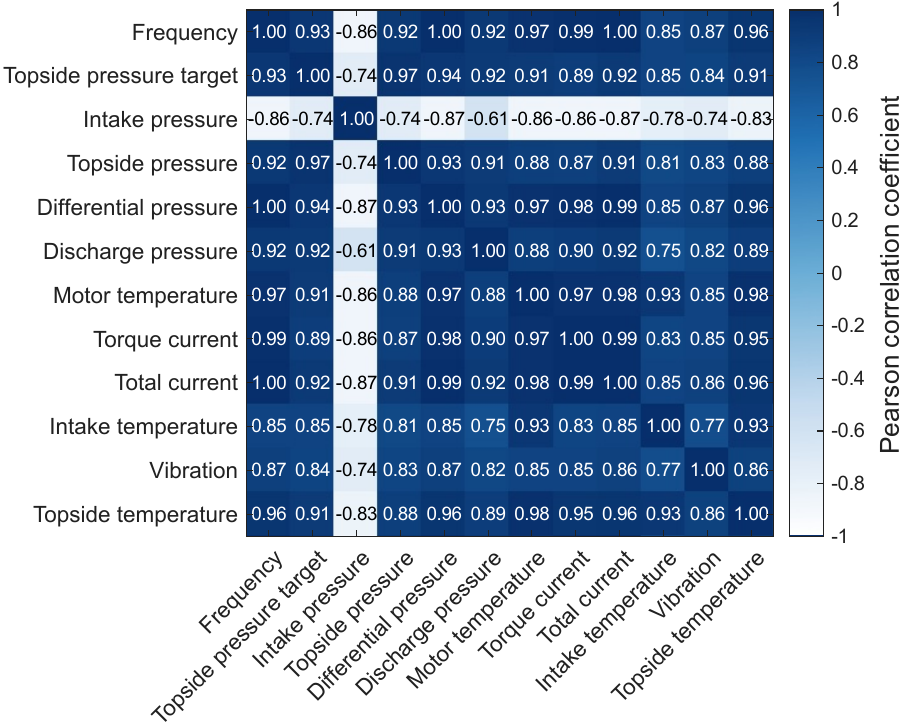}
  \caption{Correlation matrix of the ESP process variables.}
  \label{fig:corr}
\end{figure}

The continuous form of the data is shown in Figure~\ref{fig:windows}, where a dynamic window of the data can be observed in a normalised record, with the dynamic stretch used for validation expanded underneath. Each variable was normalised to the interval $[0,1]$ using the min--max limits for confidentiality reasons. These limits are the engineering ranges of each instrument rather than the observed extremes, which keeps the normalisation valid when the process visits regions absent from the historical record. Figure~\ref{fig:corr} presents the correlation matrix of the process variables. The frequency is almost perfectly correlated with the differential pressure and the total current, and anti-correlated with the intake pressure ($-0.86$), which is consistent with the pump affinity laws and anticipates the dominant control pairing. Several variable pairs are strongly correlated with one another, and Section~\ref{sec:model:lipschitz} explains why that observation led to keeping every channel rather than pruning the redundant ones.

\subsection{Data-driven dynamic modelling}
\label{sec:model}

\begin{figure}[htbp]
	\centering
	\includegraphics[width=\linewidth]{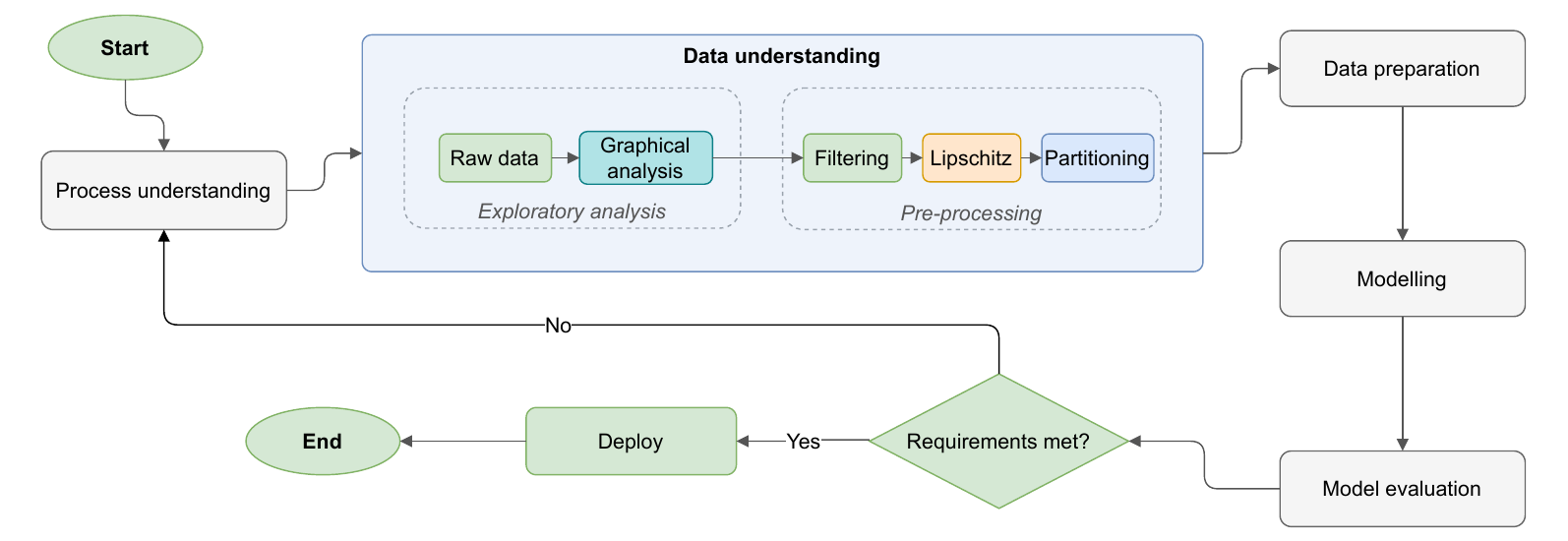}
	\caption{Methodology pipeline from process understanding and historical-data treatment to model construction and validation.}
	\label{fig:datamethod}
\end{figure}

This section presents the identification of the predictive models and the reasoning behind the structural choices. Figure~\ref{fig:datamethod} summarises the pipeline, from process understanding and historical-data treatment to model construction and validation.

\subsubsection{NARX structure and multi-step prediction}
\label{sec:model:narx}

The models follow the NARX structure \cite{Leontaritis1985,ljung1998system,keesman2011system}:
\begin{equation}
y(t)=f\!\big(y(t-1),\dots,y(t-n_b),\,u(t-1),\dots,u(t-n_a)\big),
\label{eq:narx}
\end{equation}
where $n_a$ and $n_b$ are the input and output memory lengths. Figure~\ref{fig:narx} illustrates the resulting internal-model architecture. A distinction drives the whole evaluation strategy of Section~\ref{sec:evalproto}. In identification, the model is fitted one step ahead, with measured outputs available in the regressor. Inside an MPC, the model is queried $H_p$ steps ahead in a single call, and from the second step onwards the regressor contains predictions rather than measurements. Prediction errors are fed back and accumulate along the horizon. A model with excellent one-step metrics can therefore be unusable in closed loop, which is why free-run simulation was adopted as the acceptance test throughout this work. Section~\ref{sec:eval:protocol} quantifies how large that gap is in practice.

\subsubsection{Regressor selection and variable pairing}
\label{sec:model:lipschitz}

\begin{figure}[h!]
  \centering
  \includegraphics[width=\linewidth]{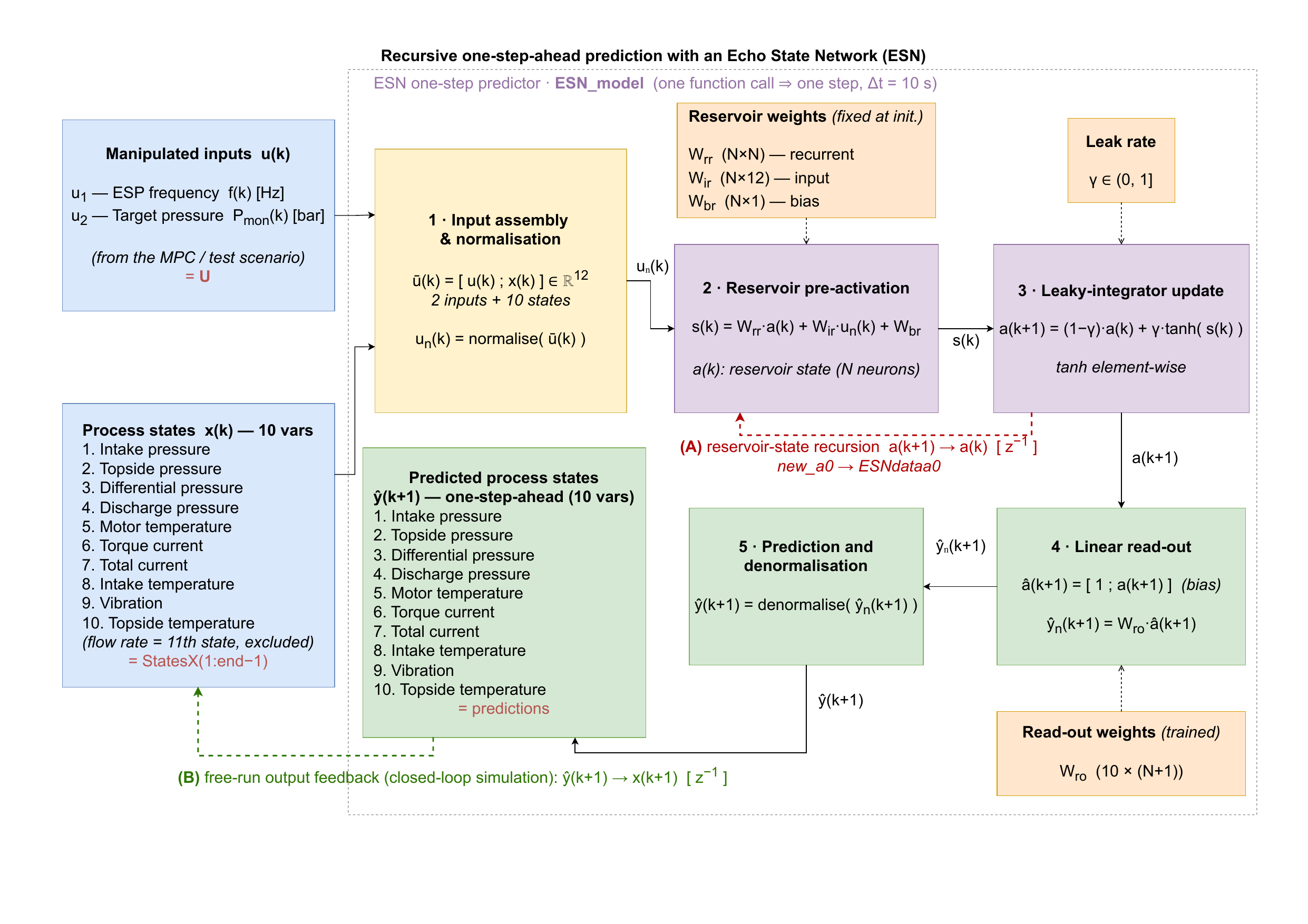}
  \caption{Internal model architectures: one-step-ahead NARX.}
  \label{fig:narx}
\end{figure}

The structure of the model departs from the usual practice of pruning a candidate regressor set. On the ESP system, only two variables are manipulated, the VSD speed (Frequency) and the topside pressure target. Ten measured process variables are predicted, and all ten are fed back as regressors. Every channel is therefore simultaneously an output of the model and an input to the prediction of the other nine, as Figure~\ref{fig:narx} shows.

The Lipschitz-coefficient method \cite{He1993}, proposed for identifying the orders of input--output models, was applied during the exploratory phase to examine how strongly the input--output pairs were coupled and how much memory the data supported, and it confirmed the strong interdependence that the correlation analysis of Section~\ref{sec:data} had already indicated. The channels were nevertheless all kept, because in free run each fed-back channel acts as a check on the others. The orders were then harmonised across channels to keep a single network structure.

The consequence matters for the controller, and it is the reason the choice is reported as a contribution rather than as a default. In free run the model receives no measurements after the first step, so a deviation in one channel propagates. With all ten channels coupled, a deviation in one is immediately constrained by the nine others through the relationships the network has learnt, and the trajectory is pulled back rather than allowed to drift. A model built on a reduced regressor set would not have this mutual restraint. The free-run results of Section~\ref{sec:eval} were obtained with this structure.

\subsubsection{Echo State Networks}
\label{sec:model:esn}

Echo State Networks belong to the reservoir-computing family of recurrent networks \cite{Jaeger2004,Lukosevicius2009,Lukosevicius2012}. The recurrent part, the reservoir, is generated randomly and never trained. Its dynamics are governed by three quantities: the reservoir size, the leak rate, which sets the memory of each unit, and the spectral radius $\rho$ of the recurrent weight matrix, which governs the echo-state property, that is, the fading of the initial reservoir state so that the states become a function of the input history alone. The condition $\rho < 1$ is the customary guideline, but it is not sufficient for the echo-state property \cite{Yildiz2012} and, for a leaky reservoir driven by inputs, not necessary either; the screening of Section~\ref{sec:evalproto} therefore treated $\rho$ as a free hyperparameter, and the selected network operates slightly above one (Section~S2 of the Supporting Information), its behaviour with output feedback being judged by the free-run validation of Section~\ref{sec:eval}. Only the linear output layer is trained, by Ridge regression on the reservoir states, so training reduces to a regularised least-squares problem that runs in seconds even for thousands of candidate networks.

Two consequences of this structure matter for the application. Training cost is low enough to make the large-scale hyperparameter screening of Section~\ref{sec:evalproto} practical. Inference cost is a single sparse matrix--vector product per step, which is what allows the model to run inside the optimisation loop of Section~\ref{sec:mpc} within the real-time budget. {For the reservoir of Section~S2 of the Supporting Information, one forward pass takes about \SI{16}{ms}, and the interior-point solver calls the model a few hundred times per instance.} Recurrent alternatives were evaluated on the same data (Section~\ref{sec:eval:protocol}). LSTM networks \cite{Hochreiter1997} were compared with ESNs on operational data of the same kind \cite{Rebello2025RNN} and trained for multi-step-ahead prediction of an offshore ESP \cite{Santana2025multistep}. It was found in the project that, while LSTM networks are capable of tracking the data, their applicability to control systems, particularly in predictive control, depends on solving an optimisation problem promptly; for the specific application addressed in this paper, LSTM networks fail to meet this requirement. The Echo State Network is the model that fits the real-time budget, and Section~\ref{sec:eval:protocol} shows that the accuracy it gives up is one-step accuracy only: in free run the reference Echo State Networks are ahead of the LSTM.

Before producing useful predictions, the reservoir state must be synchronised with the plant through a washout period. In this application the washout is 300 samples, roughly one hour at the \SI{10}{s} inference period, and it is applied once at controller start-up. %

\subsection{Model evaluation protocol}
\label{sec:evalproto}

For each candidate configuration, between 1100 and 1820 networks were generated by sweeping the reservoir size, the leak rate and the spectral radius. Evaluation followed a free-run protocol: after the washout, the network receives only the manipulated inputs and its own fed-back predictions, which is exactly the condition it faces inside the prediction horizon of the controller. One-step-ahead error was not used for ranking, because a model that is accurate one step ahead can still drift inside the controller horizon; it is reported only in Section~\ref{sec:eval:protocol}, to show how far it would have misled the selection.

\subsubsection{Error metrics and aggregation}
\label{sec:eval:metrics}

Twenty error metrics (Table~S3 of the Supporting Information) were computed for every candidate during the screening campaign. They fall into three groups: fourteen statistics of the error and prediction signals, three measures of the fraction of the run in which the percentage error exceeded a threshold, and three measures of the longest uninterrupted exceedance of the same thresholds. The last group separates a brief excursion from a sustained bias, which have very different consequences for a controller that acts on the prediction.

For ranking, the MAPE was adopted because the ten output variables carry different units and magnitudes, and a percentage measure allows a single comparable figure per channel. Values were capped at 100\,\% to limit the influence of near-zero denominators. Symmetric variants such as the SMAPE were considered and discarded. They change the ranking near zero in ways that are hard to interpret physically, and the division-by-zero pathology is better handled explicitly, as discussed in Section~\ref{sec:eval:vibration}.

Reducing ten channels to one number requires a second choice, which is usually left implicit. The mean across channels rewards a model that is good on average and can hide one badly predicted variable; the worst channel rewards uniformity and is the stricter test. Both are reported in Section~\ref{sec:eval}, and the selection follows the worst channel, since the model feeds a controller that operates against constraints and a single unreliable channel is enough to compromise that.
\begin{equation}
\mathrm{MAPE}_j=\frac{100}{N}\sum_{k=1}^{N}\left|\frac{y_j(k)-\hat y_j(k)}{y_j(k)}\right|
\label{eq:mape}
\end{equation}

\subsection{Production estimation and operating maps}
\label{sec:maps}

The economic objective of the controller is the oil flow rate, and this variable is not measured in real time. The well has no permanent multiphase flow meter, so the flow rate must be estimated, which is the virtual flow metering problem \cite{Bikmukhametov2020}. This section presents the phenomenological layer that provides this estimate together with the operating maps used to encode the axial-thrust constraints.

The flow rate is represented as a static map $Q(f, p_{\text{arr}})$, a function of the VSD frequency and of the arrival (topside) pressure. The map was built from a sensitivity analysis of the multiphase well simulator and is queried online by bilinear interpolation, which allows the map to be stored at any convenient resolution without changing the controller code. Because the underlying well conditions drift with reservoir depletion and water cut, the map is recalibrated periodically against production tests (BTP) carried out on the platform. This design decouples the economic estimate from the dynamic model: reservoir drift is absorbed by recalibrating a static map, and the neural network does not need to be retrained at each change in well conditions. {The drift is not small: a change of 0.5 in BSW displaces the flow estimate by around \SI{5}{\percent} of the operating value. Recalibrating the static map against a production test is an afternoon of work on the platform, whereas retraining the dynamic model would be a campaign. The network sees frequencies, pressures, temperatures and currents, and those relationships do not change when the reservoir depletes.}

\begin{figure}[h!tbp]
  \centering
  \includegraphics[width=\linewidth]{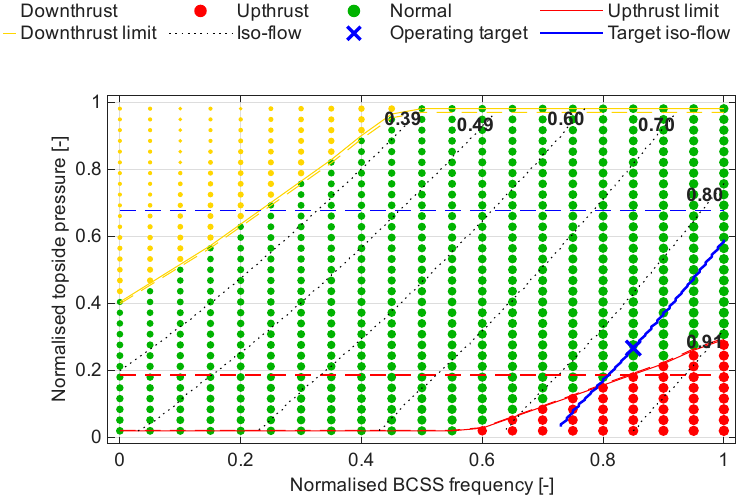}
  \caption{Reconstructed operating map in the normalised frequency--pressure plane:
  iso-flow curves, the upthrust and downthrust regions with their boundaries, and the
  operating target derived from the frequency requested by the operator (cross), together
  with the iso-flow curve that passes through it.}
  \label{fig:map}
\end{figure}

\begin{figure}[htbp]
  \centering
  \includegraphics[width=\linewidth]{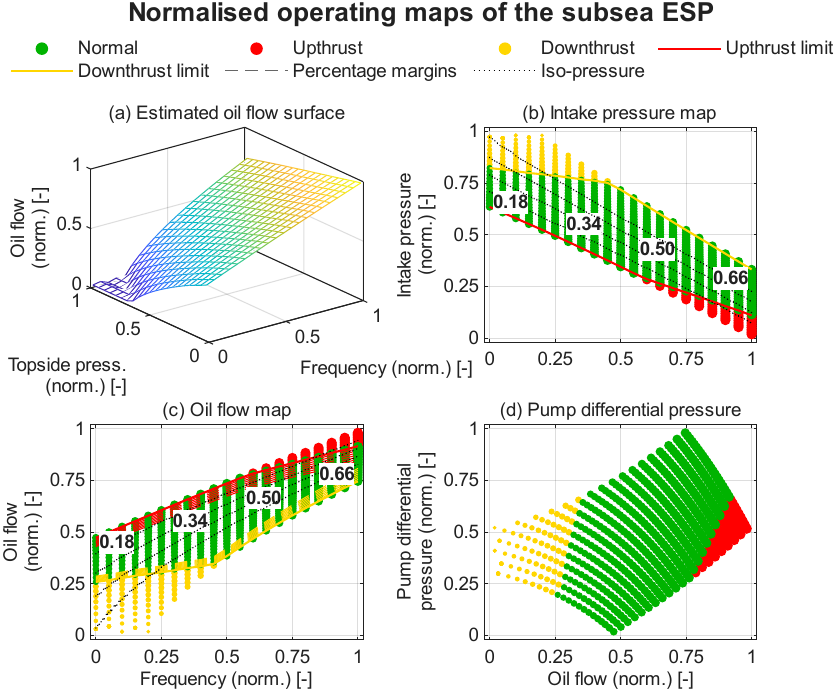}
  \caption{Complementary views of the reconstructed operating maps, all axes min--max
  normalised: the estimated oil flow surface over the frequency--pressure plane, the
  intake-pressure and flow maps against frequency with their protection limits and
  percentage margins, and the pump differential pressure against flow. The
  frequency--pressure map itself is shown in Figure~\ref{fig:map} and its flow gradient in
  Figure~\ref{fig:gradpath}, and neither is repeated here.}
  \label{fig:mapscomposite}
\end{figure}

The same map structure encodes the pump integrity limits. Upthrust and downthrust boundaries were expressed as regions of the frequency-versus-pressure plane, as Figure~\ref{fig:map} shows, together with the iso-flow curves. These boundaries enter the optimisation problem of Section~\ref{sec:mpc} as dynamic constraints, recomputed at each operating frequency. Figure~\ref{fig:mapscomposite} gathers the complementary views used to interrogate the same data set: the estimated flow surface over the frequency--pressure plane, the intake-pressure and flow maps with their protection limits, and the pump differential pressure against flow.

One consequence for the operation is worth making explicit, because it shapes how the controller is used. The operator specifies only the desired frequency. The corresponding topside pressure target is then computed from the map, using a percentage margin to the upthrust boundary agreed with the production engineering team, and is marked in Figure~\ref{fig:map} together with the iso-flow curve that passes through it. The flow set-point is the map value at the frequency requested by the operator; when that frequency lies beyond the constraint-limited region, the operating point the controller finally settles on in the frequency--pressure plane is decided by the active constraints, not imposed by the set-point.

\begin{figure}[htbp]
  \centering
  \includegraphics[width=\linewidth]{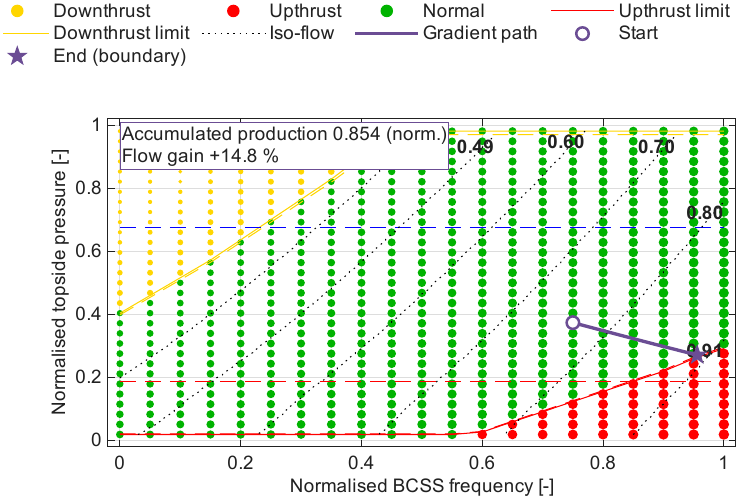}
  \caption{Steepest-ascent path of the estimated oil flow rate over the normalised operating map, with the production accumulated along the path. The ascent is an offline reading of the map: it locates the map boundary towards which the optimum is pushed, and the DDMPC does not implement it. The flow gain quoted in the figure is along this offline path and is not a field result.}
  \label{fig:gradpath}
\end{figure}

{The flow surface of Figure~\ref{fig:mapscomposite} has no interior maximum over the admissible region. Following the gradient of the estimated flow from any admissible starting point leads to a boundary, never to a stationary point inside the region. Figure~\ref{fig:gradpath} traces one such ascent path together with the production accumulated along it.}

{The path is an offline reading of the map and not the control law. The DDMPC does not follow the gradient: it solves the constrained problem of Section~\ref{sec:mpc} at every cycle, and the operating point it reaches is set by the active constraints and by the relative weight on the flow channel. The reason to present the ascent at all is that it identifies which boundary the optimum must lie on, before any closed-loop simulation is run, as far as the map alone can tell. On the map alone, the boundary the ascent reaches is the upthrust limit over most of the frequency range. In the closed-loop runs of Section~\ref{sec:tuning} a process limit, the total current, becomes active first, so the constraint the controller rides is set by the process limits as well as by the map.}

\subsection{Hybrid DDMPC formulation}
\label{sec:mpc}

\subsubsection{Prediction model in state space}
\label{sec:mpc:model}

The system is described by a state vector $x \in \mathbb{R}^{11}$, collecting the ten measured process variables and the estimated flow rate in the order fixed in Section~\ref{sec:system:vars}, a control input vector $u \in \mathbb{R}^{2}$, comprising the VSD frequency set-point and the topside pressure target, and a controlled output vector $y \in \mathbb{R}^{2}$. The prediction model is the hybrid pair
\begin{align}
\hat x(k+1) &= \hat f\!\big(x(k),u(k)\big), \label{eq:pred-x}\\
\hat y(k+1) &= h\!\big(\hat x(k+1)\big), \label{eq:pred-y}
\end{align}
where $\hat f$ combines the ESN of Section~\ref{sec:model:esn}, which propagates the ten dynamic variables, with the phenomenological flow estimator of Section~\ref{sec:maps}, and $h$ is the linear output map
\begin{equation}
h(\hat x) = H\,\hat x, \qquad H \in \{0,1\}^{2\times 11},\quad H_{1,2} = H_{2,11} = 1,
\label{eq:H}
\end{equation}
with every other entry of $H$ equal to zero. $H$ is thus a selection matrix that extracts the two controlled outputs from the state vector: the topside pressure, its second component, and the estimated oil flow rate, its eleventh (Section~\ref{sec:system:vars}). Because $h$ is linear, an error in the predicted state maps directly onto an error in the predicted output, $e_y = H e_x$, a property used in Section~\ref{sec:mpc:cost}. This split is the core of the hybrid architecture: the network carries the fast dynamics, and the physics-based map carries the economics.

\subsubsection{Multi-rate architecture and state filtering}
\label{sec:mpc:multirate}

Inference and optimisation run at different rates. The ESN produces state predictions every \SI{10}{s}. That period was not chosen for convenience: it is the fastest acquisition the platform supervisory system sustains without losing data, since the network traffic generated by the full variable set below \SI{10}{s} exceeds what the link was dimensioned for. The figure was specified by the project team as a requirement to the supervisory-platform supplier, and the whole controller architecture, including the prediction horizon of Section~\ref{sec:tuning:hp}, was then designed around it. The nonlinear programming problem (NLP) is solved every \SI{30}{s}. The dynamic response of the well to set-point changes is slow, and re-optimising at the inference rate would add computational load without benefit, while sending needlessly frequent set-point updates to the regulatory layer. The \SI{30}{s} period lets the VSD and choke PID loops accommodate each command before receiving the next one, reducing mechanical wear on the actuators.

Between the two rates sits a moving-average filter. Because three inference samples are available per optimisation cycle, the initial state fed to the NLP is the arithmetic mean of the three most recent samples. This acts as a digital low-pass filter, attenuating measurement noise and spurious spikes without introducing meaningful phase delay, and provides a cleaner starting point for the prediction horizon.

\subsubsection{Objective function and hierarchical constraints}
\label{sec:mpc:cost}

The problem is transcribed by multiple shooting: the predicted states $\hat x(k{+}i)$, the inputs $u(k{+}i)$ and the input moves $\Delta u(k{+}i)$ are all decision variables, coupled through equality constraints. The output set-point is not an external reference. It is derived from the engineering targets $u_{\text{eng}} = [f_{\text{eng}},\ p_{\text{eng}}]^{\top}$ as $y_{\text{sp}} = [\,p_{\text{eng}},\ Q(f_{\text{eng}}, p_{\text{eng}})\,]^{\top}$, using the flow map of Section~\ref{sec:maps}. The controller also receives the one-step prediction errors $e_x(k) = x(k) - \hat x(k|k{-}1)$ and $e_y(k) = H e_x(k)$, with $H$ as in Eq.~\eqref{eq:H}, both computed against the prediction made at the previous cycle.

At each optimisation instant $k$, the controller solves
\begin{equation}
\begin{split}
\min_{\hat{\mathbf{x}},\,\mathbf{u}}\;
J ={}& e_x(k)^{\top} Q_x\, e_x(k)
+\sum_{i=1}^{H_p}\big\lVert \hat y(k{+}i)-y_{\text{sp}}+e_y(k)\big\rVert^2_{Q_y} \\
&+\sum_{i=1}^{H_c}\big\lVert u(k{+}i{-}1)-u_{\text{eng}}\big\rVert^2_{Q_u}
+\sum_{i=1}^{H_c}\big\lVert \Delta u(k{+}i{-}1)\big\rVert^2_{R} \\
&+ Q_{\text{slack}}\sum_{i=1}^{H_p}\big\lVert \Phi\big(d(k{+}i)\big)\big\rVert^2_2
\end{split}
\label{eq:cost}
\end{equation}
\begin{align}
\text{s.t.}\quad
& \hat x(k)=x(k), \quad \hat x(k{+}i)=\hat f\big(\hat x(k{+}i{-}1),u(k{+}i{-}1)\big), \quad\text{(model)} \label{eq:c-model}\\
& u(k)-u(k{-}1)=\Delta u(k), \quad u(k{+}i)-u(k{+}i{-}1)=\Delta u(k{+}i), \quad\text{($\Delta u$)} \label{eq:c-du}\\
& u(k{+}i)=u(k{+}i{-}1) \ \ \text{within a supervisory period}, \quad\text{(blocking)} \label{eq:c-block}\\
& u(k{+}i)=u(k{+}H_c{-}1), \quad i \ge H_c, \quad\text{(beyond $H_c$)} \label{eq:c-hc}\\
& u_{\min}\le u(k{+}i)\le u_{\max},\quad |\Delta u(k{+}i)|\le \Delta u_{\max}, \quad\text{(actuators)} \label{eq:c-act}\\
& \Big|\textstyle\sum_{n=k-N_{\text{win}}+1}^{k}\Delta u_{\text{freq}}(n)\Big|\le \SI{1}{Hz}, \quad\text{(VSD window)} \label{eq:c-vsd}\\
& d(k{+}i) \ge 0, \quad\text{(margins)} \label{eq:c-hard}\\
& -\Delta p^{\,\text{dw}}_{\max} \le u_{p}(k{+}i) - \hat x_{p}(k{+}i) \le \Delta p^{\,\text{up}}_{\max}. \quad\text{(target band)} \label{eq:c-ptarget}
\end{align}

The first term of Eq.~\eqref{eq:cost} penalises the one-step prediction error with weight $Q_x$; it is constant with respect to the decision variables and serves as a bias-monitoring term in the reported cost. In the tracking term, the error $e_y(k)$ is added to the predicted deviation, which corrects the set-point for the current model bias and removes the steady offset that a constant plant--model mismatch would otherwise produce. The input terms are weighted by $Q_u$ and $R$ and accumulated only up to the control horizon $H_c$. Production is maximised by tracking: when the frequency requested by the operator lies above the constraint-limited region, the flow set-point is unreachable and the active constraints decide where the well settles (Section~\ref{sec:tuning:weights}).

The safety constraints are expressed as percentage distances to the frequency-dependent alarm limits. For each state $j$ and each step, with $\bar x_j(f)$ the upper (H) alarm limit returned by the protection tables of Section~\ref{sec:system:vars} and the thrust maps of Section~\ref{sec:maps} at frequency $f$,
\begin{equation}
d_j = 100\,\frac{\bar x_j(f) - \hat x_j}{\bar x_j(f)} - m,
\label{eq:margin}
\end{equation}
where $m$ is a safety margin in per cent, configured per process variable and reported in
Table~\ref{tab:margins}. Eq.~\eqref{eq:c-hard} imposes $d_j \ge 0$ as a hard constraint,
so the enforced limit is the H alarm value shrunk by $m$. The L alarm limits are treated
the same way, with the sign of the difference reversed. The soft band lies one percentage
point beyond the hard margin of the same variable. In the reference configuration $m$ is
\SI{1}{\percent} for the ten measured channels, which puts the enforced limit one per cent
inside the H alarm and the soft threshold at two per cent, and \SI{2}{\percent} for the
estimated flow rate, for the reason given in Section~\ref{sec:maps}. Expressing both
tiers as percentages of the alarm value, rather than in engineering units, is what allows
the whole scheme to be reported without disclosing the absolute Petrobras protection
settings.

\begin{table}[htbp]
  \centering
  \begin{threeparttable}
  \caption{Safety margins configured per process variable.}
  \label{tab:margins}
  \begin{tabular}{l S[table-format=1.2] S[table-format=1.2]}
\toprule
Variable & {{Hard margin [\%]}} & {{Soft band [\%]}} \\
\midrule
Intake pressure & 1.00 & 2.00 \\
Topside pressure & 1.00 & 2.00 \\
Differential pressure & 1.00 & 2.00 \\
Discharge pressure & 1.00 & 2.00 \\
Motor temperature & 1.00 & 2.00 \\
Torque current & 1.00 & 2.00 \\
Total current & 1.00 & 2.00 \\
Intake temperature & 1.00 & 2.00 \\
Vibration & 1.00 & 2.00 \\
Topside temperature & 1.00 & 2.00 \\
Oil flow rate & 2.00 & 3.00 \\
\bottomrule
\end{tabular}

  \begin{tablenotes}\footnotesize
  \item The hard margin is the percentage by which each alarm limit is shrunk to form the enforced constraint of Eq.~\eqref{eq:c-hard}. The soft band, one percentage point beyond it, is where $\Phi$ becomes active. Absolute Petrobras alarm values are withheld. Variables are listed in the order of the state vector $x$ (Section~\ref{sec:system:vars}).
  \end{tablenotes}
  \end{threeparttable}
\end{table}

The hierarchy between soft and hard protection is implemented through the last term of Eq.~\eqref{eq:cost} rather than through slack variables. $\Phi(\cdot)$ is a fourth-order polynomial fitted by least squares to an exponential in the percentage distance $d_j$, saturated outside the unit interval, and it is zero while the trajectory stays away from the soft threshold and grows steeply as $d_j$ approaches zero. Because $d_j$ is a percentage rather than an engineering quantity, one single $\Phi$ serves every variable and both the high and low alarms, which is what keeps the formulation tractable at eleven variables. The polynomial form is not cosmetic either: it keeps the term twice differentiable and cheap to evaluate inside the interior-point iterations, where a saturated exponential would not.

The distinction from a slack-variable or zone formulation is where the practical benefit lies. In those formulations the softened limit is relaxed by a decision variable that is itself penalised \cite{Scokaert1999}, so a violation of the softened limit is traded against the economic objective at a price fixed by the weight \cite{Gonzalez2009,Santana2021,Santana2022}. Here the hard constraint of Eq.~\eqref{eq:c-hard} is never relaxed, and what is priced is proximity: $Q_{\text{slack}}$ sets the cost of approaching the limit, so the optimiser is discouraged from the boundary before it reaches it instead of being charged after crossing it. The band in which $\Phi$ is active therefore behaves as the soft band, entered when a process limit is active at the optimum (Section~\ref{sec:tuning:hp}) and otherwise only transiently to absorb disturbances, and an instance that is nevertheless infeasible is handled by holding the previous action (Section~\ref{sec:mpc:nlp}).

Three input constraints reflect the actuation reality. The move-blocking constraint of Eq.~\eqref{eq:c-block} keeps the inputs constant in packets of three samples, informing the optimiser that the supervisory action is applied every \SI{30}{s} on the \SI{10}{s} prediction grid. The sliding-window constraint of Eq.~\eqref{eq:c-vsd} enforces the VSD manufacturer specification of at most \SI{1}{Hz} of cumulative frequency variation over \num{7.5} minutes, with $N_{\text{win}} = 45$ samples at $T_s = \SI{10}{s}$, the window combining the moves already applied with those being proposed. Eq.~\eqref{eq:c-ptarget} keeps the topside pressure target within a band around the measured arrival pressure, since a target far from the current measurement would act as an aggressive attractor for the choke PID rather than a realisable reference. In the implementation, computed moves smaller than a configurable dead band are also discarded before being sent to the actuators, avoiding cosmetic moves that would only produce wear.

\subsubsection{NLP solution and initialisation}
\label{sec:mpc:nlp}

The problem was implemented in CasADi \cite{Andersson2019} and solved with the interior-point solver IPOPT \cite{Wachter2006}, as in the conference paper \cite{Reges2026}. At each cycle, the decision vector is initialised with the current measurement stacked onto the state and input trajectories returned by the previous solution, with the input moves zeroed. The solver runs with a limited-memory Hessian approximation and a monotone barrier strategy initialised at a small barrier parameter, under a \SI{9}{s} wall-time cap. If an instance fails to converge, the previous action is held and a new attempt is made at the next supervisory instant. The remaining solver settings and the pseudocode of the receding-horizon loop are given in Section~S4 of the Supporting Information.

\subsection{Simulation environment and experimental protocol}
\label{sec:hil}

\subsubsection{Simulation environment}
\label{sec:hil:env}

\begin{figure}[htbp]
  \centering
  \includegraphics[width=\linewidth]{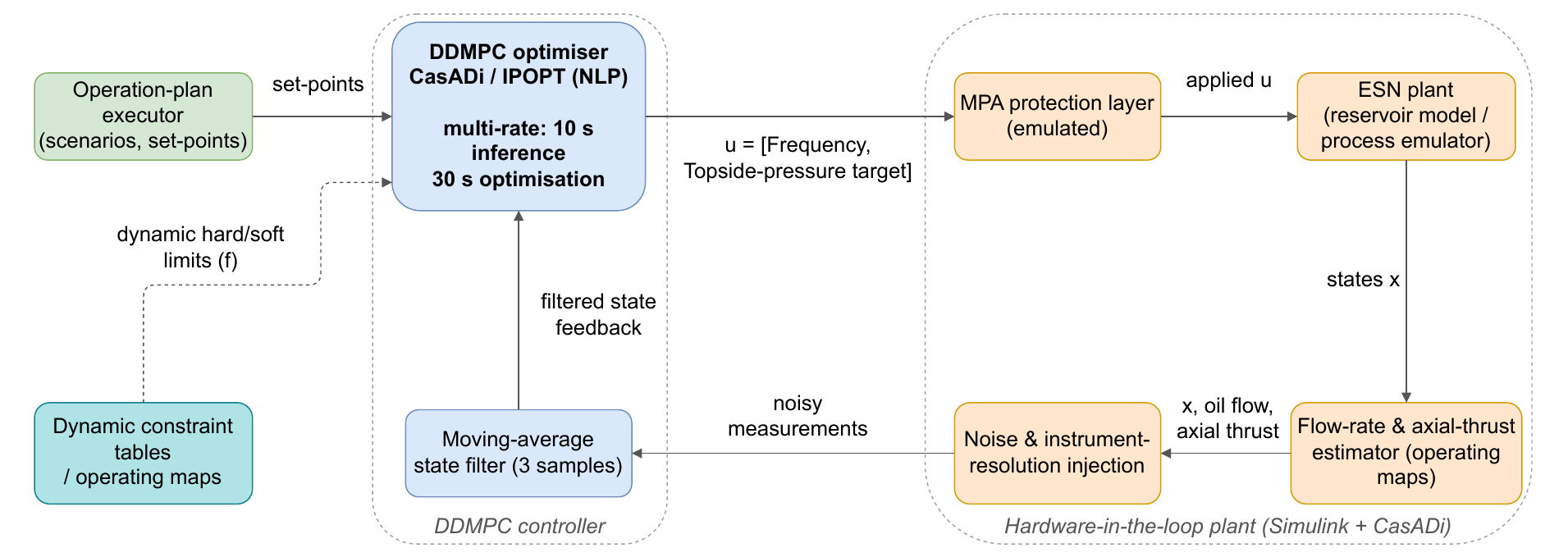}
  \caption{Simulation environment: Simulink/CasADi blocks, the ESN acting as the plant, the emulated MPA protection layer, and noise/resolution injection.}
  \label{fig:hil}
\end{figure}

Every tuning decision and every field campaign was preceded by tests in a dedicated simulation environment, shown in Figure~\ref{fig:hil}. The environment was built in Simulink with the CasADi controller embedded, and it reproduces the field configuration in three respects. The plant is emulated by an ESN acting in free run, so the controller faces the same class of dynamics it will meet in the field. The platform protection layer (MPA) is emulated, including its authority to override the controller outputs. White noise and the quantisation of the real instruments are injected into every measured channel, so the controller is never tuned on clean signals.

An operation-plan executor completes the environment, scripting sequences of targets, disturbances and protection events so that campaigns can be rehearsed end to end before the software reaches the platform. 

\subsubsection{Design of the tuning study}
\label{sec:hil:tuningdesign}

Every controller parameter was first set in this environment, under measurement noise and instrument resolution, and every later change was rehearsed there before being applied in the field. The study is organised as a set of
one-factor sweeps around a reference configuration, each simulated in
closed loop from the same initial operating point: the relative weighting $Q_y$ between the
oil-flow and arrival-pressure outputs, the hard margin $m$ of Eq.~\eqref{eq:margin} applied
to the estimated flow rate, the proximity weight $Q_{\text{slack}}$ of Eq.~\eqref{eq:cost},
the frequency target requested by the operation, and the prediction horizon $H_p$, the last
one evaluated against the solver time per call rather than against the closed-loop
trajectory alone. The two horizons are not settled by that sweep alone. They were chosen
by an exhaustive search over the region delimited by the project team,
$2 \le H_p \le 15$ and $1 \le H_c \le H_p - 1$, which gives 105 configurations. Each was
simulated in closed loop for \SI{4}{h} from the same initial operating point, without
measurement noise and with every other parameter fixed, and was scored on four quantities:
the mean oil flow over the last hour, the settled distance to the nearest alarm, the
95th-percentile solver time per call and the final VSD frequency. Sensitivity to the initial condition is assessed separately, on a grid of
three initial operating points, each simulated without and with measurement noise under two
weightings of the flow channel. Section~\ref{sec:tuning} reports the outcome.

\subsubsection{Field commissioning and test protocol}
\label{sec:field:protocol}

The field controller was implemented in Python and structured in independent modules: configuration management, OPC communication with the supervisory system, real-time data logging, and the optimisation core. This modularity mattered in practice. Individual modules were exercised and replaced between campaigns without touching the optimisation code, and the same core ran unchanged in simulation and in the field.

Validation followed a two-stage protocol. Before each field campaign, the exact software build was exercised in the simulation environment of Section~\ref{sec:hil:env} and in replay mode against historical data.
In replay mode the internal model runs in free run, which is the way it is used along the
prediction horizon (feedback path~(B) in Figure~\ref{fig:narx}). The only information supplied
is the initial state of the ten process channels and the recorded trajectories of the two
manipulated inputs. The reservoir is first driven by measured data for the washout period of
Section~S2 of the Supporting Information, which settles its internal state; from that instant on, no further
measurement of the ten channels enters the model. At every \SI{10}{s} step the predicted
states $\hat{x}(k+1)$ replace the measured states $x(k+1)$ in the regressor of the next step,
so the model evolves on its own in response to the input variations, and its trajectory is
compared with the recorded data over the whole replay, without reinitialisation. A single
replay therefore chains far more one-step predictions than the $H_p$ steps the controller
requests at each sampling instant, which makes it a harsher test than the use
the model sees inside the optimisation, and records containing decelerations and restarts
exercise the transients as well as steady operation.
Only after this stage was the controller connected to the plant, and always with the MPA active and empowered to override the DDMPC outputs. {Delivering the set-points through the protection layer rather than around it, so that the MPA retains the authority to override, was the condition under which the operation agreed to run the controller at all. The operator was also given a panel showing what the controller was proposing and why, because a set-point arriving from an optimiser with no visible justification does not survive a shift handover.}

\section{Results and discussion}
\label{sec:results}

\subsection{Model benchmarking and selection}
\label{sec:eval}

This section reports the outcome of the screening and selection protocol of
Section~\ref{sec:evalproto}, and identifies the network that was taken to the field.

The validation window reproduces a demanding field scenario: a full frequency ramp-down to shutdown followed by a restart, together with a step in the topside pressure target. 
The window is the dynamic stretch of 10--11 March 2024 identified in Section~\ref{sec:data}; it belongs to none of the training windows, since the record was split at window level rather than at sample level.
The following subsections report what the candidate networks did under this protocol.

\subsubsection{Free-run validation over the validation window}
\label{sec:eval:freerun}

\begin{figure}[htbp]
  \centering
  \includegraphics[width=\linewidth]{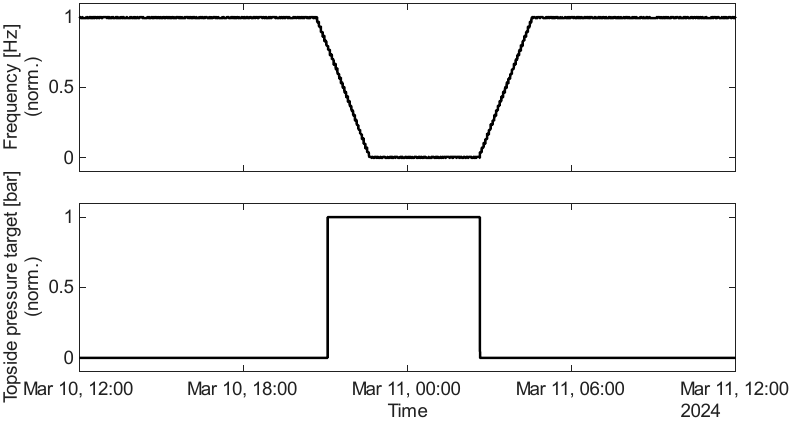}
  \caption{Measured inputs applied to the process during the validation window (VSD frequency and topside pressure target, min--max normalised).}
  \label{fig:validacaoDados_entradas}
\end{figure}

\begin{figure}[htbp]
  \centering
  \includegraphics[width=\linewidth]{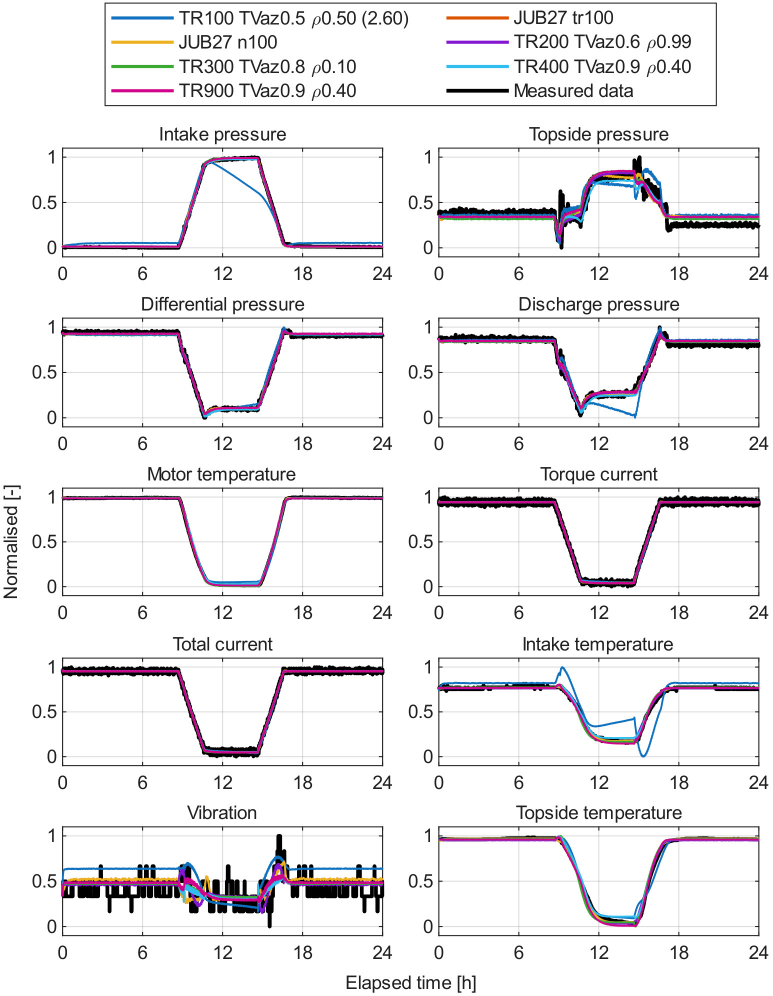}
  \caption{Benchmarking of candidate networks: free-run predictions of the ten measured process variables compared with the measured data over the validation window (10--11 March 2024; time axis in elapsed hours).}
  \label{fig:validacaoDados_variaveis}
\end{figure}

Figure~\ref{fig:validacaoDados_entradas} shows the measured inputs that were applied to the process
during the validation window, which are the only
external information the networks receive in free run.
Figure~\ref{fig:validacaoDados_variaveis} compares the free-run predictions of the candidate networks with the measured process variables over the same window. The candidate networks track the ten channels through the shutdown and restart, and the spread between them is visibly concentrated in a few channels, vibration above all.

\begin{figure}[htbp]
  \centering
  \includegraphics[width=\linewidth]{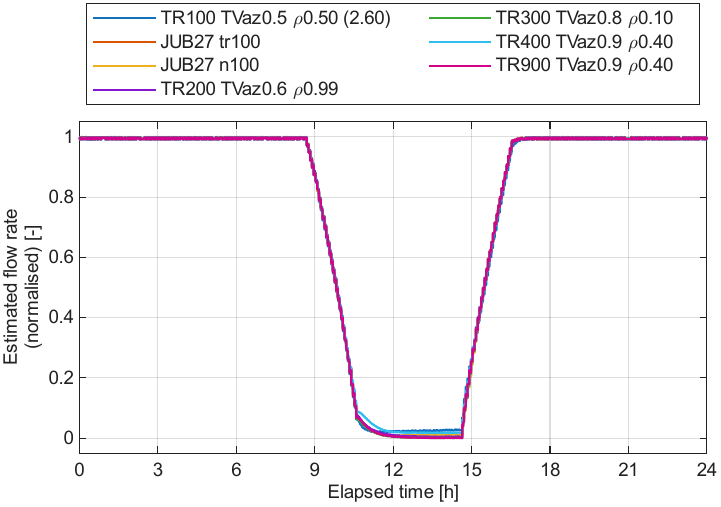}
  \caption{Flow rate estimated by each candidate model over the validation window. No direct measurement is available; the dispersion between candidates shows the part of the estimate uncertainty that comes from the dynamic model.}
  \label{fig:validacaoDados_vazao}
\end{figure}

The estimated flow rate receives a separate treatment in Figure~\ref{fig:validacaoDados_vazao}. As the well has no flow meter, there is no measured trace to compare against, and the error metrics of Section~\ref{sec:eval:metrics} do not apply to this channel. What the figure offers instead is the dispersion between the candidate models, which indicates how much of the uncertainty of the estimate comes from the dynamic model along the shutdown and restart transients.

\subsubsection{Effect of the vibration channel on model selection}
\label{sec:eval:vibration}
\begin{figure}[htbp]
	\centering
	\includegraphics[width=\linewidth]{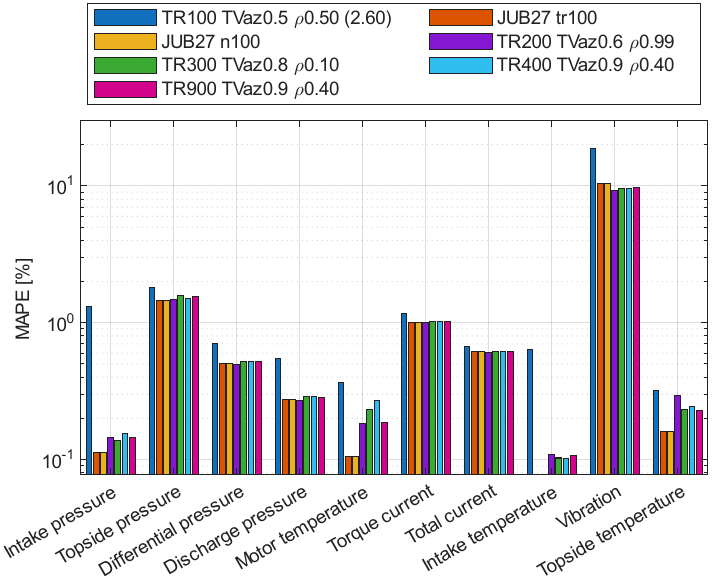}
	\caption{Free-run MAPE per channel of the candidate networks over the validation window (logarithmic scale).}
	\label{fig:validacaoDados_mape}
\end{figure}

The vibration signal operates close to zero over most of the record, so the
denominator of Eq.~\eqref{eq:mape} becomes very small and its MAPE is inflated well
beyond that of the remaining channels: across the candidate networks the
vibration MAPE lies around \SIrange{9}{19}{\percent}, six to ten times the worst
of the other channels (all at or below \SI{1.80}{\percent}), as Figure~\ref{fig:validacaoDados_mape}
shows. Because a single channel then governs the global average, it also governs
the ranking.

\begin{figure}[htbp]
  \centering
  \includegraphics[width=0.9\linewidth]{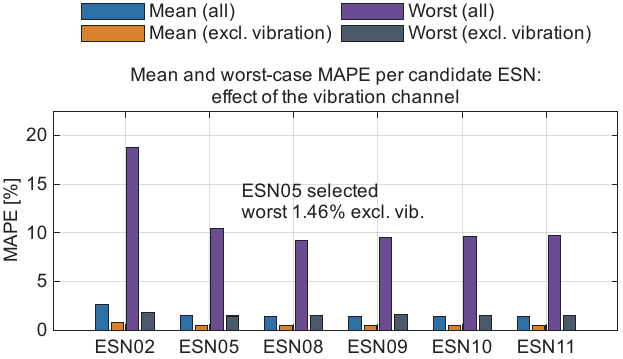}
  \caption{Mean and worst-channel MAPE per candidate ESN, with and without the vibration channel; excluding vibration changes which network ranks first.}
  \label{fig:mape_ranking}
\end{figure}

Table~\ref{tab:esnmetrics} therefore reports the per-variable MAPE together with
four summary rows, the mean and the worst channel, each computed over all ten
variables and again excluding vibration, and Table~\ref{tab:mape_comparison}
contrasts the resulting rankings. The distinction is not cosmetic, and it is
systematic rather than incidental: on both aggregations that include vibration
ESN08 comes first, with a mean MAPE of \SI{1.38}{\percent} and a worst channel of
\SI{9.20}{\percent}; on both aggregations that exclude it the leader changes and
ESN05 comes first, at \SI{0.48}{\percent} on the mean and \SI{1.46}{\percent} on
the worst channel (Figure~\ref{fig:mape_ranking}). The vibration contribution alone decides
which network the protocol returns. As that channel is not a controlled output and its percentage error is dominated by the near-zero denominator, it is excluded from the deciding number, and model selection
follows the worst remaining channel, with the full breakdown retained for
completeness. Its constraint stays enforced in the controller with the margin of Table~\ref{tab:margins}.

Two qualifications apply. The first is the size of the
margin: ESN05 leads ESN08 by \SI{0.02}{\percent} on the deciding number
(\SI{1.46}{\percent} against \SI{1.48}{\percent}), which is far below any
meaningful resolution of the metric, so the criterion should be read as
separating a group of five comparable networks from the rest rather than as
identifying a uniquely best one. What the criterion does establish firmly is the
exclusion of ESN02, whose worst non-vibration channel is \SI{1.80}{\percent} and
whose vibration MAPE is twice that of the others. The second qualification is
what the criterion turns out to measure. In every candidate network the worst
non-vibration channel is the same one, the topside arrival pressure, between
\SI{1.46}{\percent} and \SI{1.80}{\percent}, while the remaining eight channels
sit below \SI{1.2}{\percent} in every network except ESN02 (intake pressure, \SI{1.32}{\percent}). The worst-case rule is therefore not an
arbitrary tie-breaker: in this application it reduces to ranking the candidates
by how well they predict the arrival pressure, which is precisely the variable
the controller regulates. Normalising the error by the engineering
range of each variable instead of by the reading, as Section~\ref{sec:eval:histograms}
does, confirms that part of the vibration penalty is an artefact of the near-zero
denominator and part is real: on that scale the channel is still the worst predicted,
but by a factor of three to five rather than six to ten.

\begin{table}[htbp]
  \centering
  \caption{Per-variable prediction error (MAPE, \%) of the candidate Echo State Networks.}
  \label{tab:esnmetrics}
  \begin{adjustbox}{max width=\linewidth}
  \begin{tabular}{l S[table-format=2.2] S[table-format=2.2] S[table-format=1.2] S[table-format=1.2] S[table-format=1.2] S[table-format=1.2]}
\toprule
Variable & {ESN02} & {ESN05} & {ESN08} & {ESN09} & {ESN10} & {ESN11} \\
\midrule
Intake pressure & 1.32 & 0.11 & 0.14 & 0.14 & 0.15 & 0.15 \\
Topside pressure & 1.80 & 1.46 & 1.48 & 1.59 & 1.50 & 1.55 \\
Differential pressure & 0.70 & 0.51 & 0.50 & 0.52 & 0.52 & 0.52 \\
Discharge pressure & 0.55 & 0.28 & 0.27 & 0.29 & 0.29 & 0.28 \\
Motor temperature & 0.37 & 0.11 & 0.18 & 0.23 & 0.27 & 0.19 \\
Torque current & 1.18 & 1.01 & 1.00 & 1.02 & 1.02 & 1.01 \\
Total current & 0.67 & 0.62 & 0.61 & 0.62 & 0.62 & 0.61 \\
Intake temperature & 0.64 & 0.08 & 0.11 & 0.10 & 0.10 & 0.11 \\
Vibration & 18.74 & 10.43 & 9.20 & 9.50 & 9.60 & 9.74 \\
Topside temperature & 0.32 & 0.16 & 0.29 & 0.23 & 0.25 & 0.23 \\
\midrule
Mean (all) & 2.63 & 1.48 & 1.38 & 1.42 & 1.43 & 1.44 \\
Mean (excl.\ vibration) & 0.84 & 0.48 & 0.51 & 0.53 & 0.52 & 0.52 \\
Worst (all) & 18.74 & 10.43 & 9.20 & 9.50 & 9.60 & 9.74 \\
Worst (excl.\ vibration) & 1.80 & 1.46 & 1.48 & 1.59 & 1.50 & 1.55 \\
\bottomrule
\end{tabular}

  \end{adjustbox}
\end{table}

\begin{table}[htbp]
  \centering
  \caption{Per-network mean MAPE and rank, including versus excluding the vibration channel.}
  \label{tab:mape_comparison}
  \begin{adjustbox}{max width=\linewidth}
  \begin{tabular}{l S[table-format=1.2] c S[table-format=1.2] c S[table-format=2.2] c S[table-format=2.2] c}
\toprule
Network & {{Mean, all [\%]}} & {Rank} & {{Mean, excl.\ vib.\ [\%]}} & {Rank} & {{Worst, all [\%]}} & {Rank} & {{Worst, excl.\ vib.\ [\%]}} & {Rank} \\
\midrule
ESN02 & 2.63 & 6 & 0.84 & 6 & 18.74 & 6 & 1.80 & 6 \\
ESN05 & 1.48 & 5 & 0.48 & 1 & 10.43 & 5 & 1.46 & 1 \\
ESN08 & 1.38 & 1 & 0.51 & 2 & 9.20 & 1 & 1.48 & 2 \\
ESN09 & 1.42 & 2 & 0.53 & 5 & 9.50 & 2 & 1.59 & 5 \\
ESN10 & 1.43 & 3 & 0.52 & 4 & 9.60 & 3 & 1.50 & 3 \\
ESN11 & 1.44 & 4 & 0.52 & 3 & 9.74 & 4 & 1.55 & 4 \\
\bottomrule
\end{tabular}

  \end{adjustbox}
\end{table}

\subsubsection{One-step versus free-run ranking}
\label{sec:eval:protocol}

\begin{figure}[htbp]
  \centering
  \includegraphics{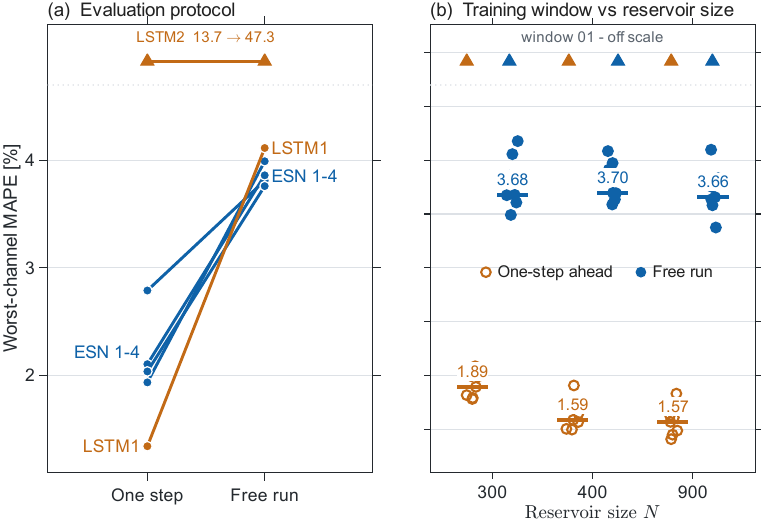}
  \caption{Why the evaluation protocol decides the ranking. The figure of merit is the
  worst channel excluding vibration, in MAPE. (a) The same reference models evaluated one
  step ahead and in free run: the free-run error is about twice the one-step error, and
  the ordering inverts, with the LSTM first under one-step evaluation and last among the
  working models in free run. (b) Three reservoir sizes, each trained on eight windows of
  the same record; bars are medians. The median free-run error is flat across a threefold
  change in reservoir size, whereas the choice of training window spans an order of
  magnitude.}
  \label{fig:benchmark}
\end{figure}

The decision to rank on free-run error was tested rather than assumed. In a separate benchmark, three Echo State Network configurations were retrained on eight windows of the same record, and each resulting network was evaluated twice, one step ahead and in free run, alongside the reference Echo State and LSTM models. The figure of merit is the worst channel excluding vibration. This benchmark uses its own validation run, so its absolute values are not directly comparable with those of Table~\ref{tab:esnmetrics}; what it establishes is the relative behaviour of the two protocols. Figure~\ref{fig:benchmark}(a) shows that free run costs a median factor of \num{2.2}
(interquartile range \numrange{1.9}{2.5} over \num{27} networks) and, more importantly,
that it reorders the models. The LSTM is the most accurate model one step ahead, at
\SI{1.34}{\percent}, and the least accurate of the working models in free run, at
\SI{4.11}{\percent}, behind each of the four reference Echo State Networks (\SIrange{3.76}{3.99}{\percent}).
Selecting on one-step-ahead accuracy would therefore have placed in the controller the
model that degrades fastest along the prediction horizon, which is the concrete form of
the argument anticipated in Section~\ref{sec:model:narx}.

Figure~\ref{fig:benchmark}(b) adds the second reason behind the structural choice. Across
reservoir sizes of \num{300}, \num{400} and \num{900}, the median free-run error is
\SI{3.68}{\percent}, \SI{3.70}{\percent} and \SI{3.66}{\percent}: flat over a threefold
change in size. One-step accuracy does improve with size, from \SI{1.89}{\percent} to
\SI{1.57}{\percent}, and that is precisely the gain that does not survive output feedback.
What does change the outcome is the training window. Of the eight windows, one yields a
divergent model in every configuration, with free-run errors between \SI{55.6}{\percent}
and \SI{90.7}{\percent}. In this benchmark the training window mattered far more than the reservoir size, which is consistent with the manual selection of dynamic windows described
in Section~\ref{sec:data}.

\subsubsection{Free-run error distributions}
\label{sec:eval:histograms}

\begin{figure}[htbp]
  \centering
  \includegraphics[width=\linewidth]{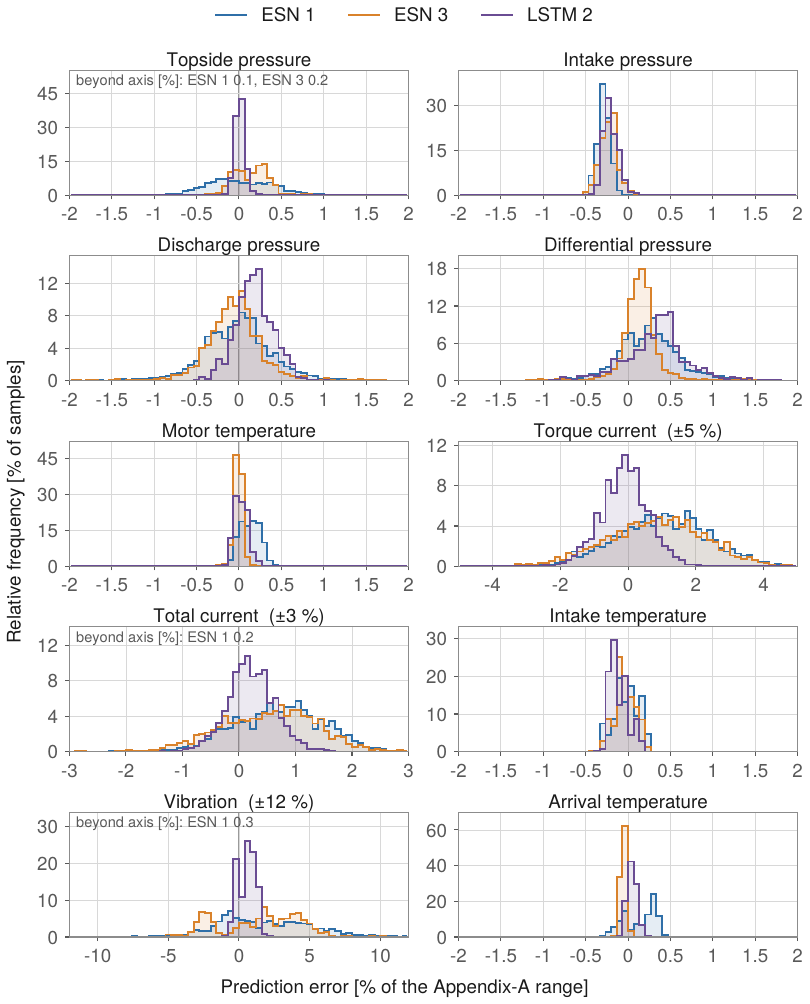}
  \caption{Free-run prediction-error distributions of the candidate networks over the
  validation run, with output feedback active. Each channel is normalised by its
  engineering range, so the panels share a comparable scale.}
  \label{fig:errhist}
\end{figure}

Since the model is queried in free run inside the horizon, what matters is not only how large the
errors are but whether they are centred, skewed or bimodal, and whether a retrained
network improves every channel or trades one against another. Figure~\ref{fig:errhist}
shows the distributions directly, for the previous and the current Echo State Network
and for the best LSTM of the same benchmark, over a common validation run with output
feedback active throughout.

The error is normalised by the engineering range of each variable, rather than by the instantaneous reading. This puts the ten heterogeneous channels on one comparable axis and removes the division-by-near-zero inflation that dominates the vibration MAPE. Seven of the nine channels other than vibration keep their free-run error inside \SI{2}{\percent} of their engineering range, and the total and torque currents, the two with the widest distributions, inside \SIrange{3}{5}{\percent}. Computed as a percentage of the reading, the quantity summarised by the MAPE of Table~\ref{tab:esnmetrics}, the same distributions (not shown) differ almost only in the vibration panel.

\subsection{Closed-loop tuning results}
\label{sec:tuning}

\subsubsection{Weights, tolerance and target sweeps}
\label{sec:tuning:weights}

\begin{figure}[htbp]
  \centering
  \includegraphics[width=\linewidth]{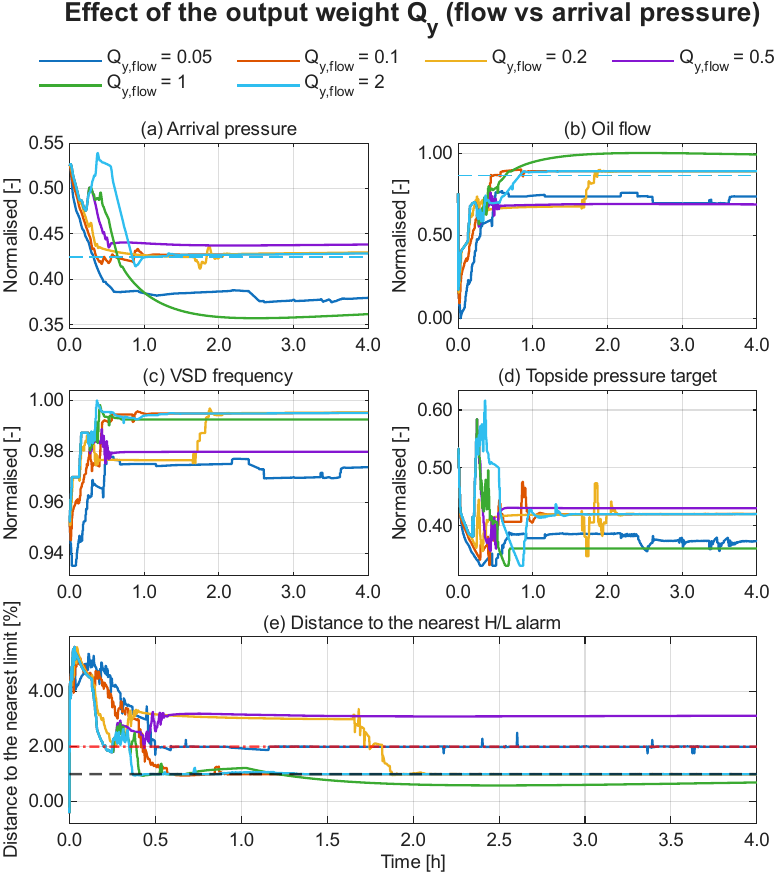}
  \caption{Closed-loop response under different output weightings $Q_y$ (oil flow versus arrival pressure). (a) arrival pressure and (b) oil flow, the two controlled outputs, with their set-points dashed; (c) VSD frequency and (d) topside pressure target, the two manipulated inputs; (e) distance from the nearest H/L alarm, in per~cent, of whichever variable is closest to one at each instant. The axis of (e) measures distance to the \emph{alarm}, so zero is the alarm itself; the dashed line marks the enforced limit at the hard margin $m$ (Table~\ref{tab:margins}) and the dash-dotted line the soft threshold one percentage point beyond it, the same convention as Figure~\ref{fig:tuning_qslack}. }
  \label{fig:tuning}
\end{figure}

Figure~\ref{fig:tuning} shows the closed-loop response for different output weightings $Q_y$, trading the oil flow term against the arrival-pressure term. Increasing the relative weight on the flow output pushes the operating point harder against the active constraints, at the cost of larger excursions in the arrival pressure. The selected weighting keeps the pressure channel regulated while leaving the flow channel free to seek the constraint boundary, which is where the economic optimum lies.

The safety margin $m$ of Eq.~\eqref{eq:margin} was expected to behave as an aggressiveness
dial, and at this operating point it does not. Sweeping it from zero to three percentage
points on the flow channel leaves the closed loop unchanged (not shown): the five trajectories are
indistinguishable in every variable. The explanation is structural rather than numerical.
The margin swept is the one applied to the estimated flow rate, whereas throughout this run
the variable nearest its limit is a process variable held at its own one-point margin, so
the binding constraint never belongs to the channel being varied. The result is reported
here as a negative one because it carries a design consequence: where the active constraint
is a process limit, the margin on the economic channel is inert, and the quantity that
shapes the approach to the limit is the proximity weight examined next. The margins are
configured per variable rather than globally, and the reference values are those
of Table~\ref{tab:margins}: the tight margins belong to the variables whose alarms trip the
well, and the wider one to the estimated flow rate, whose own uncertainty is discussed in
Section~\ref{sec:maps}.

\begin{figure}[htbp]
  \centering
  \includegraphics[width=\linewidth]{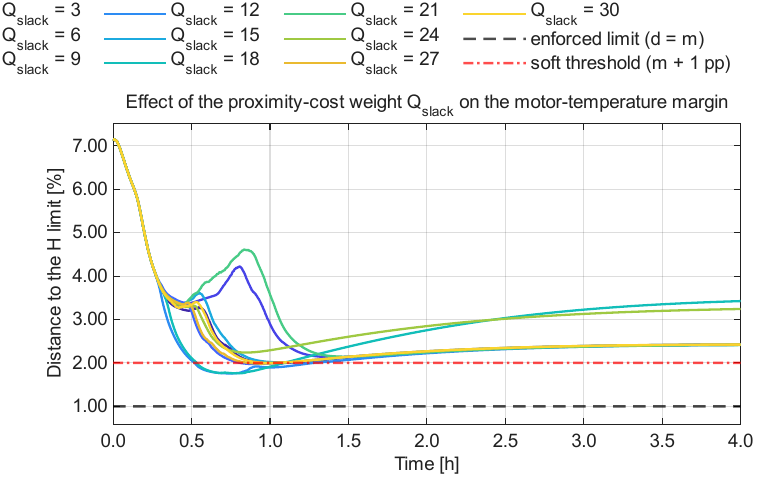}
  \caption{Effect of the constraint-proximity weight $Q_{\text{slack}}$ on the margin of the
  motor temperature to its H alarm. The vertical axis is the percentage distance to the
  \emph{alarm}, that is $d + m$ in the notation of Eq.~\eqref{eq:margin}: the dashed line at
  the hard margin $m = \SI{1}{\percent}$ is therefore the enforced limit, and the dash-dotted
  line one percentage point beyond it is the soft threshold where $\Phi$ becomes active.}
  \label{fig:tuning_qslack}
\end{figure}

The weight $Q_{\text{slack}}$ on the proximity penalty acts on the same trade-off from the
other side, and Figure~\ref{fig:tuning_qslack} isolates it over the range 3 to 30. The figure
reports the distance to the alarm rather than the temperature itself, since that distance is
the quantity $\Phi$ acts on. Three features of it matter.
First, the enforced limit is never crossed, by any of the ten weights: the closest approach
of the whole sweep is about \SI{1.75}{\percent}, against an enforced limit at $m = \SI{1.00}{\percent}$ and a soft threshold at \SI{2.00}{\percent}. Second, that closest approach lies inside the band where
$\Phi$ is active, and it is transient: every trajectory leaves the band within the first
hour and settles between \SI{2.4}{\percent} and \SI{3.4}{\percent}, outside it. This is the
behaviour the formulation of Section~\ref{sec:mpc:cost} was designed to produce, and it is
what a slack-variable formulation does not price, since the cost here is paid for being near
the limit rather than for crossing it. Third, the settled distance is not monotone in
$Q_{\text{slack}}$ over this range. The weight changes the shape of the approach, and it
keeps the trajectory out of the penalty band once settled, but within a well-behaved band of
roughly one percentage point the ordering of the settled values does not follow the ordering
of the weights. The practical reading is that any $Q_{\text{slack}}$ in this range delivers
the intended qualitative behaviour and the parameter does not need fine calibration, which
is convenient in the field.

\begin{figure}[htbp]
  \centering
  \includegraphics[width=\linewidth]{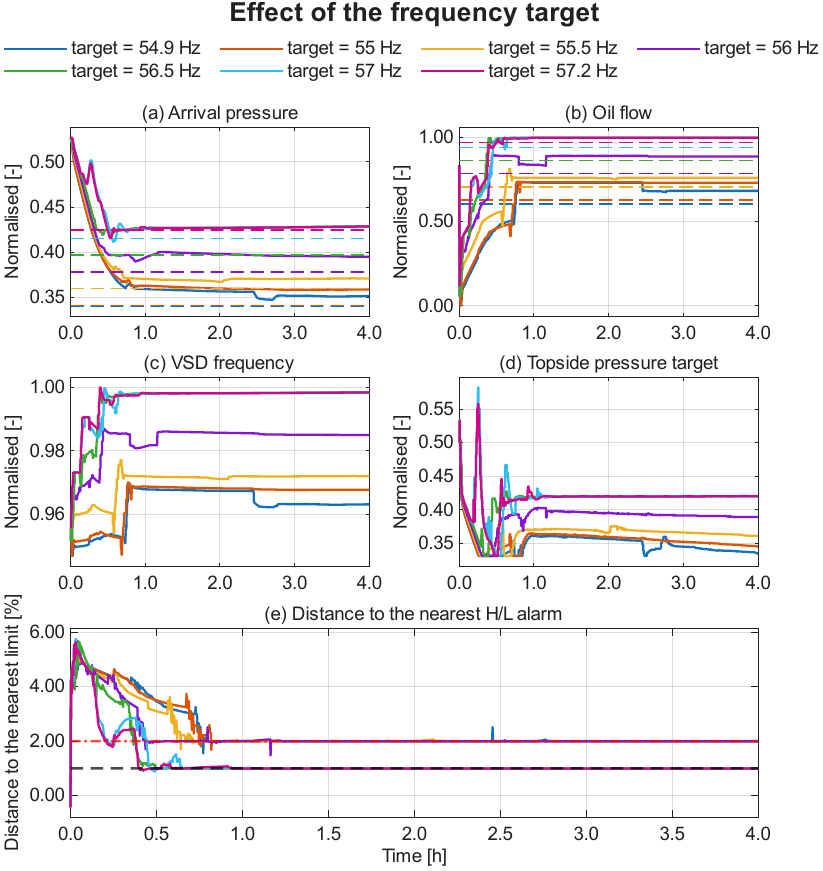}
  \caption{Closed-loop response for a sweep of the frequency target (\SIrange{54.9}{57.4}{\hertz}). Panels as in Figure~\ref{fig:tuning}.}
  \label{fig:tuning_target}
\end{figure}

Figure~\ref{fig:tuning_target} shows the response for a sensitivity analysis of the frequency target between \SI{54.9}{\hertz} and \SI{57.4}{\hertz}. The controller reaches every target below the constraint-limited region and settles on the binding constraint for the targets above it, which confirms that the target acts as an engineering preference rather than a hard set-point. The analysis was repeated with and without measurement noise and with plant--model mismatch, with no qualitative change in behaviour.

\subsubsection{Prediction horizon: cost vs solver time}
\label{sec:tuning:hp}

The prediction horizon is bounded above by the real-time budget and below by a safety requirement, and only the lower one binds. The lower bound comes from the protection logic. A variable that enters its HH alarm band trips the well after \SI{90}{s} of persistence. For the controller to be able to act on a developing violation rather than merely record it, the horizon must extend beyond that delay. With the inference period of \SI{10}{s} fixed by the platform (Section~\ref{sec:mpc:multirate}), $H_p = 10$ gives \SI{100}{s} of prediction horizon, which covers the trip delay with a margin of one sample. Anything shorter leaves the controller predicting inside a window in which the trip has already been decided.

\begin{figure}[htbp]
  \centering
  \includegraphics[width=0.8\linewidth]{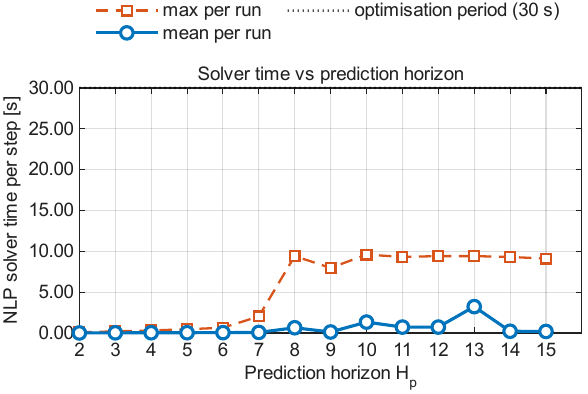}
  \caption{Trade-off between the prediction horizon $H_p$ and the NLP solver time per step; the dotted line marks the supervisory optimisation period.}
  \label{fig:solvertime}
\end{figure}

The upper bound is computational, and it turns out to be generous.
Figure~\ref{fig:solvertime} shows the NLP solver time per step against $H_p$, with the dotted
line marking the \SI{30}{s} supervisory period. The mean time per call stays below
\SI{4}{s} over the whole range. The worst case per run is negligible up to $H_p = 7$ and,
from $H_p = 8$ onwards, sits at about \SI{9.5}{s} independently of the horizon. That
plateau is not the cost of the problem growing with $H_p$: it is the wall-time cap of
\SI{9}{s} imposed on each IPOPT call (Section~S4 of the Supporting Information), together with the overhead
of the call. From $H_p = 8$ on, some calls are cut short and return the best iterate
available. The cap sits well inside the supervisory period, so the real-time budget does not
rule out the longer horizons; what the horizon changes is how often the cap is reached, and
the joint search below shows where the closed loop starts to feel it.

\begin{figure}[htbp]
  \centering
  \includegraphics[width=\linewidth]{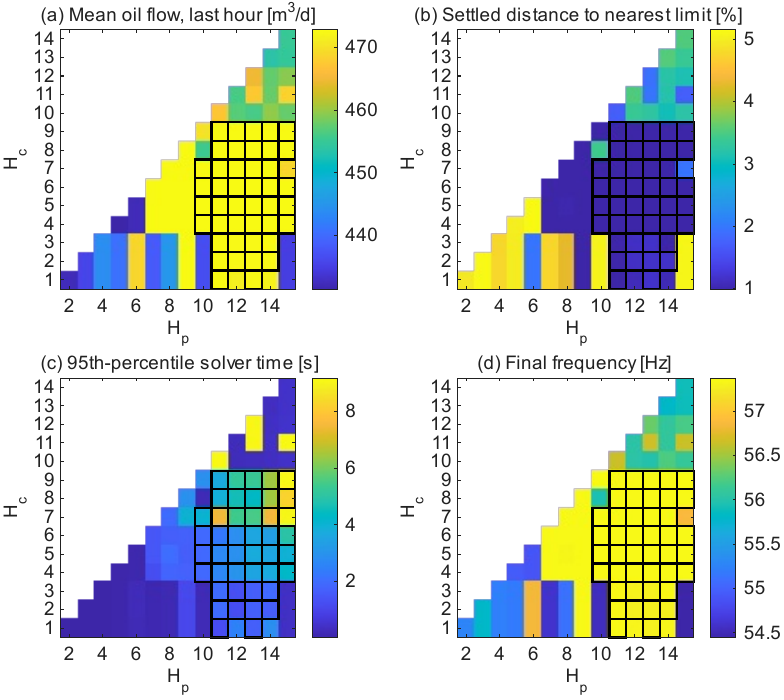}
  \caption{Exhaustive search over the prediction and control horizons: 105 closed-loop runs
  of \SI{4}{h} from the same initial operating point, one per cell. (a) Mean oil flow over
  the last hour. (b) Settled distance to the nearest alarm over the last hour. (c)
  95th-percentile NLP solver time per call; the solver is capped at \SI{9}{s}. (d) Final
  VSD frequency. Outlined cells form the admissible region: $H_p \ge 10$ and a flow within
  \SI{0.1}{\percent} of the best run.}
  \label{fig:hphcgrid}
\end{figure}

The two horizons were then chosen jointly, by the exhaustive search of
Section~\ref{sec:hil:tuningdesign}, summarised in Figure~\ref{fig:hphcgrid}. Every
configuration that succeeds reaches the same operating point, \SI{472.8}{m^3/d} at
\SI{57.36}{Hz} (panels a and d), where the total current settles at \SI{1.0}{\percent}
from its alarm (panel b), that is, exactly on the hard margin. The optimum is therefore set
by the constraints, not by the horizons; the horizons only decide whether the controller
gets there, and three regions of the search space do not. With $H_p \le 6$, no control
horizon reaches it. With $H_c \le 3$ the outcome changes irregularly with the prediction
horizon: the loop fails at $H_p = 7$, 8, 10 and 15, where the frequency stays between
\num{54.4} and \SI{55.1}{Hz} and the flow settles \SIrange{27}{38}{m^3/d} below the
optimum, and reaches it, or comes within \SI{1.2}{m^3/d}, at the neighbouring horizons, which makes short control horizons
unreliable whatever $H_p$ is. With $H_c$ close to $H_p$ ($H_c \ge 10$, which exists only for $H_p \ge 11$),
the loop stops at \SIrange{55.8}{56.7}{Hz}. The solver budget explains only part of this
last failure: four of those fifteen runs reach the \SI{9}{s} cap at the 95th percentile
(panel c), while the others converge quickly to a solution that stops short of the target.
Combined with the safety bound $H_p \ge 10$, the search leaves the admissible region
outlined in the figure, in which the optimum is reached. Its robust core is
$4 \le H_c \le 6$, the only band of control horizons that reaches the optimum at every
prediction horizon from 10 to 15, with a 95th-percentile solver time between \num{1.9} and
\SI{4.9}{s}.

The search establishes how the horizons are chosen and which configurations are acceptable;
it is not meant to single out the pair used on the platform. Within the admissible region,
the final tuning is an operational choice of the asset operator, and, for confidentiality,
the configuration implemented in the field is not reported. That many pairs lead to the
same operating point is itself a result: it makes the closed loop robust to the choice of
horizons and leaves the operation flexibility in that choice.

\subsubsection{Robustness to initialisation}
\label{sec:tuning:init}

\begin{figure}[htbp]
  \centering
  \includegraphics[width=\linewidth]{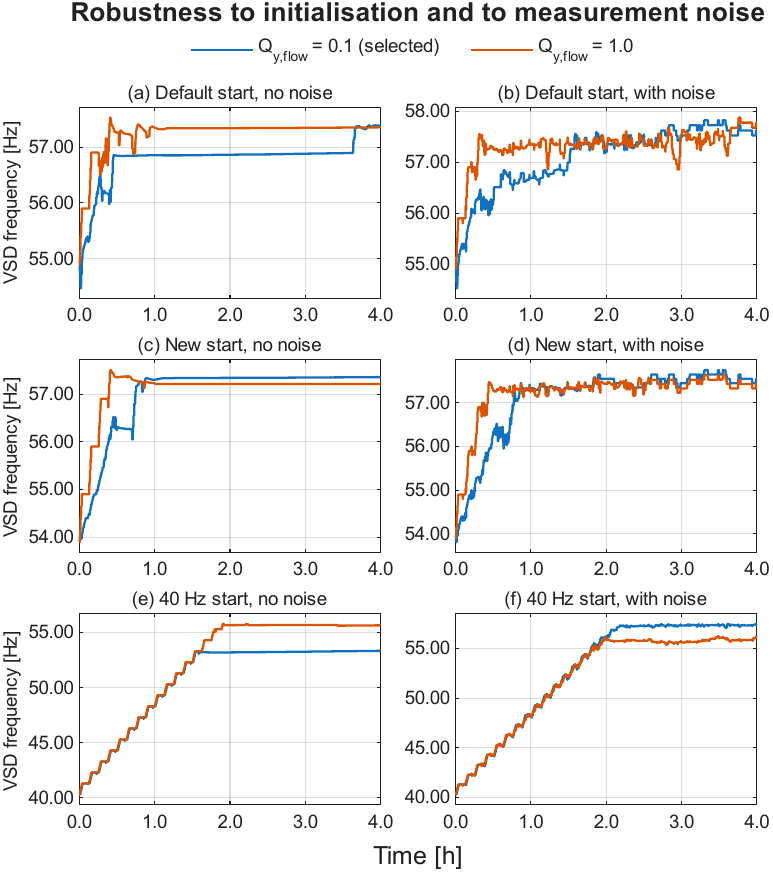}
  \caption{Robustness to initialisation: three initial operating points, each without and with measurement noise, under two weightings of the flow channel. The starting conditions are \SI{54.9}{\hertz} / \SI{31.5}{bar} (default, panels (a) and (b)), \SI{53.9}{\hertz} / \SI{34.2}{bar} (new, (c) and (d)) and \SI{40.3}{\hertz} / \SI{32.25}{bar} (low frequency, (e) and (f)).}
  \label{fig:tuning_scenarios}
\end{figure}

A nonlinear programme solved in receding horizon inherits some sensitivity to its initial
guess. To assess this, the closed loop was simulated from three initial operating points,
each with and without measurement noise and under two weightings of the flow channel,
giving the six scenarios of Figure~\ref{fig:tuning_scenarios}. Two of the initialisations sit
close to the intended operating region and the third starts the well at \SI{40}{Hz}, far
below it.

The two near initialisations behave as intended. Both reach the same neighbourhood of the
operating point under either weighting, with and without noise, and the differences are
confined to the approach: the selected weighting is the slower of the two, taking over
three hours to close the last half hertz in the noise-free case, which is of no operational
consequence at this timescale. The distant initialisation behaves differently. The ascent from \SI{40}{Hz} is
paced by the sliding-window constraint on the drive, which admits one hertz every
\num{7.5} minutes and therefore occupies close to two hours on its own; the loop then
settles wherever the first constraint it meets becomes active, and that settled frequency
differs between the two weightings by \SI{2.4}{\hertz} without noise and \SI{1.5}{\hertz}
with it, in opposite directions. What the scenario grid establishes is therefore weaker
than convergence to a common point, and more useful for commissioning: started inside the
region the identification data covers, the settled point is reproducible and the tuning
decides it; started well outside that region, the settled point is decided by the first
constraint met during the ramp. Warm-starting from the previous solution gave the shortest
solver times and was retained for the field implementation.

An earlier and broader version of the same experiment supports the conclusion with a
larger sample. Five initial operating points were run as eight four-hour cases, three of
them starting at the foot of an acceleration ramp and two near normal operation, with
measurement noise at \SI{20}{dB} throughout. Across the five initialisations sharing a
common target, the mean solver time per call varied between \SI{1.27}{s} and
\SI{1.90}{s}, and the solver returned a feasible solution in
\SIrange{97.3}{99.8}{\percent} of the calls. The closed-loop cost varied by a factor of about twenty across the cases, against about one and a half for the solver time,
which is the signature of initialisations starting at genuinely different distances from
the same optimum rather than of a solver struggling to converge. One case in that set was
configured to reject its own start-up condition, so as to exercise the protection that
holds the previous action when an instance is declared infeasible. It behaved as intended.

\begin{table}[htbp]
  \centering
  \caption{Reference controller tuning of the simulation study. The configuration
  implemented in the field is not disclosed (Section~\ref{sec:tuning:hp}).}
  \label{tab:tuning}
  \begin{tabular}{@{}>{\raggedright\arraybackslash}p{0.50\textwidth} >{\raggedright\arraybackslash}p{0.44\textwidth}@{}}
\toprule
Parameter & Value \\
\midrule
Output weight $Q_y$ (topside press., flow) & [1.0,\ 0.1] \\
Input weight $Q_u$ (freq., press. target) & [0,\ 0] \\
Move-suppression $R$ (freq., press. target) & [1,\ 1] \\
Prediction-error weight $Q_x$ & 1 \\
Constraint-proximity penalty $Q_{\mathrm{slack}}$ & 30 \\
Hard margin $m$ & per variable (Table~\ref{tab:margins}) \\
Soft band beyond the hard margin & 1 percentage point \\
Prediction / control horizons & $H_p \ge 10$, $4 \le H_c \le 6$ (robust core of the admissible region, Section~\ref{sec:tuning:hp}) \\
Inference period $T_s$ / optimisation period $3T_s$ & 10 / 30 s \\
VSD sliding-window limit & 1 Hz / 7.5 min \\
\bottomrule
\end{tabular}

\end{table}

\subsection{Field validation: multi-campaign results}
\label{sec:field}

This section reports the field validation of the controller on the subsea well over eighteen months of campaigns, culminating in the production-gain test of December 2025.
The controller build, the commissioning sequence and the
test protocol are those of Section~\ref{sec:field:protocol}.

\subsubsection{Campaign chronology}
\label{sec:field:chrono}

Table~\ref{tab:campaigns} summarises the six field campaigns. The July 2024 campaign performed open-loop step tests on the manipulated variables and validated the OPC link. In August 2024, the ESN ran online for the first time, confirming that inference and optimisation fit within the \SI{10}{s} and \SI{30}{s} budgets on the platform hardware. The December 2024 campaign integrated the DDMPC with the MPA protection layer. In March 2025, a retrained ESN was re-validated in the field, following the procedure of Section~\ref{sec:evalproto}. The October--November 2025 campaign focused on set-point stability at high frequency targets, and the final campaign of 17 December 2025 was dedicated to the production-gain test. OPC data logging at \SI{10}{s} continued after this campaign and supports the sustained-operation analysis of Section~\ref{sec:field:sustained}.

\begin{table}[htbp]
  \centering
  \caption{Quantitative summary of the field campaigns (2024--2025).}
  \label{tab:campaigns}
  \begin{adjustbox}{max width=\linewidth}
  \begin{tabular}{@{}l p{3.4cm} l p{4.8cm}@{}}
\toprule
Date & Objective & Freq.\ target & Main outcome \\
\midrule
Jul 2024 & MV step tests (open loop) & n/a & Commissioning, OPC link validated \\
Aug 2024 & On-line prediction + timing & n/a & Real-time ESN inference confirmed \\
Dec 2024 & DDMPC-MPA integration & n/a & Protection layer integrated \\
Mar 2025 & Model re-validation & 55--56 Hz & New ESN validated in the field \\
Oct--Nov 2025 & Stability tuning iterations & 56.4--57 Hz & $\sigma_f$ reduced from 0.14 to 0.06 Hz \\
17 Dec 2025 & Production-gain test & 56.7--57 Hz & +6.2\% vs 55 Hz (same day); $\sigma_f=0.02$--$0.057$ Hz \\
\bottomrule
\end{tabular}

  \end{adjustbox}
\end{table}

\subsubsection{Set-point variability and disturbance response}
\label{sec:field:stability}

A primary indicator of control quality in the field is the variability of the frequency command. The October--November 2025 campaign reduced the standard deviation of the frequency command progressively, from \SI{0.14}{Hz} to \SI{0.06}{Hz}, through successive tuning iterations. The December test consolidated the improvement: with the \SI{57}{Hz} target, the deviation was \SI{0.02}{Hz}, despite the persistent multiphase-flow variability in the measured states, and \SI{0.057}{Hz} in the secondary scenario at \SI{56.7}{Hz}. These are the lowest standard deviations recorded in the project, and they were obtained while the controller operated close to the constraint boundary rather than in a conservative interior point.

\begin{figure}[htbp]
	\centering
	\includegraphics[width=\linewidth,height=0.82\textheight,keepaspectratio]{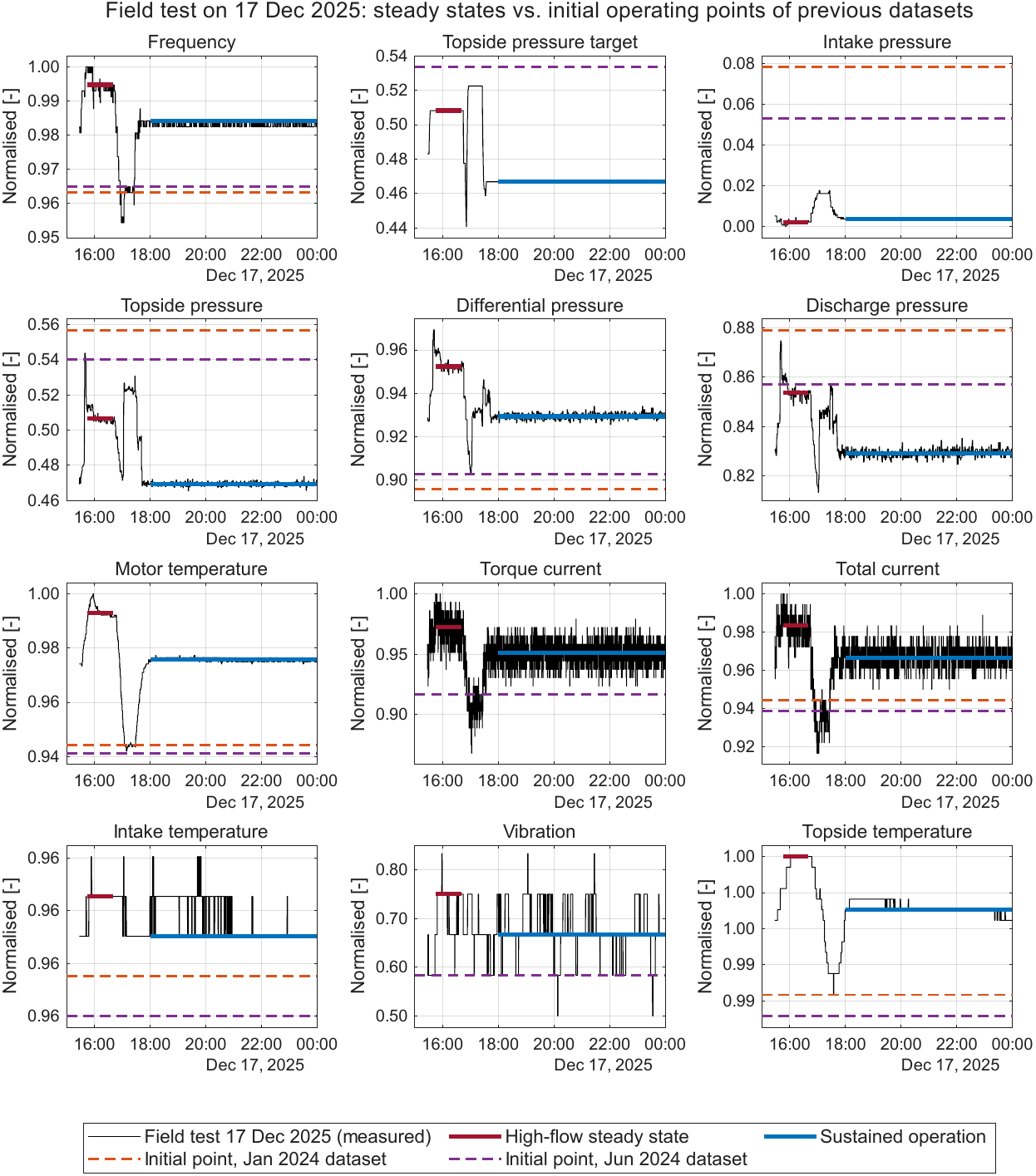}
	\caption{Dynamic behaviour of critical ESP variables under DDMPC control (field test, 17 Dec 2025, min--max normalised): the high-flow steady state and the sustained operation, against the initial operating points of the January and June 2024 data sets.}
	\label{fig:fig_field_test_20251217_ss_comparison}
\end{figure}

Figure~\ref{fig:fig_field_test_20251217_ss_comparison} shows the dynamic behaviour of the critical variables during the December 2025 test, against the initial operating points of the 2024 data sets. The measured ESP frequency exhibits occasional sudden drops of \SIrange{0.1}{0.3}{Hz}. These drops were not commanded by the controller. They originate in the variable-speed drive, whose frequency resolution and actuation delay set a floor on how finely a command can be realised. It should not be inferred that a step of \SI{0.1}{Hz} is negligible: at the operating point of the test, a change of that size is visible in the controlled variables. The DDMPC absorbed them without divergent behaviour: the move-suppression term $R$ acts as an algorithmic low-pass filter, and the sliding-window constraint of Eq.~\eqref{eq:c-vsd} bounds any corrective action, so the frequency returned to its target asymptotically instead of being fought sample by sample.

\subsubsection{Operational safety and validated production gain}
\label{sec:field:gain}

Throughout the closed-loop campaigns, every monitored variable remained strictly within the process constraints, including the indirectly computed thrust limits, even under aggressive maximisation at the \SI{57}{Hz} target. The soft bands worked as designed: transient excursions into the buffer between soft and hard limits occurred during noise rejection, and none propagated to a protection limit.

\begin{figure}[htbp]
  \centering
  \includegraphics{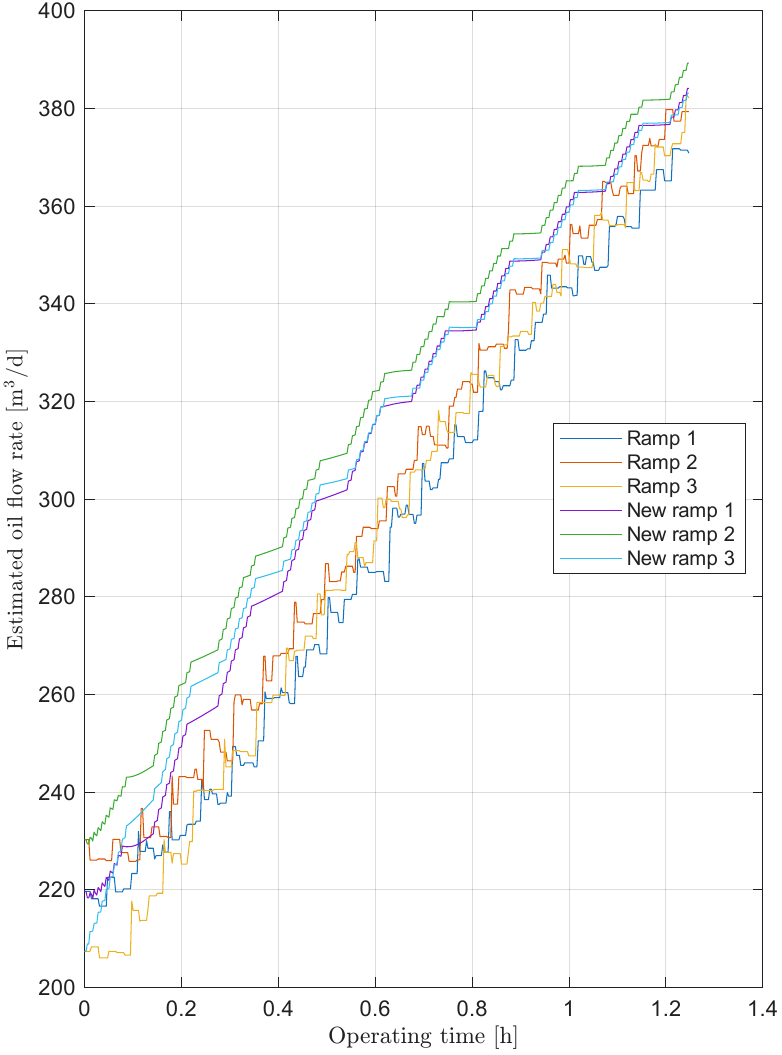}
  \caption{Estimated oil flow rate during the start-up ramps: original versus revised ramp profiles.}
  \label{fig:baseline}
\end{figure}

\begin{figure}[htbp]
  \centering
  \includegraphics{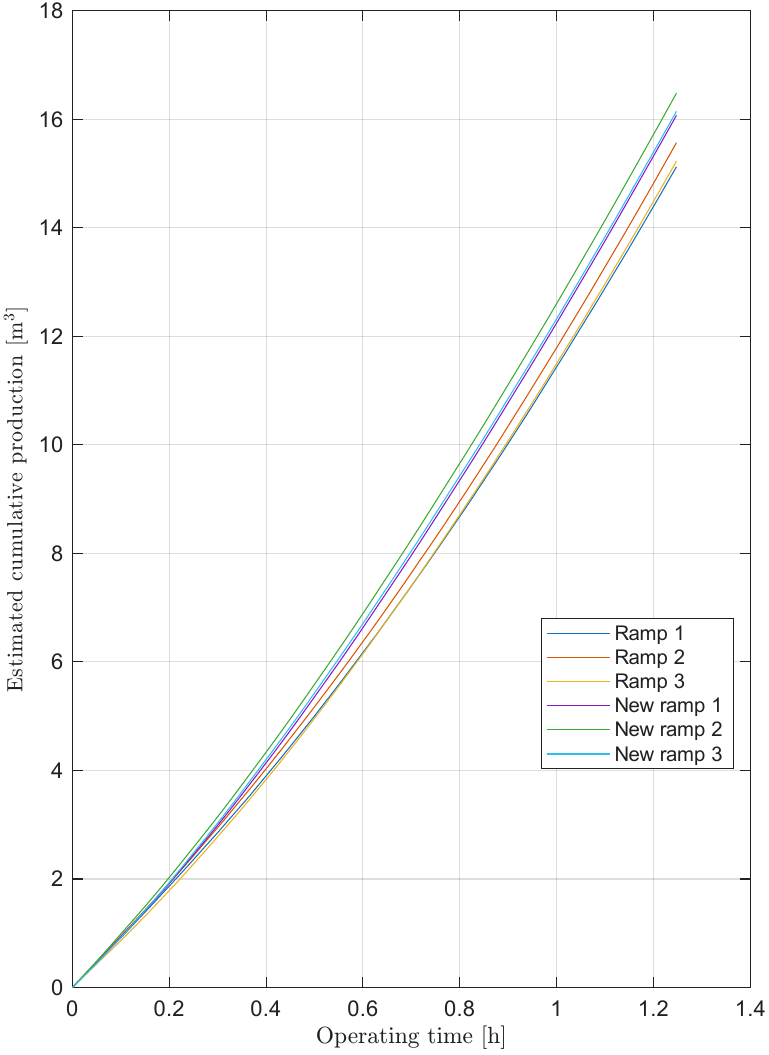}
  \caption{Estimated cumulative production during the start-up ramps: original versus revised ramp profiles.}
  \label{fig:cumulative}
\end{figure}

The production result was obtained under a same-day comparison protocol. The controller sustained an operating point close to the \SI{57}{Hz} target with an average estimated flow rate of \SI{457.75}{\cubic\metre\per\day}. On the same day, under the same well and reservoir conditions, the historical baseline operating point (\SI{55}{Hz} with the choke at its three-year typical position) was reproduced and its flow rate estimated with the same map, yielding \SI{430.94}{\cubic\metre\per\day}. The same-day, like-for-like comparison (Table~\ref{tab:campaigns}) therefore validates a net production gain of 6.2\,\% over the operating point of the preceding three years, and of approximately 2\,\% over operation at \SI{56}{Hz}. Figures~\ref{fig:baseline} and~\ref{fig:cumulative} present the estimated flow rate and the cumulative production during the start-up ramps, where the revised ramp profiles recover additional volume during the transient itself. The gain was achieved purely through the control algorithm, with no physical intervention in the well.

{The economic objective is evaluated on an estimate. There is no flow meter on this well, so the \SI{6.2}{\percent} gain is a gain in estimated flow, measured against a baseline estimated the same way on the same day with the same map. The comparison is internally consistent.}

\subsubsection{Sustained operation after the production-gain test}
\label{sec:field:sustained}

\begin{figure}[htbp]
  \centering
  \includegraphics[width=\linewidth]{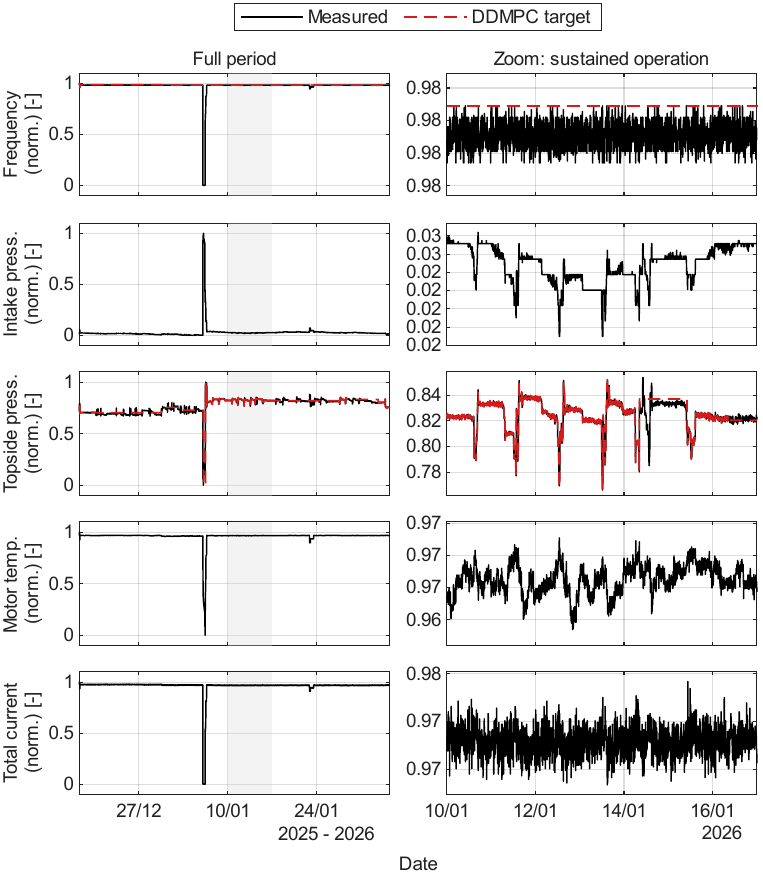}
  \caption{Sustained operation over the seven weeks following the production-gain test (17 December 2025 to 4 February 2026): VSD frequency with the DDMPC target, intake and topside pressures, motor temperature and total current (\num{5}-minute averages of \SI{10}{s} OPC records; each variable min--max normalised over the displayed period). Left: full period, with an unplanned shutdown and restart; the dotted vertical line marks the production-gain test and the shaded band marks the zoom window. Right: zoom into the sustained-operation stretch; note the expanded vertical scale.}
  \label{fig:sustained}
\end{figure}

A production-gain figure measured on a single day answers the efficacy question but not the endurance one. Figure~\ref{fig:sustained} extends the record to seven weeks of OPC records logged on the platform, from the production-gain test to early February 2026, sampled at \SI{10}{s} and averaged over \num{5}-minute windows. The left column shows the full period, including an unplanned shutdown and the subsequent restart, and the right column zooms into a representative week of sustained operation, shaded in the overview. Within the sustained stretch, the frequency remained regulated around the high-production targets, the intake and topside pressures stayed inside their operating bands, and the motor temperature and total current show no upward drift over the period.

{The same record bears on the model, which does not adapt online. The Echo State Network is trained offline and its output weights are fixed thereafter, so a slow change in the dynamics is corrected only at the next retraining, and the correction $e_y(k)$ in the tracking term of Eq.~\eqref{eq:cost} is what absorbs the bias of the controlled outputs in the meantime. No drift is visible over the seven weeks.}

\FloatBarrier

\section{Conclusions}
\label{sec:conc}

This article has followed one subsea ESP well from three years of raw historian records to a field-validated production gain. Two of the decisions carried more weight than the rest. The first concerns the data. A
plant-information historian applies deadband compression independently to each tag, and the
surviving timestamps can therefore invert the order of two coupled events, which disqualifies
the record for dynamic identification however large it is. Separating the exploratory record
from a dedicated uniform-rate acquisition was what made the models usable. The second concerns
the constraints. Pricing the approach to a limit through a proximity penalty, rather than
relaxing the limit with a slack variable, discourages the optimiser from a limit before it is reached
and lets the controller ride an active constraint instead of retreating from it. Since the
economic optimum of this well lies on a boundary and not in the interior, that is not a
refinement but a requirement.

The rest of the contribution is accumulation of evidence. A multi-output NARX structure in
which all ten measured channels are fed back, which keeps the free-run error inside
\SI{2}{\percent} of the engineering range on seven of them and inside \SI{5}{\percent} on the two motor currents. A selection protocol ranked in free run on the
worst channel, with vibration treated separately, since a single unreliable variable is enough to compromise
a controller that operates against limits. Operating maps reconstructed from a simulator
sensitivity study and interrogated through iso-flow curves and the flow gradient. A
simulation environment with the protection layer, measurement noise and instrument resolution
emulated, in which every parameter change was rehearsed before reaching the platform. Eighteen months
and six field campaigns, the last two of which brought the standard deviation of the frequency command
from \SI{0.14}{Hz} to \SI{0.02}{Hz}. And a same-day, like-for-like production gain of
\SI{6.2}{\percent} over the \SI{55}{Hz} operating point of the preceding three years, obtained with no physical intervention in the
well.

{Four directions follow from the limitations noted in the article.} Online
adaptation of the output layer would let the model track reservoir decline and water-cut
changes without a retraining campaign, as in the online learning schemes of digital twins \cite{Rebello2025DT}, and the Ridge structure of the readout makes recursive
updating cheap. The residuals of the model are already computed at every cycle and are
a natural input to fault detection, which the current implementation discards. Dissemination
to the remaining ESP wells of the same platform would test how much of the calibration is
asset-specific. Whether the horizon chosen against the trip delay (Section~\ref{sec:tuning:hp}) lowers the trip rate remains to be measured, and settling it needs
months of continuous operation rather than weeks.

\begin{acknowledgement}
The authors thank CNPq, CAPES (Finance Code 001), FAPESB, and Petrobras SA for their financial support. During the preparation of this work the authors used Claude (Anthropic) to assist in the revision of the text and in the preparation of the table of contents graphic. The authors reviewed and edited the content and take full responsibility for it.
\end{acknowledgement}

\begin{suppinfo}
Dynamic constraint bands at representative frequencies, hyperparameters of the selected Echo State Network, error metrics computed in the model screening, pseudocode of the multi-rate control loop and solver settings (PDF)
\end{suppinfo}

\bibliography{references}

\providecommand{\latin}[1]{#1}
\makeatletter
\providecommand{\doi}
  {\begingroup\let\do\@makeother\dospecials
  \catcode`\{=1 \catcode`\}=2 \doi@aux}
\providecommand{\doi@aux}[1]{\endgroup\texttt{#1}}
\makeatother
\providecommand*\mcitethebibliography{\thebibliography}
\csname @ifundefined\endcsname{endmcitethebibliography}
  {\let\endmcitethebibliography\endthebibliography}{}
\begin{mcitethebibliography}{87}
\providecommand*\natexlab[1]{#1}
\providecommand*\mciteSetBstSublistMode[1]{}
\providecommand*\mciteSetBstMaxWidthForm[2]{}
\providecommand*\mciteBstWouldAddEndPuncttrue
  {\def\EndOfBibitem{\unskip.}}
\providecommand*\mciteBstWouldAddEndPunctfalse
  {\let\EndOfBibitem\relax}
\providecommand*\mciteSetBstMidEndSepPunct[3]{}
\providecommand*\mciteSetBstSublistLabelBeginEnd[3]{}
\providecommand*\EndOfBibitem{}
\mciteSetBstSublistMode{f}
\mciteSetBstMaxWidthForm{subitem}{(\alph{mcitesubitemcount})}
\mciteSetBstSublistLabelBeginEnd
  {\mcitemaxwidthsubitemform\space}
  {\relax}
  {\relax}

\bibitem[Foss(2012)]{Foss2012}
Foss,~B. {Process control in conventional oil and gas fields--Challenges and
  opportunities}. \emph{Control Engineering Practice} \textbf{2012}, \emph{20},
  1058--1064\relax
\mciteBstWouldAddEndPuncttrue
\mciteSetBstMidEndSepPunct{\mcitedefaultmidpunct}
{\mcitedefaultendpunct}{\mcitedefaultseppunct}\relax
\EndOfBibitem
\bibitem[Foss \latin{et~al.}(2018)Foss, Knudsen, and Grimstad]{Foss2018}
Foss,~B.; Knudsen,~B.~R.; Grimstad,~B. {Petroleum production optimization -- A
  static or dynamic problem?} \emph{Computers \& Chemical Engineering}
  \textbf{2018}, \emph{114}, 245--253\relax
\mciteBstWouldAddEndPuncttrue
\mciteSetBstMidEndSepPunct{\mcitedefaultmidpunct}
{\mcitedefaultendpunct}{\mcitedefaultseppunct}\relax
\EndOfBibitem
\bibitem[Takacs(2018)]{Takacs2018}
Takacs,~G. \emph{{Electrical Submersible Pumps Manual}}; Elsevier, 2018\relax
\mciteBstWouldAddEndPuncttrue
\mciteSetBstMidEndSepPunct{\mcitedefaultmidpunct}
{\mcitedefaultendpunct}{\mcitedefaultseppunct}\relax
\EndOfBibitem
\bibitem[Qin and Badgwell(2003)Qin, and Badgwell]{Qin2003}
Qin,~S.; Badgwell,~T.~A. {A survey of industrial model predictive control
  technology}. \emph{Control Engineering Practice} \textbf{2003}, \emph{11},
  733--764\relax
\mciteBstWouldAddEndPuncttrue
\mciteSetBstMidEndSepPunct{\mcitedefaultmidpunct}
{\mcitedefaultendpunct}{\mcitedefaultseppunct}\relax
\EndOfBibitem
\bibitem[Pavlov \latin{et~al.}(2014)Pavlov, Krishnamoorthy, Fjalestad, Aske,
  and Fredriksen]{Pavlov2014}
Pavlov,~A.; Krishnamoorthy,~D.; Fjalestad,~K.; Aske,~E.; Fredriksen,~M.
  {Modelling and model predictive control of oil wells with Electric
  Submersible Pumps}. 2014 IEEE Conference on Control Applications (CCA). 2014;
  pp 586--592\relax
\mciteBstWouldAddEndPuncttrue
\mciteSetBstMidEndSepPunct{\mcitedefaultmidpunct}
{\mcitedefaultendpunct}{\mcitedefaultseppunct}\relax
\EndOfBibitem
\bibitem[Coulson \latin{et~al.}(2019)Coulson, Lygeros, and
  Dorfler]{Coulson2019}
Coulson,~J.; Lygeros,~J.; Dorfler,~F. {Data-Enabled Predictive Control: In the
  Shallows of the DeePC}. 2019 18th European Control Conference (ECC). 2019; pp
  307--312\relax
\mciteBstWouldAddEndPuncttrue
\mciteSetBstMidEndSepPunct{\mcitedefaultmidpunct}
{\mcitedefaultendpunct}{\mcitedefaultseppunct}\relax
\EndOfBibitem
\bibitem[Berberich \latin{et~al.}(2021)Berberich, Kohler, Muller, and
  Allgower]{Berberich2021}
Berberich,~J.; Kohler,~J.; Muller,~M.~A.; Allgower,~F. {Data-Driven Model
  Predictive Control With Stability and Robustness Guarantees}. \emph{IEEE
  Transactions on Automatic Control} \textbf{2021}, \emph{66}, 1702--1717\relax
\mciteBstWouldAddEndPuncttrue
\mciteSetBstMidEndSepPunct{\mcitedefaultmidpunct}
{\mcitedefaultendpunct}{\mcitedefaultseppunct}\relax
\EndOfBibitem
\bibitem[Jaeger and Haas(2004)Jaeger, and Haas]{Jaeger2004}
Jaeger,~H.; Haas,~H. {Harnessing Nonlinearity: Predicting Chaotic Systems and
  Saving Energy in Wireless Communication}. \emph{Science} \textbf{2004},
  \emph{304}, 78--80\relax
\mciteBstWouldAddEndPuncttrue
\mciteSetBstMidEndSepPunct{\mcitedefaultmidpunct}
{\mcitedefaultendpunct}{\mcitedefaultseppunct}\relax
\EndOfBibitem
\bibitem[Lukoševičius and Jaeger(2009)Lukoševičius, and
  Jaeger]{Lukosevicius2009}
Lukoševičius,~M.; Jaeger,~H. {Reservoir computing approaches to recurrent
  neural network training}. \emph{Computer Science Review} \textbf{2009},
  \emph{3}, 127--149\relax
\mciteBstWouldAddEndPuncttrue
\mciteSetBstMidEndSepPunct{\mcitedefaultmidpunct}
{\mcitedefaultendpunct}{\mcitedefaultseppunct}\relax
\EndOfBibitem
\bibitem[Jordanou \latin{et~al.}(2018)Jordanou, Camponogara, Antonelo, and
  Schmitz~Aguiar]{Jordanou2018}
Jordanou,~J.~P.; Camponogara,~E.; Antonelo,~E.~A.; Schmitz~Aguiar,~M.~A.
  {Nonlinear Model Predictive Control of an Oil Well with Echo State Networks}.
  \emph{IFAC-PapersOnLine} \textbf{2018}, \emph{51}, 13--18\relax
\mciteBstWouldAddEndPuncttrue
\mciteSetBstMidEndSepPunct{\mcitedefaultmidpunct}
{\mcitedefaultendpunct}{\mcitedefaultseppunct}\relax
\EndOfBibitem
\bibitem[Jordanou \latin{et~al.}(2022)Jordanou, Antonelo, and
  Camponogara]{Jordanou2022}
Jordanou,~J.~P.; Antonelo,~E.~A.; Camponogara,~E. {Echo State Networks for
  Practical Nonlinear Model Predictive Control of Unknown Dynamic Systems}.
  \emph{IEEE Transactions on Neural Networks and Learning Systems}
  \textbf{2022}, \emph{33}, 2615--2629\relax
\mciteBstWouldAddEndPuncttrue
\mciteSetBstMidEndSepPunct{\mcitedefaultmidpunct}
{\mcitedefaultendpunct}{\mcitedefaultseppunct}\relax
\EndOfBibitem
\bibitem[Rebello \latin{et~al.}(2022)Rebello, Marrocos, Costa, Santana,
  Rodrigues, Ribeiro, and Nogueira]{Rebello2022guideline}
Rebello,~C.~M.; Marrocos,~P.~H.; Costa,~E.~A.; Santana,~V.~V.;
  Rodrigues,~A.~E.; Ribeiro,~A.~M.; Nogueira,~I. B.~R. {Machine Learning-Based
  Dynamic Modeling for Process Engineering Applications: A Guideline for
  Simulation and Prediction from Perceptron to Deep Learning}. \emph{Processes}
  \textbf{2022}, \emph{10}, 250\relax
\mciteBstWouldAddEndPuncttrue
\mciteSetBstMidEndSepPunct{\mcitedefaultmidpunct}
{\mcitedefaultendpunct}{\mcitedefaultseppunct}\relax
\EndOfBibitem
\bibitem[Santana \latin{et~al.}(2025)Santana, Rebello, Costa, Abreu, Reges,
  Mendes, Schnitman, Ribeiro, Fontana, and Nogueira]{Santana2025multistep}
Santana,~V.~V.; Rebello,~C.~M.; Costa,~E.~A.; Abreu,~O. S.~L.; Reges,~G.;
  Mendes,~T. P.~G.; Schnitman,~L.; Ribeiro,~M.~P.; Fontana,~M.; Nogueira,~I.
  {Recurrent Deep Learning Models for Multi-step Ahead Prediction: Comparison
  and Evaluation for Real Electrical Submersible Pump (ESP) System}. Systems
  and Control Transactions Proceedings of the 35th European Symposium on
  Computer Aided Process Engineering (ESCAPE 35). 2025; pp 1208--1213\relax
\mciteBstWouldAddEndPuncttrue
\mciteSetBstMidEndSepPunct{\mcitedefaultmidpunct}
{\mcitedefaultendpunct}{\mcitedefaultseppunct}\relax
\EndOfBibitem
\bibitem[Reges \latin{et~al.}(2026)Reges, Mendes, Costa, Rebello, Abreu, Lima,
  Santana, Bacellar, Nogueira, Fontana, Ribeiro, and Schnitman]{Reges2026}
Reges,~G.; Mendes,~T.; Costa,~E.; Rebello,~C.; Abreu,~O.~S.; Lima,~L.;
  Santana,~V.; Bacellar,~F.; Nogueira,~I.; Fontana,~M.; Ribeiro,~M.~P.;
  Schnitman,~L. {A Hybrid Predictive Control Approach with Echo State Networks
  for Real-Time Optimization of Electrical Submersible Pump Systems}. SPE Latin
  American and Caribbean Petroleum Engineering Conference. 2026; p
  D031S021R003\relax
\mciteBstWouldAddEndPuncttrue
\mciteSetBstMidEndSepPunct{\mcitedefaultmidpunct}
{\mcitedefaultendpunct}{\mcitedefaultseppunct}\relax
\EndOfBibitem
\bibitem[Jordanou \latin{et~al.}(2019)Jordanou, Antonelo, and
  Camponogara]{Jordanou2019}
Jordanou,~J.~P.; Antonelo,~E.~A.; Camponogara,~E. {Online learning control with
  Echo State Networks of an oil production platform}. \emph{Engineering
  Applications of Artificial Intelligence} \textbf{2019}, \emph{85},
  214--228\relax
\mciteBstWouldAddEndPuncttrue
\mciteSetBstMidEndSepPunct{\mcitedefaultmidpunct}
{\mcitedefaultendpunct}{\mcitedefaultseppunct}\relax
\EndOfBibitem
\bibitem[Antonelo \latin{et~al.}(2024)Antonelo, Camponogara, Seman, Jordanou,
  de~Souza, and Hübner]{Antonelo2024}
Antonelo,~E.~A.; Camponogara,~E.; Seman,~L.~O.; Jordanou,~J.~P.;
  de~Souza,~E.~R.; Hübner,~J.~F. {Physics-informed neural nets for control of
  dynamical systems}. \emph{Neurocomputing} \textbf{2024}, \emph{579},
  127419\relax
\mciteBstWouldAddEndPuncttrue
\mciteSetBstMidEndSepPunct{\mcitedefaultmidpunct}
{\mcitedefaultendpunct}{\mcitedefaultseppunct}\relax
\EndOfBibitem
\bibitem[Armenio \latin{et~al.}(2019)Armenio, Terzi, Farina, and
  Scattolini]{Armenio2019}
Armenio,~L.~B.; Terzi,~E.; Farina,~M.; Scattolini,~R. {Model Predictive Control
  Design for Dynamical Systems Learned by Echo State Networks}. \emph{IEEE
  Control Systems Letters} \textbf{2019}, \emph{3}, 1044--1049\relax
\mciteBstWouldAddEndPuncttrue
\mciteSetBstMidEndSepPunct{\mcitedefaultmidpunct}
{\mcitedefaultendpunct}{\mcitedefaultseppunct}\relax
\EndOfBibitem
\bibitem[Bonassi \latin{et~al.}(2022)Bonassi, Farina, Xie, and
  Scattolini]{Bonassi2022}
Bonassi,~F.; Farina,~M.; Xie,~J.; Scattolini,~R. {On Recurrent Neural Networks
  for learning-based control: Recent results and ideas for future
  developments}. \emph{Journal of Process Control} \textbf{2022}, \emph{114},
  92--104\relax
\mciteBstWouldAddEndPuncttrue
\mciteSetBstMidEndSepPunct{\mcitedefaultmidpunct}
{\mcitedefaultendpunct}{\mcitedefaultseppunct}\relax
\EndOfBibitem
\bibitem[Ren \latin{et~al.}(2022)Ren, Alhajeri, Luo, Chen, Abdullah, Wu, and
  Christofides]{Ren2022}
Ren,~Y.~M.; Alhajeri,~M.~S.; Luo,~J.; Chen,~S.; Abdullah,~F.; Wu,~Z.;
  Christofides,~P.~D. {A tutorial review of neural network modeling approaches
  for model predictive control}. \emph{Computers \& Chemical Engineering}
  \textbf{2022}, \emph{165}, 107956\relax
\mciteBstWouldAddEndPuncttrue
\mciteSetBstMidEndSepPunct{\mcitedefaultmidpunct}
{\mcitedefaultendpunct}{\mcitedefaultseppunct}\relax
\EndOfBibitem
\bibitem[Wu \latin{et~al.}(2019)Wu, Tran, Rincon, and Christofides]{Wu2019}
Wu,~Z.; Tran,~A.; Rincon,~D.; Christofides,~P.~D. {Machine learning-based
  predictive control of nonlinear processes. Part I: Theory}. \emph{AIChE
  Journal} \textbf{2019}, \emph{65}, e16729\relax
\mciteBstWouldAddEndPuncttrue
\mciteSetBstMidEndSepPunct{\mcitedefaultmidpunct}
{\mcitedefaultendpunct}{\mcitedefaultseppunct}\relax
\EndOfBibitem
\bibitem[Hewing \latin{et~al.}(2020)Hewing, Wabersich, Menner, and
  Zeilinger]{Hewing2020}
Hewing,~L.; Wabersich,~K.~P.; Menner,~M.; Zeilinger,~M.~N. {Learning-Based
  Model Predictive Control: Toward Safe Learning in Control}. \emph{Annual
  Review of Control, Robotics, and Autonomous Systems} \textbf{2020}, \emph{3},
  269--296\relax
\mciteBstWouldAddEndPuncttrue
\mciteSetBstMidEndSepPunct{\mcitedefaultmidpunct}
{\mcitedefaultendpunct}{\mcitedefaultseppunct}\relax
\EndOfBibitem
\bibitem[Costa \latin{et~al.}(2021)Costa, de~Abreu, Silva, Ribeiro, and
  Schnitman]{Costa2021Bayes}
Costa,~E.; de~Abreu,~O.; Silva,~T.~O.; Ribeiro,~M.; Schnitman,~L. {A Bayesian
  approach to the dynamic modeling of ESP-lifted oil well systems: An
  experimental validation on an ESP prototype}. \emph{Journal of Petroleum
  Science and Engineering} \textbf{2021}, \emph{205}, 108880\relax
\mciteBstWouldAddEndPuncttrue
\mciteSetBstMidEndSepPunct{\mcitedefaultmidpunct}
{\mcitedefaultendpunct}{\mcitedefaultseppunct}\relax
\EndOfBibitem
\bibitem[Almeida~Costa \latin{et~al.}(2024)Almeida~Costa, de~Menezes~Rebello,
  Viena~Santana, Reges, de~Oliveira~Silva, Santana Luiz~de Abreu,
  Pellegrini~Ribeiro, Pereira~Foresti, Fontana, Bessa~dos Reis~Nogueira, and
  Schnitman]{AlmeidaCosta2024}
Almeida~Costa,~E.; de~Menezes~Rebello,~C.; Viena~Santana,~V.; Reges,~G.;
  de~Oliveira~Silva,~T.; Santana Luiz~de Abreu,~O.; Pellegrini~Ribeiro,~M.;
  Pereira~Foresti,~B.; Fontana,~M.; Bessa~dos Reis~Nogueira,~I.; Schnitman,~L.
  {An uncertainty approach for Electric Submersible Pump modeling through Deep
  Neural Network}. \emph{Heliyon} \textbf{2024}, \emph{10}, e24047\relax
\mciteBstWouldAddEndPuncttrue
\mciteSetBstMidEndSepPunct{\mcitedefaultmidpunct}
{\mcitedefaultendpunct}{\mcitedefaultseppunct}\relax
\EndOfBibitem
\bibitem[de~Abreu \latin{et~al.}(2024)de~Abreu, Ribeiro, Schnitman, and
  Martins]{deAbreu2024}
de~Abreu,~O.~S.; Ribeiro,~M.~P.; Schnitman,~L.; Martins,~M.~A. {Experimental
  evaluation of soft sensors in an electrical submersible pump system:
  Validation in an ESP-lifted oil well pilot plant}. \emph{Geoenergy Science
  and Engineering} \textbf{2024}, \emph{240}, 213033\relax
\mciteBstWouldAddEndPuncttrue
\mciteSetBstMidEndSepPunct{\mcitedefaultmidpunct}
{\mcitedefaultendpunct}{\mcitedefaultseppunct}\relax
\EndOfBibitem
\bibitem[Rebello \latin{et~al.}(2025)Rebello, Costa, Luiz~de Abreu, Reges,
  Viena, Guimares~Mendes, Fontana, Ribeiro~Bacellar, Ribeiro, Foresti,
  Schnitman, and Nogueira]{Rebello2025RNN}
Rebello,~C.~M.; Costa,~E.~A.; Luiz~de Abreu,~O.~S.; Reges,~G.; Viena,~V.;
  Guimares~Mendes,~T.~P.; Fontana,~M.; Ribeiro~Bacellar,~F.~R.; Ribeiro,~M.~P.;
  Foresti,~B.~P.; Schnitman,~L.; Nogueira,~I.~B. {Application of Recurrent
  Neural Networks for Predictive Modeling of Electrical Submersible Pumps in
  Oil Extraction}. \emph{IFAC-PapersOnLine} \textbf{2025}, \emph{59},
  49--54\relax
\mciteBstWouldAddEndPuncttrue
\mciteSetBstMidEndSepPunct{\mcitedefaultmidpunct}
{\mcitedefaultendpunct}{\mcitedefaultseppunct}\relax
\EndOfBibitem
\bibitem[Krishnamoorthy \latin{et~al.}(2016)Krishnamoorthy, Bergheim, Pavlov,
  Fredriksen, and Fjalestad]{Krishnamoorthy2016}
Krishnamoorthy,~D.; Bergheim,~E.~M.; Pavlov,~A.; Fredriksen,~M.; Fjalestad,~K.
  {Modelling and Robustness Analysis of Model Predictive Control for Electrical
  Submersible Pump Lifted Heavy Oil Wells}. \emph{IFAC-PapersOnLine}
  \textbf{2016}, \emph{49}, 544--549\relax
\mciteBstWouldAddEndPuncttrue
\mciteSetBstMidEndSepPunct{\mcitedefaultmidpunct}
{\mcitedefaultendpunct}{\mcitedefaultseppunct}\relax
\EndOfBibitem
\bibitem[Binder \latin{et~al.}(2014)Binder, Kufoalor, Pavlov, and
  Johansen]{Binder2014}
Binder,~B. J.~T.; Kufoalor,~D. K.~M.; Pavlov,~A.; Johansen,~T.~A. {Embedded
  Model Predictive Control for an Electric Submersible Pump on a Programmable
  Logic Controller}. 2014 IEEE Conference on Control Applications (CCA). 2014;
  pp 579--585\relax
\mciteBstWouldAddEndPuncttrue
\mciteSetBstMidEndSepPunct{\mcitedefaultmidpunct}
{\mcitedefaultendpunct}{\mcitedefaultseppunct}\relax
\EndOfBibitem
\bibitem[Patel \latin{et~al.}(2019)Patel, Bakhurji, Salloum, Kim, Winarno, and
  Mubarak]{Patel2019}
Patel,~K.~M.; Bakhurji,~A.~S.; Salloum,~H.~S.; Kim,~H.~K.; Winarno,~M.~B.;
  Mubarak,~S. {Use of Advanced Process Control for Automating Conventional
  Oilfield Operations}. \emph{SPE Production \& Operations} \textbf{2019},
  \emph{34}, 678--692\relax
\mciteBstWouldAddEndPuncttrue
\mciteSetBstMidEndSepPunct{\mcitedefaultmidpunct}
{\mcitedefaultendpunct}{\mcitedefaultseppunct}\relax
\EndOfBibitem
\bibitem[Delou \latin{et~al.}(2019)Delou, de~Azevedo, Krishnamoorthy, de~Souza,
  and Secchi]{Delou2019}
Delou,~P. d.~A.; de~Azevedo,~J.~P.; Krishnamoorthy,~D.; de~Souza,~M.~B.;
  Secchi,~A.~R. {Model Predictive Control with Adaptive Strategy Applied to an
  Electric Submersible Pump in a Subsea Environment}. \emph{IFAC-PapersOnLine}
  \textbf{2019}, \emph{52}, 784--789\relax
\mciteBstWouldAddEndPuncttrue
\mciteSetBstMidEndSepPunct{\mcitedefaultmidpunct}
{\mcitedefaultendpunct}{\mcitedefaultseppunct}\relax
\EndOfBibitem
\bibitem[Santana \latin{et~al.}(2021)Santana, Fontes, Schnitman, and
  Martins]{Santana2021}
Santana,~B.~A.; Fontes,~R.~M.; Schnitman,~L.; Martins,~M.~A. {An Adaptive
  Infinite Horizon Model Predictive Control Strategy Applied to an ESP-lifted
  Oil Well System}. \emph{IFAC-PapersOnLine} \textbf{2021}, \emph{54},
  176--181\relax
\mciteBstWouldAddEndPuncttrue
\mciteSetBstMidEndSepPunct{\mcitedefaultmidpunct}
{\mcitedefaultendpunct}{\mcitedefaultseppunct}\relax
\EndOfBibitem
\bibitem[Santana \latin{et~al.}(2022)Santana, Fontes, and Martins]{Santana2022}
Santana,~B.~A.; Fontes,~R.~M.; Martins,~M.~A. {An implementable zone NMPC
  applied to an ESP-lifted oil well system: Handling the lack of measurements
  with nonlinear state estimator coupling}. \emph{Journal of Petroleum Science
  and Engineering} \textbf{2022}, \emph{216}, 110816\relax
\mciteBstWouldAddEndPuncttrue
\mciteSetBstMidEndSepPunct{\mcitedefaultmidpunct}
{\mcitedefaultendpunct}{\mcitedefaultseppunct}\relax
\EndOfBibitem
\bibitem[Costa \latin{et~al.}(2025)Costa, de~Abreu, Reges, Rebello, Fontana,
  Ribeiro, Nogueira, and Schnitman]{Costa2025IFACMPC}
Costa,~E.~A.; de~Abreu,~O. S.~L.; Reges,~G.; Rebello,~C.~M.; Fontana,~M.;
  Ribeiro,~M.~P.; Nogueira,~I.~B.; Schnitman,~L. {Standard MPC and
  Inputs-Target MPC implementation comparison in ESP systems}.
  \emph{IFAC-PapersOnLine} \textbf{2025}, \emph{59}, 91--96\relax
\mciteBstWouldAddEndPuncttrue
\mciteSetBstMidEndSepPunct{\mcitedefaultmidpunct}
{\mcitedefaultendpunct}{\mcitedefaultseppunct}\relax
\EndOfBibitem
\bibitem[Costa \latin{et~al.}(2026)Costa, de~Abreu, Reges, Silva, Rebello,
  Fontana, Ribeiro, Nogueira, and Schnitman]{Costa2026legacy}
Costa,~E.~A.; de~Abreu,~O. S.~L.; Reges,~G.; Silva,~T. d.~O.; Rebello,~C.~M.;
  Fontana,~M.; Ribeiro,~M.~P.; Nogueira,~I. B.~R.; Schnitman,~L. {Comparative
  Real-Time Implementation of Standard and Input-Target MPC for ESP-Based
  Artificial Lift Systems in a Legacy Automation Architecture}.
  \emph{Automation} \textbf{2026}, \emph{7}, 70\relax
\mciteBstWouldAddEndPuncttrue
\mciteSetBstMidEndSepPunct{\mcitedefaultmidpunct}
{\mcitedefaultendpunct}{\mcitedefaultseppunct}\relax
\EndOfBibitem
\bibitem[Rebello \latin{et~al.}(2025)Rebello, Costa, Ribeiro, Fontana,
  Schnitman, and Nogueira]{Rebello2025NMPC}
Rebello,~C.~M.; Costa,~E.~A.; Ribeiro,~M.~P.; Fontana,~M.; Schnitman,~L.;
  Nogueira,~I. B. d.~R. {Non-Linear Model Predictive Control for Oil Production
  in Wells Using Electric Submersible Pumps}. Systems and Control Transactions
  Proceedings of the 35th European Symposium on Computer Aided Process
  Engineering (ESCAPE 35). 2025; pp 1145--1150\relax
\mciteBstWouldAddEndPuncttrue
\mciteSetBstMidEndSepPunct{\mcitedefaultmidpunct}
{\mcitedefaultendpunct}{\mcitedefaultseppunct}\relax
\EndOfBibitem
\bibitem[Menezes~Rebello \latin{et~al.}(2026)Menezes~Rebello, Almeida~Costa,
  and Nogueira]{MenezesRebello2026}
Menezes~Rebello,~C.; Almeida~Costa,~E.; Nogueira,~I.~B. {Offset-free
  Lyapunov-based stabilising robust nonlinear control using data-driven model:
  A nonlinear multi-model computationally efficient approach}. \emph{Control
  Engineering Practice} \textbf{2026}, \emph{173}, 107013\relax
\mciteBstWouldAddEndPuncttrue
\mciteSetBstMidEndSepPunct{\mcitedefaultmidpunct}
{\mcitedefaultendpunct}{\mcitedefaultseppunct}\relax
\EndOfBibitem
\bibitem[Rebello \latin{et~al.}(2026)Rebello, Costa, Rapello~dos Santos, and
  Nogueira]{Rebello2026minmax}
Rebello,~C.~M.; Costa,~E.~A.; Rapello~dos Santos,~A.; Nogueira,~I. B.~R.
  {Robust Min--Max MPC with Symbolic Regression Ensembles: Constraint
  Satisfaction for Uncertain Nonlinear Process Control}. \emph{Industrial \&
  Engineering Chemistry Research} \textbf{2026}, \emph{65}, 17277--17297\relax
\mciteBstWouldAddEndPuncttrue
\mciteSetBstMidEndSepPunct{\mcitedefaultmidpunct}
{\mcitedefaultendpunct}{\mcitedefaultseppunct}\relax
\EndOfBibitem
\bibitem[Costa \latin{et~al.}(2026)Costa, Rebello, and Nogueira]{Costa2026RTO}
Costa,~E.~A.; Rebello,~C. d.~M.; Nogueira,~I.~B. {RTO and advanced process
  control for electric submersible pumps using surrogate models}. \emph{Journal
  of Process Control} \textbf{2026}, \emph{161}, 103683\relax
\mciteBstWouldAddEndPuncttrue
\mciteSetBstMidEndSepPunct{\mcitedefaultmidpunct}
{\mcitedefaultendpunct}{\mcitedefaultseppunct}\relax
\EndOfBibitem
\bibitem[de~Oliveira and Biegler(1994)de~Oliveira, and Biegler]{deOliveira1994}
de~Oliveira,~N. M.~C.; Biegler,~L.~T. {Constraint handing and stability
  properties of model-predictive control}. \emph{AIChE Journal} \textbf{1994},
  \emph{40}, 1138--1155\relax
\mciteBstWouldAddEndPuncttrue
\mciteSetBstMidEndSepPunct{\mcitedefaultmidpunct}
{\mcitedefaultendpunct}{\mcitedefaultseppunct}\relax
\EndOfBibitem
\bibitem[Scokaert and Rawlings(1999)Scokaert, and Rawlings]{Scokaert1999}
Scokaert,~P. O.~M.; Rawlings,~J.~B. {Feasibility issues in linear model
  predictive control}. \emph{AIChE Journal} \textbf{1999}, \emph{45},
  1649--1659\relax
\mciteBstWouldAddEndPuncttrue
\mciteSetBstMidEndSepPunct{\mcitedefaultmidpunct}
{\mcitedefaultendpunct}{\mcitedefaultseppunct}\relax
\EndOfBibitem
\bibitem[González and Odloak(2009)González, and Odloak]{Gonzalez2009}
González,~A.~H.; Odloak,~D. {A stable MPC with zone control}. \emph{Journal of
  Process Control} \textbf{2009}, \emph{19}, 110--122\relax
\mciteBstWouldAddEndPuncttrue
\mciteSetBstMidEndSepPunct{\mcitedefaultmidpunct}
{\mcitedefaultendpunct}{\mcitedefaultseppunct}\relax
\EndOfBibitem
\bibitem[Ferramosca \latin{et~al.}(2010)Ferramosca, Limon, González, Odloak,
  and Camacho]{Ferramosca2010}
Ferramosca,~A.; Limon,~D.; González,~A.; Odloak,~D.; Camacho,~E. {MPC for
  tracking zone regions}. \emph{Journal of Process Control} \textbf{2010},
  \emph{20}, 506--516\relax
\mciteBstWouldAddEndPuncttrue
\mciteSetBstMidEndSepPunct{\mcitedefaultmidpunct}
{\mcitedefaultendpunct}{\mcitedefaultseppunct}\relax
\EndOfBibitem
\bibitem[Wills and Heath(2004)Wills, and Heath]{Wills2004}
Wills,~A.~G.; Heath,~W.~P. {Barrier function based model predictive control}.
  \emph{Automatica} \textbf{2004}, \emph{40}, 1415--1422\relax
\mciteBstWouldAddEndPuncttrue
\mciteSetBstMidEndSepPunct{\mcitedefaultmidpunct}
{\mcitedefaultendpunct}{\mcitedefaultseppunct}\relax
\EndOfBibitem
\bibitem[Feller and Ebenbauer(2017)Feller, and Ebenbauer]{Feller2017}
Feller,~C.; Ebenbauer,~C. {Relaxed Logarithmic Barrier Function Based Model
  Predictive Control of Linear Systems}. \emph{IEEE Transactions on Automatic
  Control} \textbf{2017}, \emph{62}, 1223--1238\relax
\mciteBstWouldAddEndPuncttrue
\mciteSetBstMidEndSepPunct{\mcitedefaultmidpunct}
{\mcitedefaultendpunct}{\mcitedefaultseppunct}\relax
\EndOfBibitem
\bibitem[Krishnamoorthy \latin{et~al.}(2019)Krishnamoorthy, Fjalestad, and
  Skogestad]{Krishnamoorthy2019}
Krishnamoorthy,~D.; Fjalestad,~K.; Skogestad,~S. {Optimal operation of oil and
  gas production using simple feedback control structures}. \emph{Control
  Engineering Practice} \textbf{2019}, \emph{91}, 104107\relax
\mciteBstWouldAddEndPuncttrue
\mciteSetBstMidEndSepPunct{\mcitedefaultmidpunct}
{\mcitedefaultendpunct}{\mcitedefaultseppunct}\relax
\EndOfBibitem
\bibitem[Skogestad(2000)]{Skogestad2000}
Skogestad,~S. {Plantwide control: the search for the self-optimizing control
  structure}. \emph{Journal of Process Control} \textbf{2000}, \emph{10},
  487--507\relax
\mciteBstWouldAddEndPuncttrue
\mciteSetBstMidEndSepPunct{\mcitedefaultmidpunct}
{\mcitedefaultendpunct}{\mcitedefaultseppunct}\relax
\EndOfBibitem
\bibitem[Skogestad(2023)]{Skogestad2023}
Skogestad,~S. {Advanced control using decomposition and simple elements}.
  \emph{Annual Reviews in Control} \textbf{2023}, \emph{56}, 100903\relax
\mciteBstWouldAddEndPuncttrue
\mciteSetBstMidEndSepPunct{\mcitedefaultmidpunct}
{\mcitedefaultendpunct}{\mcitedefaultseppunct}\relax
\EndOfBibitem
\bibitem[Engell(2007)]{Engell2007}
Engell,~S. {Feedback control for optimal process operation}. \emph{Journal of
  Process Control} \textbf{2007}, \emph{17}, 203--219\relax
\mciteBstWouldAddEndPuncttrue
\mciteSetBstMidEndSepPunct{\mcitedefaultmidpunct}
{\mcitedefaultendpunct}{\mcitedefaultseppunct}\relax
\EndOfBibitem
\bibitem[Ellis \latin{et~al.}(2014)Ellis, Durand, and Christofides]{Ellis2014}
Ellis,~M.; Durand,~H.; Christofides,~P.~D. {A tutorial review of economic model
  predictive control methods}. \emph{Journal of Process Control} \textbf{2014},
  \emph{24}, 1156--1178\relax
\mciteBstWouldAddEndPuncttrue
\mciteSetBstMidEndSepPunct{\mcitedefaultmidpunct}
{\mcitedefaultendpunct}{\mcitedefaultseppunct}\relax
\EndOfBibitem
\bibitem[Angeli \latin{et~al.}(2012)Angeli, Amrit, and Rawlings]{Angeli2012}
Angeli,~D.; Amrit,~R.; Rawlings,~J.~B. {On Average Performance and Stability of
  Economic Model Predictive Control}. \emph{IEEE Transactions on Automatic
  Control} \textbf{2012}, \emph{57}, 1615--1626\relax
\mciteBstWouldAddEndPuncttrue
\mciteSetBstMidEndSepPunct{\mcitedefaultmidpunct}
{\mcitedefaultendpunct}{\mcitedefaultseppunct}\relax
\EndOfBibitem
\bibitem[Martins \latin{et~al.}(2021)Martins, Rodrigues, Loureiro, Ribeiro, and
  Nogueira]{Martins2021}
Martins,~M.~A.; Rodrigues,~A.~E.; Loureiro,~J.~M.; Ribeiro,~A.~M.;
  Nogueira,~I.~B. {Artificial Intelligence-oriented economic non-linear model
  predictive control applied to a pressure swing adsorption unit: Syngas
  purification as a case study}. \emph{Separation and Purification Technology}
  \textbf{2021}, \emph{276}, 119333\relax
\mciteBstWouldAddEndPuncttrue
\mciteSetBstMidEndSepPunct{\mcitedefaultmidpunct}
{\mcitedefaultendpunct}{\mcitedefaultseppunct}\relax
\EndOfBibitem
\bibitem[Darby \latin{et~al.}(2011)Darby, Nikolaou, Jones, and
  Nicholson]{Darby2011}
Darby,~M.~L.; Nikolaou,~M.; Jones,~J.; Nicholson,~D. {RTO: An overview and
  assessment of current practice}. \emph{Journal of Process Control}
  \textbf{2011}, \emph{21}, 874--884\relax
\mciteBstWouldAddEndPuncttrue
\mciteSetBstMidEndSepPunct{\mcitedefaultmidpunct}
{\mcitedefaultendpunct}{\mcitedefaultseppunct}\relax
\EndOfBibitem
\bibitem[Bikmukhametov and Jäschke(2020)Bikmukhametov, and
  Jäschke]{Bikmukhametov2020}
Bikmukhametov,~T.; Jäschke,~J. {First Principles and Machine Learning Virtual
  Flow Metering: A Literature Review}. \emph{Journal of Petroleum Science and
  Engineering} \textbf{2020}, \emph{184}, 106487\relax
\mciteBstWouldAddEndPuncttrue
\mciteSetBstMidEndSepPunct{\mcitedefaultmidpunct}
{\mcitedefaultendpunct}{\mcitedefaultseppunct}\relax
\EndOfBibitem
\bibitem[Grimstad \latin{et~al.}(2021)Grimstad, Hotvedt, Sandnes, Kolbjørnsen,
  and Imsland]{Grimstad2021}
Grimstad,~B.; Hotvedt,~M.; Sandnes,~A.~T.; Kolbjørnsen,~O.; Imsland,~L.~S.
  {Bayesian neural networks for virtual flow metering: An empirical study}.
  \emph{Applied Soft Computing} \textbf{2021}, \emph{112}, 107776\relax
\mciteBstWouldAddEndPuncttrue
\mciteSetBstMidEndSepPunct{\mcitedefaultmidpunct}
{\mcitedefaultendpunct}{\mcitedefaultseppunct}\relax
\EndOfBibitem
\bibitem[Hotvedt \latin{et~al.}(2022)Hotvedt, Grimstad, Ljungquist, and
  Imsland]{Hotvedt2022}
Hotvedt,~M.; Grimstad,~B.; Ljungquist,~D.; Imsland,~L. {On gray-box modeling
  for virtual flow metering}. \emph{Control Engineering Practice}
  \textbf{2022}, \emph{118}, 104974\relax
\mciteBstWouldAddEndPuncttrue
\mciteSetBstMidEndSepPunct{\mcitedefaultmidpunct}
{\mcitedefaultendpunct}{\mcitedefaultseppunct}\relax
\EndOfBibitem
\bibitem[Binder \latin{et~al.}(2015)Binder, Pavlov, and Johansen]{Binder2015}
Binder,~B.~J.; Pavlov,~A.; Johansen,~T.~A. {Estimation of Flow Rate and
  Viscosity in a Well with an Electric Submersible Pump using Moving Horizon
  Estimation}. \emph{IFAC-PapersOnLine} \textbf{2015}, \emph{48},
  140--146\relax
\mciteBstWouldAddEndPuncttrue
\mciteSetBstMidEndSepPunct{\mcitedefaultmidpunct}
{\mcitedefaultendpunct}{\mcitedefaultseppunct}\relax
\EndOfBibitem
\bibitem[Thornhill \latin{et~al.}(2004)Thornhill, Shoukat~Choudhury, and
  Shah]{Thornhill2004}
Thornhill,~N.~F.; Shoukat~Choudhury,~M.; Shah,~S.~L. {The impact of compression
  on data-driven process analyses}. \emph{Journal of Process Control}
  \textbf{2004}, \emph{14}, 389--398\relax
\mciteBstWouldAddEndPuncttrue
\mciteSetBstMidEndSepPunct{\mcitedefaultmidpunct}
{\mcitedefaultendpunct}{\mcitedefaultseppunct}\relax
\EndOfBibitem
\bibitem[Badgwell \latin{et~al.}(2018)Badgwell, Lee, and Liu]{Badgwell2018}
Badgwell,~T.~A.; Lee,~J.~H.; Liu,~K.-H. {Reinforcement Learning -- Overview of
  Recent Progress and Implications for Process Control}. In \emph{13th
  International Symposium on Process Systems Engineering (PSE 2018)}; Elsevier,
  2018; pp 71--85\relax
\mciteBstWouldAddEndPuncttrue
\mciteSetBstMidEndSepPunct{\mcitedefaultmidpunct}
{\mcitedefaultendpunct}{\mcitedefaultseppunct}\relax
\EndOfBibitem
\bibitem[Nian \latin{et~al.}(2020)Nian, Liu, and Huang]{Nian2020}
Nian,~R.; Liu,~J.; Huang,~B. {A review On reinforcement learning: Introduction
  and applications in industrial process control}. \emph{Computers \& Chemical
  Engineering} \textbf{2020}, \emph{139}, 106886\relax
\mciteBstWouldAddEndPuncttrue
\mciteSetBstMidEndSepPunct{\mcitedefaultmidpunct}
{\mcitedefaultendpunct}{\mcitedefaultseppunct}\relax
\EndOfBibitem
\bibitem[Spielberg \latin{et~al.}(2019)Spielberg, Tulsyan, Lawrence, Loewen,
  and Gopaluni]{Spielberg2019}
Spielberg,~S.; Tulsyan,~A.; Lawrence,~N.~P.; Loewen,~P.~D.; Gopaluni,~R.~B.
  {Toward self-driving processes: A deep reinforcement learning approach to
  control}. \emph{AIChE Journal} \textbf{2019}, \emph{65}, e16689\relax
\mciteBstWouldAddEndPuncttrue
\mciteSetBstMidEndSepPunct{\mcitedefaultmidpunct}
{\mcitedefaultendpunct}{\mcitedefaultseppunct}\relax
\EndOfBibitem
\bibitem[Bloor \latin{et~al.}(2025)Bloor, Ahmed, Kotecha, Mercangöz, Tsay, and
  del Río-Chanona]{Bloor2025}
Bloor,~M.; Ahmed,~A.; Kotecha,~N.; Mercangöz,~M.; Tsay,~C.; del
  Río-Chanona,~E.~A. {Control-Informed Reinforcement Learning for Chemical
  Processes}. \emph{Industrial \& Engineering Chemistry Research}
  \textbf{2025}, \emph{64}, 4966--4978\relax
\mciteBstWouldAddEndPuncttrue
\mciteSetBstMidEndSepPunct{\mcitedefaultmidpunct}
{\mcitedefaultendpunct}{\mcitedefaultseppunct}\relax
\EndOfBibitem
\bibitem[Psichogios and Ungar(1992)Psichogios, and Ungar]{Psichogios1992}
Psichogios,~D.~C.; Ungar,~L.~H. {A hybrid neural network-first principles
  approach to process modeling}. \emph{AIChE Journal} \textbf{1992}, \emph{38},
  1499--1511\relax
\mciteBstWouldAddEndPuncttrue
\mciteSetBstMidEndSepPunct{\mcitedefaultmidpunct}
{\mcitedefaultendpunct}{\mcitedefaultseppunct}\relax
\EndOfBibitem
\bibitem[Thompson and Kramer(1994)Thompson, and Kramer]{Thompson1994}
Thompson,~M.~L.; Kramer,~M.~A. {Modeling chemical processes using prior
  knowledge and neural networks}. \emph{AIChE Journal} \textbf{1994},
  \emph{40}, 1328--1340\relax
\mciteBstWouldAddEndPuncttrue
\mciteSetBstMidEndSepPunct{\mcitedefaultmidpunct}
{\mcitedefaultendpunct}{\mcitedefaultseppunct}\relax
\EndOfBibitem
\bibitem[von Stosch \latin{et~al.}(2014)von Stosch, Oliveira, Peres, and
  Feyo~de Azevedo]{vonStosch2014}
von Stosch,~M.; Oliveira,~R.; Peres,~J.; Feyo~de Azevedo,~S. {Hybrid
  semi-parametric modeling in process systems engineering: Past, present and
  future}. \emph{Computers \& Chemical Engineering} \textbf{2014}, \emph{60},
  86--101\relax
\mciteBstWouldAddEndPuncttrue
\mciteSetBstMidEndSepPunct{\mcitedefaultmidpunct}
{\mcitedefaultendpunct}{\mcitedefaultseppunct}\relax
\EndOfBibitem
\bibitem[Schweidtmann \latin{et~al.}(2024)Schweidtmann, Zhang, and von
  Stosch]{Schweidtmann2024}
Schweidtmann,~A.~M.; Zhang,~D.; von Stosch,~M. {A review and perspective on
  hybrid modeling methodologies}. \emph{Digital Chemical Engineering}
  \textbf{2024}, \emph{10}, 100136\relax
\mciteBstWouldAddEndPuncttrue
\mciteSetBstMidEndSepPunct{\mcitedefaultmidpunct}
{\mcitedefaultendpunct}{\mcitedefaultseppunct}\relax
\EndOfBibitem
\bibitem[Costa \latin{et~al.}(2024)Costa, Rebello, Santana, and
  Nogueira]{Costa2024multistep}
Costa,~E.~A.; Rebello,~C.~M.; Santana,~V.~V.; Nogueira,~I.~B. {Machine learning
  multi-step-ahead modelling with uncertainty assessment}.
  \emph{IFAC-PapersOnLine} \textbf{2024}, \emph{58}, 25--30\relax
\mciteBstWouldAddEndPuncttrue
\mciteSetBstMidEndSepPunct{\mcitedefaultmidpunct}
{\mcitedefaultendpunct}{\mcitedefaultseppunct}\relax
\EndOfBibitem
\bibitem[Costa \latin{et~al.}(2023)Costa, Rebello, Fontana, Schnitman, and
  Nogueira]{Costa2022robust}
Costa,~E.~A.; Rebello,~C. d.~M.; Fontana,~M.; Schnitman,~L.; Nogueira,~I. B.
  d.~R. {A Robust Learning Methodology for Uncertainty-Aware Scientific Machine
  Learning Models}. \emph{Mathematics} \textbf{2023}, \emph{11}, 74\relax
\mciteBstWouldAddEndPuncttrue
\mciteSetBstMidEndSepPunct{\mcitedefaultmidpunct}
{\mcitedefaultendpunct}{\mcitedefaultseppunct}\relax
\EndOfBibitem
\bibitem[Nogueira \latin{et~al.}(2022)Nogueira, Santana, Ribeiro, and
  Rodrigues]{Nogueira2022UDE}
Nogueira,~I. B.~R.; Santana,~V.~V.; Ribeiro,~A.~M.; Rodrigues,~A.~E. {Using
  scientific machine learning to develop universal differential equation for
  multicomponent adsorption separation systems}. \emph{The Canadian Journal of
  Chemical Engineering} \textbf{2022}, \emph{100}, 2279--2290\relax
\mciteBstWouldAddEndPuncttrue
\mciteSetBstMidEndSepPunct{\mcitedefaultmidpunct}
{\mcitedefaultendpunct}{\mcitedefaultseppunct}\relax
\EndOfBibitem
\bibitem[Lima \latin{et~al.}(2023)Lima, Rebello, Costa, Santana, de~Moares,
  Barreto, Secchi, de~Souza, and Nogueira]{Lima2023UDE}
Lima,~F. A.~R.; Rebello,~C.~M.; Costa,~E.~A.; Santana,~V.~V.; de~Moares,~M.~G.;
  Barreto,~A.~G.; Secchi,~A.~R.; de~Souza,~M.~B.; Nogueira,~I.~B. {Improved
  modeling of crystallization processes by Universal Differential Equations}.
  \emph{Chemical Engineering Research and Design} \textbf{2023}, \emph{200},
  538--549\relax
\mciteBstWouldAddEndPuncttrue
\mciteSetBstMidEndSepPunct{\mcitedefaultmidpunct}
{\mcitedefaultendpunct}{\mcitedefaultseppunct}\relax
\EndOfBibitem
\bibitem[Rebello \latin{et~al.}(2024)Rebello, Costa, Fontana, Schnitman, and
  Nogueira]{Rebello2024interp}
Rebello,~C.~M.; Costa,~E.~A.; Fontana,~M.; Schnitman,~L.; Nogueira,~I. B.~R.
  {Interpretable Scientific Machine Learning Approach for Correcting
  Phenomenological Models: Methodology Validation on an ESP Prototype}.
  \emph{Industrial \& Engineering Chemistry Research} \textbf{2024}, \emph{63},
  19030--19050\relax
\mciteBstWouldAddEndPuncttrue
\mciteSetBstMidEndSepPunct{\mcitedefaultmidpunct}
{\mcitedefaultendpunct}{\mcitedefaultseppunct}\relax
\EndOfBibitem
\bibitem[Bemporad \latin{et~al.}(2002)Bemporad, Morari, Dua, and
  Pistikopoulos]{Bemporad2002}
Bemporad,~A.; Morari,~M.; Dua,~V.; Pistikopoulos,~E.~N. {The explicit linear
  quadratic regulator for constrained systems}. \emph{Automatica}
  \textbf{2002}, \emph{38}, 3--20\relax
\mciteBstWouldAddEndPuncttrue
\mciteSetBstMidEndSepPunct{\mcitedefaultmidpunct}
{\mcitedefaultendpunct}{\mcitedefaultseppunct}\relax
\EndOfBibitem
\bibitem[Karg and Lucia(2020)Karg, and Lucia]{Karg2020}
Karg,~B.; Lucia,~S. {Efficient Representation and Approximation of Model
  Predictive Control Laws via Deep Learning}. \emph{IEEE Transactions on
  Cybernetics} \textbf{2020}, \emph{50}, 3866--3878\relax
\mciteBstWouldAddEndPuncttrue
\mciteSetBstMidEndSepPunct{\mcitedefaultmidpunct}
{\mcitedefaultendpunct}{\mcitedefaultseppunct}\relax
\EndOfBibitem
\bibitem[Mendes \latin{et~al.}(2022)Mendes, Schnitman, Nogueira, Ribeiro,
  Rodrigues, Loureiro, and Martins]{Mendes2022TSK}
Mendes,~T. P.~G.; Schnitman,~L.; Nogueira,~I. B.~R.; Ribeiro,~A. M. A.~P.;
  Rodrigues,~A.~E.; Loureiro,~J.~M.; Martins,~M. A.~F. {A new
  Takagi-Sugeno-Kang model-based stabilizing explicit MPC formulation: An
  experimental case study with implementation embedded in a PLC}. \emph{Expert
  Systems with Applications} \textbf{2022}, \emph{210}, 118369\relax
\mciteBstWouldAddEndPuncttrue
\mciteSetBstMidEndSepPunct{\mcitedefaultmidpunct}
{\mcitedefaultendpunct}{\mcitedefaultseppunct}\relax
\EndOfBibitem
\bibitem[Queiroz \latin{et~al.}(2023)Queiroz, Rebello, Costa, Santana,
  Rodrigues, Rodrigues, Ribeiro, and Nogueira]{Queiroz2023RL}
Queiroz,~L.~P.; Rebello,~C.~M.; Costa,~E.~A.; Santana,~V.~V.; Rodrigues,~B.
  C.~L.; Rodrigues,~A.~E.; Ribeiro,~A.~M.; Nogueira,~I. B.~R. {A Reinforcement
  Learning Framework to Discover Natural Flavor Molecules}. \emph{Foods}
  \textbf{2023}, \emph{12}, 1147\relax
\mciteBstWouldAddEndPuncttrue
\mciteSetBstMidEndSepPunct{\mcitedefaultmidpunct}
{\mcitedefaultendpunct}{\mcitedefaultseppunct}\relax
\EndOfBibitem
\bibitem[Nogueira \latin{et~al.}(2019)Nogueira, Martins, Requião, Oliveira,
  Viena, Koivisto, Rodrigues, Loureiro, and Ribeiro]{Nogueira2019SPSO}
Nogueira,~I.~B.; Martins,~M.~A.; Requião,~R.; Oliveira,~A.~R.; Viena,~V.;
  Koivisto,~H.; Rodrigues,~A.~E.; Loureiro,~J.~M.; Ribeiro,~A.~M. {Optimization
  of a True Moving Bed unit and determination of its feasible operating region
  using a novel Sliding Particle Swarm Optimization}. \emph{Computers \&
  Industrial Engineering} \textbf{2019}, \emph{135}, 368--381\relax
\mciteBstWouldAddEndPuncttrue
\mciteSetBstMidEndSepPunct{\mcitedefaultmidpunct}
{\mcitedefaultendpunct}{\mcitedefaultseppunct}\relax
\EndOfBibitem
\bibitem[Nogueira \latin{et~al.}(2022)Nogueira, Dias, Rebello, Costa, Santana,
  Rodrigues, Ferreira, and Ribeiro]{Nogueira2022nested}
Nogueira,~I.~B.; Dias,~R.~O.; Rebello,~C.~M.; Costa,~E.~A.; Santana,~V.~V.;
  Rodrigues,~A.~E.; Ferreira,~A.; Ribeiro,~A.~M. {A novel nested loop
  optimization problem based on deep neural networks and feasible operation
  regions definition for simultaneous material screening and process
  optimization}. \emph{Chemical Engineering Research and Design} \textbf{2022},
  \emph{180}, 243--253\relax
\mciteBstWouldAddEndPuncttrue
\mciteSetBstMidEndSepPunct{\mcitedefaultmidpunct}
{\mcitedefaultendpunct}{\mcitedefaultseppunct}\relax
\EndOfBibitem
\bibitem[Sánchez \latin{et~al.}(2020)Sánchez, Oliveira-Esquerre, dos
  Reis~Nogueira, de~Jong, and Filho]{Sanchez2020water}
Sánchez,~A.~S.; Oliveira-Esquerre,~K.~P.; dos Reis~Nogueira,~I.~B.;
  de~Jong,~P.; Filho,~A.~A. {Water Loss Management Through Smart Water
  Systems}. In \emph{Smart Village Technology}; Springer International
  Publishing, 2020; pp 233--266\relax
\mciteBstWouldAddEndPuncttrue
\mciteSetBstMidEndSepPunct{\mcitedefaultmidpunct}
{\mcitedefaultendpunct}{\mcitedefaultseppunct}\relax
\EndOfBibitem
\bibitem[Leontaritis and Billings(1985)Leontaritis, and
  Billings]{Leontaritis1985}
Leontaritis,~I.~J.; Billings,~S.~A. {Input-output parametric models for
  non-linear systems Part I: deterministic non-linear systems}.
  \emph{International Journal of Control} \textbf{1985}, \emph{41},
  303--328\relax
\mciteBstWouldAddEndPuncttrue
\mciteSetBstMidEndSepPunct{\mcitedefaultmidpunct}
{\mcitedefaultendpunct}{\mcitedefaultseppunct}\relax
\EndOfBibitem
\bibitem[Ljung(1998)]{ljung1998system}
Ljung,~L. \emph{{System Identification: Theory for the User}}; Pearson
  Education, 1998\relax
\mciteBstWouldAddEndPuncttrue
\mciteSetBstMidEndSepPunct{\mcitedefaultmidpunct}
{\mcitedefaultendpunct}{\mcitedefaultseppunct}\relax
\EndOfBibitem
\bibitem[Keesman(2011)]{keesman2011system}
Keesman,~K.~J. \emph{{System Identification: An Introduction}}; Advanced
  Textbooks in Control and Signal Processing; Springer: London, 2011\relax
\mciteBstWouldAddEndPuncttrue
\mciteSetBstMidEndSepPunct{\mcitedefaultmidpunct}
{\mcitedefaultendpunct}{\mcitedefaultseppunct}\relax
\EndOfBibitem
\bibitem[He and Asada(1993)He, and Asada]{He1993}
He,~X.; Asada,~H. {A New Method for Identifying Orders of Input-Output Models
  for Nonlinear Dynamic Systems}. 1993 American Control Conference. 1993; pp
  2520--2523\relax
\mciteBstWouldAddEndPuncttrue
\mciteSetBstMidEndSepPunct{\mcitedefaultmidpunct}
{\mcitedefaultendpunct}{\mcitedefaultseppunct}\relax
\EndOfBibitem
\bibitem[Lukoševičius(2012)]{Lukosevicius2012}
Lukoševičius,~M. {A Practical Guide to Applying Echo State Networks}. In
  \emph{Neural Networks: Tricks of the Trade}; Springer Berlin Heidelberg,
  2012; pp 659--686\relax
\mciteBstWouldAddEndPuncttrue
\mciteSetBstMidEndSepPunct{\mcitedefaultmidpunct}
{\mcitedefaultendpunct}{\mcitedefaultseppunct}\relax
\EndOfBibitem
\bibitem[Yildiz \latin{et~al.}(2012)Yildiz, Jaeger, and Kiebel]{Yildiz2012}
Yildiz,~I.~B.; Jaeger,~H.; Kiebel,~S.~J. {Re-visiting the echo state property}.
  \emph{Neural Networks} \textbf{2012}, \emph{35}, 1--9\relax
\mciteBstWouldAddEndPuncttrue
\mciteSetBstMidEndSepPunct{\mcitedefaultmidpunct}
{\mcitedefaultendpunct}{\mcitedefaultseppunct}\relax
\EndOfBibitem
\bibitem[Hochreiter and Schmidhuber(1997)Hochreiter, and
  Schmidhuber]{Hochreiter1997}
Hochreiter,~S.; Schmidhuber,~J. {Long Short-Term Memory}. \emph{Neural
  Computation} \textbf{1997}, \emph{9}, 1735--1780\relax
\mciteBstWouldAddEndPuncttrue
\mciteSetBstMidEndSepPunct{\mcitedefaultmidpunct}
{\mcitedefaultendpunct}{\mcitedefaultseppunct}\relax
\EndOfBibitem
\bibitem[Andersson \latin{et~al.}(2019)Andersson, Gillis, Horn, Rawlings, and
  Diehl]{Andersson2019}
Andersson,~J. A.~E.; Gillis,~J.; Horn,~G.; Rawlings,~J.~B.; Diehl,~M. {CasADi:
  a software framework for nonlinear optimization and optimal control}.
  \emph{Mathematical Programming Computation} \textbf{2019}, \emph{11},
  1--36\relax
\mciteBstWouldAddEndPuncttrue
\mciteSetBstMidEndSepPunct{\mcitedefaultmidpunct}
{\mcitedefaultendpunct}{\mcitedefaultseppunct}\relax
\EndOfBibitem
\bibitem[Wächter and Biegler(2006)Wächter, and Biegler]{Wachter2006}
Wächter,~A.; Biegler,~L.~T. {On the implementation of an interior-point filter
  line-search algorithm for large-scale nonlinear programming}.
  \emph{Mathematical Programming} \textbf{2006}, \emph{106}, 25--57\relax
\mciteBstWouldAddEndPuncttrue
\mciteSetBstMidEndSepPunct{\mcitedefaultmidpunct}
{\mcitedefaultendpunct}{\mcitedefaultseppunct}\relax
\EndOfBibitem
\bibitem[Rebello and Nogueira(2025)Rebello, and Nogueira]{Rebello2025DT}
Rebello,~C.~M.; Nogueira,~I.~B. {Digital twins in chemical engineering: An
  integrated framework for identification, implementation, online learning, and
  uncertainty assessment}. \emph{Computers \& Chemical Engineering}
  \textbf{2025}, \emph{200}, 109178\relax
\mciteBstWouldAddEndPuncttrue
\mciteSetBstMidEndSepPunct{\mcitedefaultmidpunct}
{\mcitedefaultendpunct}{\mcitedefaultseppunct}\relax
\EndOfBibitem
\end{mcitethebibliography}

\clearpage
\section*{Supporting Information}
\setcounter{section}{0}\setcounter{table}{0}\setcounter{figure}{0}\setcounter{algorithm}{0}
\renewcommand{\thesection}{S\arabic{section}}
\renewcommand{\thetable}{S\arabic{table}}
\renewcommand{\thefigure}{S\arabic{figure}}
\renewcommand{\thealgorithm}{S\arabic{algorithm}}
\renewcommand{\thepage}{S\arabic{page}}

\FloatBarrier
\section{Dynamic constraint tables}
\label{app:constraints}
\begin{table}[htbp]
  \centering
  \begin{threeparttable}
  \caption{Dynamic hard-constraint bands (min--max normalised) at representative set-point frequencies.}
  \label{tab:constraints}
  \begin{tabular}{l l l l}
\toprule
Variable & 55 Hz & 56 Hz & 57 Hz \\
\midrule
Intake pressure & -0.11--0.30 & -0.11--0.30 & -0.11--0.30 \\
Topside pressure & 0.27--1.09 & 0.30--1.09 & 0.33--1.09 \\
Differential pressure & 0.42--1.19 & 0.42--1.19 & 0.42--1.19 \\
Discharge pressure & 0.25--1.08 & 0.25--1.08 & 0.25--1.08 \\
Motor temperature & 0.05--1.02 & 0.05--1.02 & 0.05--1.02 \\
Torque current & 0.70--0.97 & 0.72--1.00 & 0.75--1.03 \\
Total current & 0.10--1.02 & 0.10--1.02 & 0.10--1.02 \\
Intake temperature & -0.15--2.87 & -0.15--2.87 & -0.15--2.87 \\
Vibration & 0.08--1.67 & 0.08--1.67 & 0.08--1.67 \\
Topside temperature & 0.52--1.39 & 0.52--1.39 & 0.52--1.39 \\
\bottomrule
\end{tabular}

  \begin{tablenotes}\footnotesize
  \item Each cell is the normalised lower--upper bound, with $0$ and $1$ the normalisation limits of each variable; alarm limits can lie outside that interval. Absolute Petrobras protection limits are withheld.
  \end{tablenotes}
  \end{threeparttable}
\end{table}

\FloatBarrier
\section{Selected ESN hyperparameters}
\label{app:esn}
\begin{table}[htbp]
  \centering
  \caption{Hyperparameters of the selected Echo State Network.}
  \label{tab:esnhyper}
  \begin{tabular}{l l}
\toprule
Hyperparameter & Value \\
\midrule
Reservoir size $N$ & 100 \\
Leak rate $\gamma$ & 0.12 \\
Spectral radius $\rho$ & 1.03 \\
Input scaling & 1.0 \\
Ridge regularisation $\lambda$ & 1e-6 \\
Washout & 300 samples ($\approx$1 h) \\
\bottomrule
\end{tabular}

\end{table}

\FloatBarrier
\section{Screening error metrics}
\label{app:metrics}
\begin{table}[htbp]
  \centering
  \begin{threeparttable}
  \caption{The twenty error metrics computed for every candidate model.}
  \label{tab:errmetrics}
  \begin{tabular}{@{}l p{6.4cm} l@{}}
\toprule
Metric & Definition & Unit \\
\midrule
Minimum error & $\min_k e_k$ & EU \\
Maximum error & $\max_k e_k$ & EU \\
Mean error & $\frac{1}{N}\sum_k e_k$ & EU \\
Mean absolute error & $\frac{1}{N}\sum_k |e_k|$ & EU \\
Mean squared error & $\frac{1}{N}\sum_k e_k^2$ & EU$^2$ \\
Root mean squared error & $\sqrt{\frac{1}{N}\sum_k e_k^2}$ & EU \\
Mean percentage error & $\frac{100}{N}\sum_k e_k/y_k$ & \% \\
Mean absolute percentage error & $\min\!\left(100,\ \frac{100}{N}\sum_k |e_k/y_k|\right)$ & \% \\
Symmetric MAPE & $\frac{100}{N}\sum_k \frac{|e_k|}{(|y_k|+|\hat y_k|)/2}$ & \% \\
Range-to-mean ratio & $(\max_k \hat y_k-\min_k \hat y_k)/\overline{\hat y}$ & -- \\
Error standard deviation & $\sigma(e)$ & EU \\
Error variance & $\sigma^2(e)$ & EU$^2$ \\
Error mode & $\mathrm{mode}(e)$ & EU \\
Error median & $\mathrm{median}(e)$ & EU \\
Time above 5\% error & share of samples with $|e_k/y_k|>5\,\%$ & \% \\
Time above 3\% error & share of samples with $|e_k/y_k|>3\,\%$ & \% \\
Time above 1\% error & share of samples with $|e_k/y_k|>1\,\%$ & \% \\
Longest run above 5\% & longest uninterrupted exceedance of $5\,\%$ & \% \\
Longest run above 3\% & longest uninterrupted exceedance of $3\,\%$ & \% \\
Longest run above 1\% & longest uninterrupted exceedance of $1\,\%$ & \% \\
\bottomrule
\end{tabular}

  \begin{tablenotes}\footnotesize
  \item $e_k = y_k - \hat y_k$ is the prediction error at sample $k$, $N$ the number of samples, and EU the engineering unit of the variable. Only the MAPE was used for ranking (Section~\ref{sec:eval:metrics}).
  \end{tablenotes}
  \end{threeparttable}
\end{table}

\FloatBarrier
\section{NLP pseudocode and solver parameters}
\label{app:nlp}

Algorithm~\ref{alg:mpc} states the receding-horizon loop as implemented. The inference
task runs every $T_s = \SI{10}{s}$ and the optimisation task every $3T_s = \SI{30}{s}$, so
one optimisation consumes three inference samples, which is also the width of the
move-blocking packet of Eq.~\eqref{eq:c-block}.

\begin{algorithm}[htbp]
\caption{Multi-rate DDMPC loop as implemented in the field.}
\label{alg:mpc}
\begin{algorithmic}[1]
\State \textbf{Offline:} load ESN weights $(W^{\mathrm{in}},W^{\mathrm{res}},W^{\mathrm{out}})$, flow and thrust maps, protection tables; build the CasADi NLP once and keep the solver object
\State \textbf{Start-up:} run the reservoir for 300 samples of measured data (washout) so that the internal state is synchronised with the plant
\State $\mathbf{u}^{\star} \gets$ current plant inputs; $\hat{\mathbf{x}}^{\star} \gets$ current measurements repeated over the horizon
\For{each inference instant $k$}
  \State read the 10 measured channels; estimate $\hat Q$ from the flow map
  \State propagate the reservoir one step and store $\hat x(k{+}1|k)$
  \If{$k \bmod 3 = 0$} \Comment{supervisory instant}
    \State $x(k) \gets$ mean of the three most recent samples \Comment{moving-average filter}
    \State $e_x(k) \gets x(k) - \hat x(k|k{-}1)$; \ $e_y(k) \gets H e_x(k)$
    \State recompute the H/L limits at the current frequency
    \State shrink each limit by the per-variable margin $m$ of Table~\ref{tab:margins}
    \State $y_{\mathrm{sp}} \gets [\,p_{\mathrm{eng}},\ Q(f_{\mathrm{eng}},p_{\mathrm{eng}})\,]^{\top}$ from the map
    \State warm start: shift $(\hat{\mathbf{x}}^{\star},\mathbf{u}^{\star})$ one step, append the last element, set $\Delta\mathbf{u} \gets 0$, overwrite $\hat x(k)$ with $x(k)$
    \State solve \eqref{eq:cost}--\eqref{eq:c-ptarget} with IPOPT under the wall-time cap
    \If{converged or acceptable point reached}
      \State $(\hat{\mathbf{x}}^{\star},\mathbf{u}^{\star}) \gets$ solution; $u_{\mathrm{cmd}} \gets u^{\star}(k)$
    \Else
      \State hold the previous command \Comment{previous set-points remain in force}
    \EndIf
    \If{$|u_{\mathrm{cmd}} - u(k{-}1)| >$ actuation dead band}
      \State send $u_{\mathrm{cmd}}$ to the regulatory layer through the MPA
    \EndIf
  \EndIf
\EndFor
\end{algorithmic}
\end{algorithm}

Two details of the loop are easy to miss in the equations. The warm start overwrites the
first state of the shifted trajectory with the current filtered measurement, so the initial
condition of Eq.~\eqref{eq:c-model} is always the plant and never the previous prediction.
And a failed instance is not retried within the same supervisory period. The previous action
is held and the next attempt happens at the following cycle.

Table~\ref{tab:solver} lists the solver settings and the remaining implementation
parameters. The tolerances are loose for an interior-point method. The
objective is expressed in normalised units and the decision variables are min--max scaled,
so a tight tolerance would spend iterations refining an operating point far inside the
uncertainty of the flow estimate itself. The acceptable-point criteria exist so that a
usable solution is returned when the wall-time cap bites, rather than nothing at all.

\begin{table}[htbp]
  \centering
  \begin{threeparttable}
  \caption{Solver settings and implementation parameters.}
  \label{tab:solver}
  \begin{tabular}{l p{6.9cm}}
  \toprule
  Parameter & Value \\
  \midrule
  Transcription & multiple shooting: $\hat{\mathbf x}$, $\mathbf u$ and $\Delta\mathbf u$ are decision variables \\
  Framework / solver & CasADi with IPOPT \\
  Hessian & limited-memory (L-BFGS) approximation \\
  Barrier strategy & monotone, initialised at a small $\mu$ \\
  \texttt{max\_wall\_time} & \SI{9}{s} \\
  \texttt{tol} / \texttt{acceptable\_tol} & 0.4 / 0.6 \\
  \texttt{constr\_viol\_tol} & 0.01 \\
  \texttt{acceptable\_constr\_viol\_tol} & 0.01 to 0.10 (swept) \\
  \texttt{bound\_relax\_factor} & $10^{-3}$ \\
  \midrule
  Inference / optimisation period & \SI{10}{s} / \SI{30}{s} \\
  Move blocking & packets of 3 samples \\
  Input bounds $u_{\min}$, $u_{\max}$ & \SIrange{40}{60}{Hz}, \SIrange{20}{50}{bar} \\
  Move bounds $\Delta u_{\max}$ & \SI{0.25}{Hz}, \SI{2}{bar} per supervisory step \\
  VSD sliding window & \SI{1}{Hz} over $N_{\mathrm{win}} = 45$ samples (\SI{7.5}{min}) \\
  Pressure-target band & $\pm\SI{6}{bar}$ around the measured arrival pressure \\
  Proximity penalty $\Phi$ & $0.5973$, $-2.3091$, $3.8479$, $-3.1350$, $0.9994$ \\
  Reservoir washout & 300 samples ($\approx$\SI{1}{h}) \\
  \bottomrule
  \end{tabular}
  \end{threeparttable}
\end{table}

\FloatBarrier

\end{document}